\documentclass[fleqn,usenatbib]{mnras}

\usepackage{amsmath}
\usepackage[no-math]{fontspec}

\IfFontExistsTF{texgyretermes-regular.otf}{
    \setmainfont{texgyretermes-regular.otf}[
        BoldFont       = texgyretermes-bold.otf,
        ItalicFont     = texgyretermes-italic.otf,
        BoldItalicFont = texgyretermes-bolditalic.otf,
        Ligatures      = TeX
    ]
}{
    \setmainfont{NimbusRoman-Regular.otf}[
        BoldFont       = NimbusRoman-Bold.otf,
        ItalicFont     = NimbusRoman-Italic.otf,
        BoldItalicFont = NimbusRoman-BoldItalic.otf,
        Ligatures      = TeX
    ]
}

\usepackage{newtxmath}

\usepackage{xeCJK}

\newCJKfontfamily\zhname{gbsn00lp.ttf}

\newCJKfontfamily\zhrare{bsmi00lp.ttf}

\DeclareRobustCommand{\cjkname}[2]{%
    #1\space({\zhname #2})%
}

\DeclareRobustCommand{\diliname}{%
    Di Li\space({\zhname 李}{\zhrare 菂})%
}

\usepackage{graphicx}
\usepackage{xcolor}
\usepackage{pdflscape}
\usepackage{array}
\usepackage{capt-of}
\usepackage{wrapfig}
\usepackage{xspace}
\usepackage{siunitx}
\usepackage[normalem]{ulem}
\usepackage{float}
\usepackage{microtype}
\usepackage{booktabs}
\usepackage{multirow}
\usepackage{longtable}
\usepackage[version=4]{mhchem}

\DeclareRobustCommand{\bluecorrstar}{%
    \textsuperscript{\textcolor{blue}{$\star$}}%
}

\hypersetup{
    colorlinks=true,
    linkcolor=blue,
    citecolor=blue,
    urlcolor=blue,
    pdfauthor={Jinjin Xie et al.}
}

\begin{document}

\title[Infall Studies towards 56 ALOHA IRDCs]
{ALOHA IRDCs molecular line follow-up: II. Multi-tracer infall profile parameters}

\author[Xie et al.]{%
\cjkname{Jinjin Xie}{谢津津}$^{1}$,
Jarken Esimbek$^{1,2,3}$\thanks{E-mail: dili@mail.tsinghua.edu.cn (DL); jarken@xao.ac.cn (JE); g.fuller@manchester.ac.uk (GAF)},
\cjkname{Qian-Ru He}{何茜茹}$^{4,5}$,
\cjkname{Yan Duan}{段言}$^{6}$,
\newauthor
Gary A. Fuller$^{7,8}$\bluecorrstar,
Nicolas Peretto$^{9}$,
Glenn J. White$^{10,11}$,
\cjkname{Xue-Peng Chen}{陈学鹏}$^{5,6}$,
\newauthor
Chang Won Lee$^{12,13}$,
Ruili Wang$^{5,6}$,
Xindi Tang$^{1,2,3}$,
Wenjin Yang$^{14}$,
Christian Henkel$^{15}$,
\newauthor
Zhiyuan Ren$^{16}$,
Hauyu Baobab Liu$^{17,18}$,
Long-Fei Chen$^{19}$,
Willem A. Baan$^{1,20}$,
David Eden$^{21,22}$,
\newauthor
Jianjun Zhou$^{1,2,3}$,
Ken'ichi Tatematsu$^{23,24}$,
Yaoting Yan$^{25}$,
Patricio Sanhueza$^{26}$,
Kee-Tae Kim$^{12,13}$,
\newauthor
Yuxin He$^{1,2,3}$,
Huei-Ru Vivien Chen$^{27}$,
Yongxiong Wang$^{7}$,
Gordon MacLoed$^{1,28}$,
\newauthor
Andrej Sobolev$^{1,29}$,
Dmitry A. Ladeyschikov$^{29}$,
Maria Jesus Jimenez Donaire$^{30,31}$,
Jingwen Wu$^{3}$,
\newauthor
Jifeng Xia$^{16,3}$,
Jiawei Liu$^{16,3}$,
Xin Lyu$^{16,3}$,
and \diliname$^{32}$\bluecorrstar
\\
Affiliations are listed at the end of the paper
}

\date{Accepted XXX. Received YYY; in original form ZZZ}

\pubyear{2026}

\label{firstpage}
\pagerange{\pageref{firstpage}--\pageref{lastpage}}

\maketitle

\begin{abstract}
We investigate infall signatures towards 56 clumps in the ALOHA IRDCs sample using single-pointing W-band observations from the Yebes 40-m telescope. The spectra cover three optically thick infall tracers, H$_2$CO, HCO$^+$, and HNC, together with the optically thinner lines H$^{13}$CO$^+$ and HN$^{13}$C. We identify 37 clumps as infall candidates in at least one tracer, corresponding to 49 self-absorbed blue-asymmetric profiles: 6 in H$_2$CO, 22 in HCO$^+$, and 21 in HNC. Two radiative-transfer-model-free approaches are used. First, we apply Fuller's measurements to self-absorbed spectra, characterising the blue-to-red peak contrast, peak separation, profile width, and absorption-dip depth. Secondly, we use the normalised velocity difference, $\delta V$, and the blue excess, $E$, to quantify the population-level preference for blue-asymmetric emission. HCO$^+$ shows the strongest median self-absorbed morphology, with the largest median peak contrast and deepest absorption dips. HNC gives the strongest statistical blue excess, with $E=0.30$, followed by HCO$^+$ with $E=0.26$. Comparisons with NH$_2$D detections and H$_2$O maser associations do not show a systematic enhancement of blue excess on either cold-gas chemistry or embedded activity. Instead, the observed differences in blue asymmetry appear more strongly associated with tracer choice, opacity, excitation, and beam-averaged velocity structure than with the adopted NH$_{2}$D or H$_{2}$O-maser groupings. The prominent HCO$^+$ and HNC blue excesses indicate that nearby, high-density ALOHA IRDCs provide a promising sample for investigating candidate inward motions during the early mass-assembly phase of massive star formation. 
\end{abstract}




\section{Introduction}
Infrared Dark Clouds (IRDCs), characterised by low temperatures and high densities, are prime sites for studying the earliest stages of massive star and cluster formation \citep[e.g.,][]{2006rathborne,2009perettofuller,2018motte}. Infall-driven mass accumulation is a key process in star-formation theory \citep[e.g.][]{1969larson,1977shu,1977elmegreen,2001bonnell,2003mckee,2007zinnecker,2017vazquezsemadeni,2020padoan,2021hennebelle}. Observationally, infall is traced by optically thick molecular lines showing blue-asymmetric self-absorbed profiles, while optically thinner lines peak near the absorption dip and trace the systemic velocity \citep[e.g.][]{1977snell,1993zhou,1996myers}.

A wide range of molecular transitions has been used to search for infall signatures, including CS, HCO$^+$, HCN, HNC, H$_2$CO, and CO \citep[e.g.][]{1995myers,1998tafalla,1999lee,2000gregersen,2001lee,2003wu,2005fuller,2007sohn,2013rygl,2015he,2018contreras,2019jackson,2021xieraa,2023yang,2025morii,2026jackson,2026morii}. Additional probes include red-shifted absorption against bright continuum sources, such as NH$_3$ and CH$_3$OH \citep{2012wyrowski,2016wyrowski,2022yang}. These studies show that infall signatures are not unique and their detectability depends on optical depth, excitation conditions, velocity structure, and viewing geometry. In particular, HCO$^+$ J$\,=1-0$ is widely used as a primary tracer of blue-asymmetric infall in dense clumps \citep{2005fuller,2013smith,2018yoo,2021xieraa}. Simulations further demonstrate that infall does not always produce a simple blue-asymmetric profile due to radiative transfer effects and kinematic complexity \citep{2012smith,2013smith,2014chira}. Consequently, multi-tracer analyses are essential for constructing statistically robust samples of infall candidates.

The ALOHA IRDCs sample analysed here consists of 56 clumps selected from the JCMT ALOHA IRDCs (A Lei Of the Habitat and Assembly towards Infrared Dark Clouds)\footnote{\url{https://www.eaobservatory.org/jcmt/science/large-programs/aloha-irdcs/}} project. Targets were drawn from the IRDC catalogue of \citet{2009perettofuller}, identified via mid-infrared extinction against the Galactic background. The original ALOHA selection prioritised nearby, high-column-density IRDCs for JCMT continuum mapping. Early analyses have demonstrated resolved filament–clump substructure in these regions \citep{2024shen}, while full continuum properties will be presented in companion studies (\cite{2026liu}; \cite{2026aloha}; Ren et al., in prep.; Liu et al., in prep.).

To assess possible links between infall signatures and source evolution, we use NH$_2$D detections and H$_2$O masers. NH$_2$D traces cold, dense gas where deuterium fractionation is efficient \citep[e.g.][]{2007pillai,2011pillai,2013sanhueza,2016lackington,2020zhang,2021wienen}, but its detectability also depends on density, temperature, depletion, chemical timescales, and beam-averaged structure. H$_2$O masers trace shocks and embedded star formation activity \citep{1995codella,2019kim,2023yang}. These indicators are therefore used only for statistical comparison with line asymmetry properties, not for defining evolutionary stages.

We present a multi-tracer census of infall signatures in the ALOHA IRDCs. We do not derive physical infall velocities or mass infall rates, as these require assumptions about excitation, optical depth structure, geometry, and velocity fields \citep{1996myers,2005hill,2026xie}. Instead, we combine direct measurements, i.e., Fuller's $D_{r}$ and $P_{r}$ measurements (Fuller's measurements hereafter), of self-absorbed double-peaked line profiles \citep{2026xie} and a statistical analysis using the normalised velocity difference, $\delta V$, and the blue-excess statistic \citep{1997mardones}. This approach characterises both individual candidate infall morphologies and the ensemble prevalence of blue asymmetry across tracers.

The paper is organised as follows. Section~\ref{sec:Observation} describes the observations and molecular transitions. Section~\ref{sec: Results} presents the Gaussian fitting, Fuller's measurements, and $\delta V$ analysis. Section~\ref{sec: Discussion} discusses tracer dependence, source-by-source consistency, and comparisons with NH$_2$D and maser properties, as well as previous infall surveys. Section~\ref{sec:summary} summarises the results.

\section{Observations}
\label{sec:Observation}

W-band single-pointing observations covering 72.8--91.3 GHz were carried out towards 56 clumps in the ALOHA IRDCs sample with the Yebes 40-m radio telescope \citep{2021tercero} located in Spain in February, March, and May 2022 \citep{2026aloha}. The observations used eight spectral windows, with 200 MHz overlaps between adjacent windows, to cover the full frequency range. The off-source positions were selected at +30 arcmin in right ascension from the target positions. The spectra were inspected for possible OFF-position contamination, and no systematic contamination was found for the lines used in the infall analysis. The system temperature was typically $\sim 130$ K. The total instantaneous bandwidth was 18.5 GHz, with an original channel spacing of about 38 kHz. To improve the signal-to-noise ratio in individual spectral channels, the spectra were smoothed by two channels, giving a final frequency resolution of about 76 kHz. This corresponds to velocity resolutions of $\sim 0.31$ km s$^{-1}$ at 72.8 GHz and $\sim 0.25$ km s$^{-1}$ near 90 GHz.

The half-power beam width (HPBW) varies from about 24.3 arcsec at 72.8 GHz to about 19.9 arcsec near 90 GHz. The main beam efficiencies are 0.30, 0.29, 0.28, 0.27, 0.25, 0.24, 0.23, and 0.22 for the eight spectral windows, in order of increasing frequency. The calibrated antenna temperatures were converted to the main beam brightness temperature scale using the main-beam efficiency appropriate to each spectral window and transition frequency. The efficiency correction was therefore applied uniformly to all transitions before the line parameters were measured. Throughout this work, all line intensities and the rms values listed in Table~\ref{tab:observed_lines} are reported on the $T_{\rm MB}$ scale.

The molecular transitions used in the analysis are listed in Table~\ref{tab:observed_lines}. The primary optically thick tracers for the infall study are H$_2$CO, HCO$^+$, and HNC, while H$^{13}$CO$^+$ and HN$^{13}$C are adopted as optically thinner reference lines where available. The spectral setup also includes the NH$_2$D $1_{11}-1_{01}$ transition at 85.926 GHz, which is used as an additional tracer of cold and dense gas.

The H$_2$O $6_{16}-5_{23}$ maser detections at 22.235\,GHz are taken from the Effelsberg 100-m observations presented in \citet{2026aloha}, which cover the H$_2$O $6_{16}-5_{23}$ maser transition at 22.235 GHz. These data are used to assess whether the inferred infall signatures show any dependence on embedded star formation activity.

\begin{table*}
\centering
\caption{Molecular transitions used in the infall analysis.}
\label{tab:observed_lines}
\begin{tabular}{lcccc}
\hline\hline
Molecular transition & Frequency & rms & Velocity resolution & HPBW \\
 & (MHz) & (mK) & (km s$^{-1}$) & (arcsec) \\
\hline
H$_2$CO $1_{01}-0_{00}$ & 72837.951 & 52 & 0.31 & 24.2 \\
H$^{13}$CO$^+$ $1-0$ & 86754.288 & 51 & 0.26 & 20.3 \\
HN$^{13}$C $1-0^{a}$ & 87090.859 & 53 & 0.26 & 20.2 \\
HCO$^+$ $1-0$ & 89188.526 & 53 & 0.25 & 19.7 \\
HNC $1-0$ & 90663.564 & 53 & 0.25 & 19.4 \\
\hline
\end{tabular}
\begin{minipage}{0.95\textwidth}
\footnotesize Notes. $^{a}$ For HN$^{13}$C, the tabulated rest frequency corresponds to the $F\,=2-1$ hyperfine component. The frequencies are taken from Splatalogue (https://www.cv.nrao.edu/php/splat/). The rms is measured on the $T_{\rm MB}$ scale after spectral smoothing. HPBW values are approximate and correspond to the observing frequency of each transition.
\end{minipage}
\end{table*}

\section{Results}
\label{sec: Results}

\subsection{Candidate infall profiles and profile parameters}
\label{sec:profile_parameters_results}

\begin{figure*}
\centering
\includegraphics[width=0.31\textwidth]{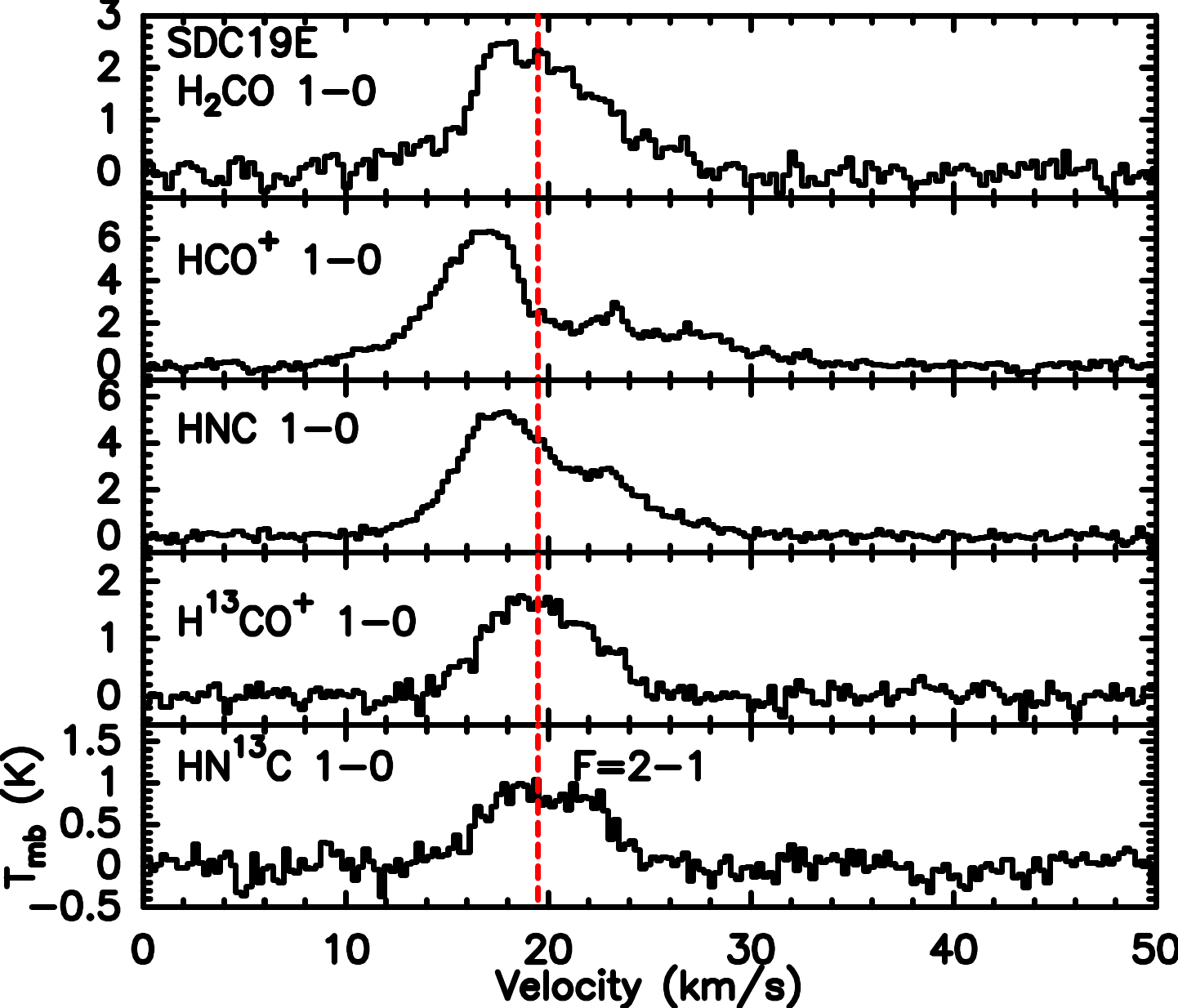}
\hfill
\includegraphics[width=0.31\textwidth]{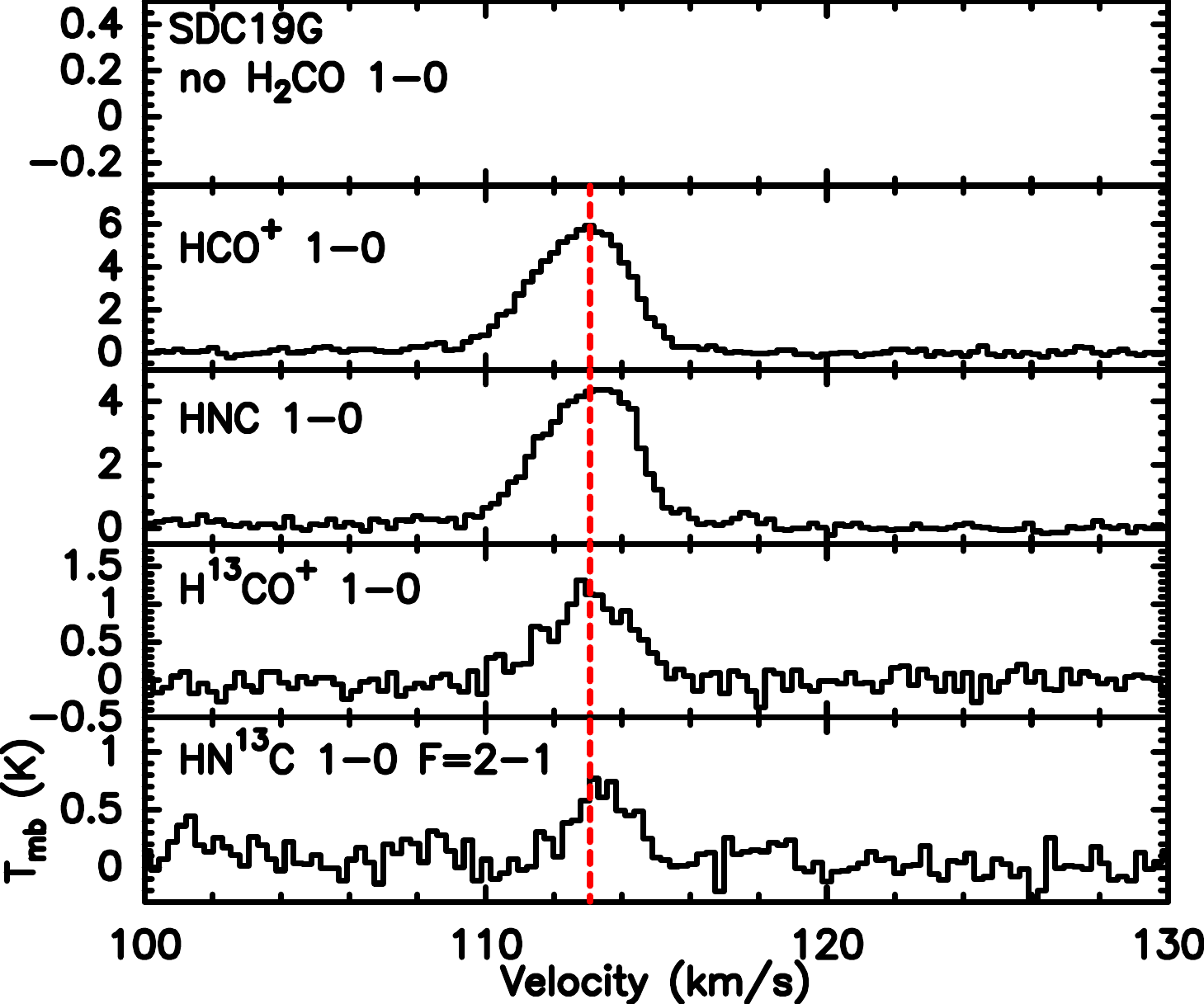}
\hfill
\includegraphics[width=0.31\textwidth]{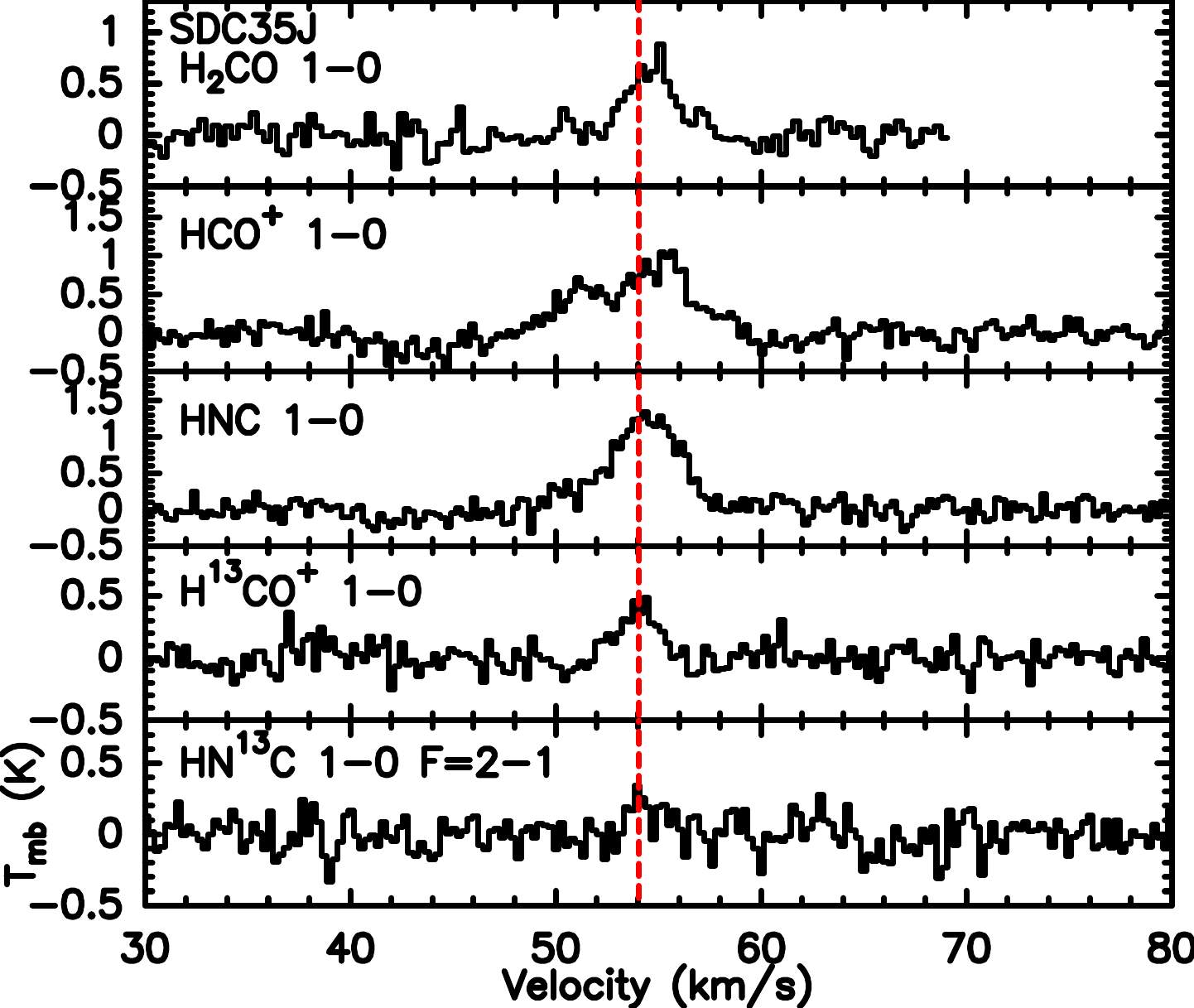}
\caption{
Representative examples of the line shape morphologies and $\delta V$ classifications used in this work. From left to right, the panels show SDC19E, SDC19G, and SDC35J as examples classified as blue, red, and neutral according to the thresholds $\delta V<-0.25$, $\delta V>+0.25$, and $-0.25\leq\delta V\leq+0.25$, respectively. The blue-classified source also illustrates the self-absorbed, double-peaked morphology adopted for candidate infall profiles, with a stronger blue-shifted peak, a weaker red-shifted peak, and an absorption dip between the two peaks. The molecular transitions are indicated in the individual panels.
}
\label{fig:examples}
\end{figure*}

Following the standard interpretation of blue-asymmetric self-absorbed profiles as candidate infall signatures, we identify a candidate infall profile as an optically thick, double-peaked line shape in which the blue-shifted peak is stronger than the red-shifted peak, with the absorption dip located between the two peaks and close to the velocity traced by an optically thinner line \citep[e.g.][]{1993zhou,1996myers,1999lee}. This morphology is illustrated for SDC19E in Fig.~\ref{fig:examples}, where all tracers show clear infall signatures. The H$_2$CO, HCO$^+$, HNC, H$^{13}$CO$^+$, and HN$^{13}$C spectra for all remaining sources are shown in Appendix~\ref{app:spectra}.

For the empirical profile shape analysis, we included only profiles in which both emission peaks were detected above the local rms noise and the absorption dip lay between the two peaks. Spectra affected by severe blending, poorly defined peaks, or ambiguous velocity-component assignment were excluded from the measurements. Based on these criteria, 37 clumps are identified as infall candidates in at least one tracer, corresponding to 49 self-absorbed blue-asymmetric profiles: 6 in H$_2$CO, 22 in HCO$^+$, and 21 in HNC.

\begin{figure}
\centering
\includegraphics[width=0.49\textwidth]{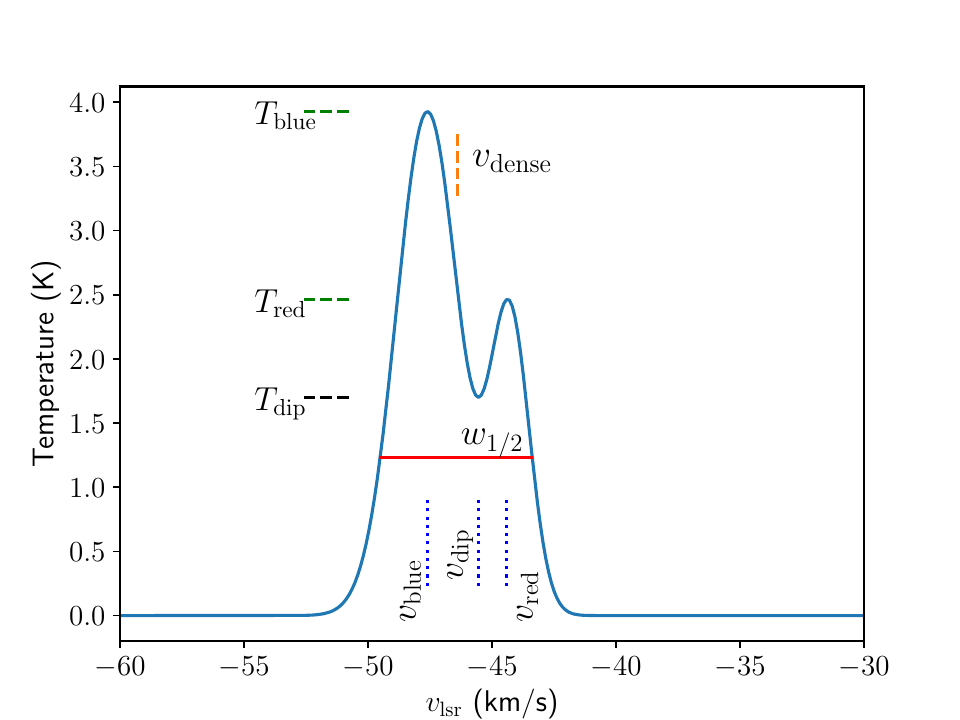}
\caption{
Schematic definition of the profile parameters used in Fuller's measurements of self-absorbed double-peaked spectra. The blue and red local maxima define $(V_{\rm blue}, T_{\rm blue})$ and $(V_{\rm red}, T_{\rm red})$, respectively, while the absorption minimum between them defines $(V_{\rm dip}, T_{\rm dip})$. The vertical dashed line marks $V_{\rm dense}$, the velocity traced by an optically thinner dense-gas line. The quantity $w_{1/2}$ is measured directly from the observed spectra as the full velocity width at half of the red-peak intensity. These measurements are used to derive the profile shape parameters $P_{\rm r}$, $v_{\rm sep}$, and $D_{\rm r}$. Adapted from \citet{2026xie}. 
}
\label{fig:linechara}
\end{figure}

For spectra with double-peaked self-absorbed morphologies, we measured the empirical line shape parameters following Fuller's empirical measurements described in \citet{2026xie} (see Fig.~\ref{fig:linechara}). This approach is designed to characterise the classical self-absorbed morphology, while other methods have been developed in the absence of prominent infall profiles \citep{2018traficanteb,2019jackson,2026jackson}. Following the profile-parameter definitions in Appendix A in \citet{2026xie} (see also Fig.~\ref{fig:linechara}), we identified velocity intervals containing the blue emission peak, the intervening absorption dip, and the red emission peak in each spectrally smoothed profile. Within each peak interval, the channel with the highest brightness temperature supplied $T_{\rm blue}$ or $T_{\rm red}$ and its channel-centre velocity, $V_{\rm blue}$ or $V_{\rm red}$. The channel with the lowest brightness temperature in the dip interval supplied $T_{\rm dip}$ and $V_{\rm dip}$. The velocity traced by the optically thinner reference line is denoted by $V_{\rm dense}$ in Fig.~\ref{fig:linechara}.

From these quantities, we define
\begin{equation} \hspace*{\fill} 
P_{\rm r} = \frac{T_{\rm blue}}{T_{\rm red}}, \hspace*{\fill} \label{eq:pr} 
\end{equation} 
\begin{equation} \hspace*{\fill} 
 v_{\rm sep} = V_{\rm red} - V_{\rm blue},\hspace*{\fill} \label{eq:vsep} \end{equation}
and 
\begin{equation}\hspace*{\fill} 
D_{\rm r} = \frac{T_{\rm dip}}{T_{\rm blue}}. \hspace*{\fill} \label{eq:dr} \end{equation} 
The parameter $P_{\rm r}$ measures the blue-to-red peak contrast, $v_{\rm sep}$ gives the velocity separation between the two emission peaks, and $D_{\rm r}$ quantifies the relative depth of the absorption dip. The width $w_{1/2}$ is not the Gaussian full width at half maximum (FWHM). It is measured directly from the observed spectra as the full velocity width at half of the red-peak intensity,
\begin{equation}\hspace*{\fill} 
\begin{aligned} w_{1/2} &= V_{1/2,{\rm r}} - V_{1/2,{\rm b}},\\ T(V_{1/2,{\rm r}}) &= T(V_{1/2,{\rm b}}) = \frac{1}{2}T_{\rm red}, \end{aligned} \hspace*{\fill} \label{eq:w12} \end{equation} 
where $V_{1/2,{\rm b}}$ and $V_{1/2,{\rm r}}$ are the blue- and red-side velocities at which the spectral emission reaches half of $T_{\rm red}$. This definition ensures that both sides of the self-absorbed profile are included in the empirical width measurement. These parameters describe the observed line morphology and should not be interpreted as direct measurements of the physical infall velocity. Appendix~\ref{app:profile_parameters} lists the full set of measured profile shape parameters for all candidate infall profiles.

We estimated the uncertainties in the empirical profile-shape parameters using 20,000 Monte Carlo noise realisations of each processed spectrum. The noise level was determined from the line-free rms of the processed spectrum (representative values are listed in Table~\ref{tab:observed_lines}). Simulated white noise was passed through the same Hanning-smoothing and factor-of-two channel-sampling procedure as the observations and normalised so that the resulting processed noise reproduced the measured line-free rms. The fitted baseline was held fixed because its uncertainty was negligible compared with the channel rms. For each source and tracer, the blue-peak, dip, and red-peak velocity intervals were held fixed across all noise realisations. Within each realisation, we identified the maximum-temperature channel in each peak interval and the minimum-temperature channel in the dip interval, and recalculated $P_{\rm r}$, $v_{\rm sep}$, and $D_{\rm r}$. We also redetermined the red-peak intensity and the two outer crossings at half that intensity to obtain $w_{1/2}$. The $1\sigma$ uncertainty in each parameter was taken as half the difference between the 84th and 16th percentiles of its distribution. The central values in Table~\ref{tab:fuller_settled} were measured from the smoothed spectra using the same procedure, rather than taken from the Monte Carlo medians.

 HCO$^+$ shows the strongest median blue profile morphology among the three main tracers, with median $P_{\rm r}=1.69$ and median $D_{\rm r}=0.35$, similar to those derived towards one of the most massive IRDCs, SDC335 \citep{2026xie}. HNC shows a slightly weaker blue-to-red contrast and shallower absorption dips than HCO$^+$, with median $P_{\rm r}=1.57$ and median $D_{\rm r}=0.50$. The H$_2$CO sample is smaller and shows weaker self-absorption on average, with median $P_{\rm r}=1.43$ and median $D_{\rm r}=0.57$. The median $w_{1/2}$ is largest for H$_2$CO, 5.25 km s$^{-1}$, compared with 3.86 km s$^{-1}$ for HCO$^+$ and 3.27 km s$^{-1}$ for HNC.

\begin{figure}
\centering
\includegraphics[width=0.45\textwidth]{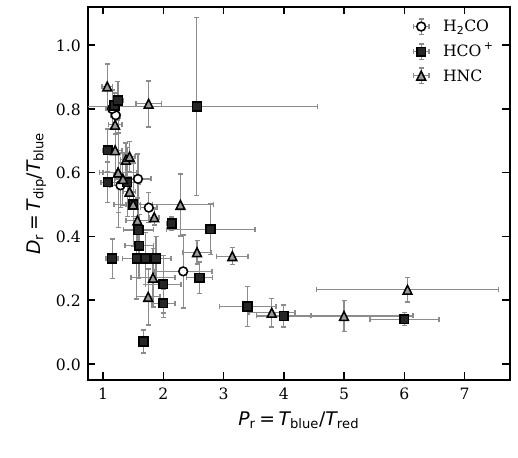}
\caption{Comparison of the empirical profile shape parameters for self-absorbed blue-asymmetric profiles in H$_2$CO, HCO$^+$, and HNC. The horizontal axis shows the blue-to-red peak contrast, $P_{\rm r}=T_{\rm blue}/T_{\rm red}$, and the vertical axis shows the relative dip depth, $D_{\rm r}=T_{\rm dip}/T_{\rm blue}$. The horizontal and vertical error bars show the corresponding $1\sigma$ uncertainties obtained from noise resampling of the processed spectra. Larger $P_{\rm r}$ and smaller $D_{\rm r}$ correspond to stronger blue-asymmetric self-absorption. H$_2$CO, HCO$^+$, and HNC are shown as open circles, dark filled squares, and grey filled triangles, respectively. The three tracers occupy overlapping regions, HCO$^+$ has the highest median $P_{\rm r}$ and lowest median $D_{\rm r}$, while individual HCO$^+$ and HNC profiles extend to similarly large $P_{\rm r}$.}
\label{fig:profile_shape_pr_dr}
\end{figure}

Figure~\ref{fig:profile_shape_pr_dr} compares the parameters $P_{\rm r}$ and $D_{\rm r}$ derived from Fuller's measurements for individual blue-asymmetric profiles. The distributions overlap substantially. HCO$^+$ has the highest median $P_{\rm r}$ and the lowest median $D_{\rm r}$, while individual HCO$^+$ and HNC profiles extend to similarly large $P_{\rm r}$. These parameters describe spectral morphology only and are not direct kinematic measurements.

Gaussian fitting was applied only to the optically thinner isotopologue spectra, H$^{13}$CO$^+$ and HN$^{13}$C. For each detected velocity component, we derived the integrated intensity, centroid velocity $V_{\rm lsr}$, linewidth $\Delta V$, and peak brightness temperature $T_{\rm peak}$. The fitted centroid velocities and linewidths provide the velocity references used in the $\delta V$ analysis, with H$^{13}$CO$^+$ used for HCO$^+$ and H$_2$CO, and HN$^{13}$C used for HNC (for their velocity difference, see the next subsection). The full source-by-source Gaussian fitting results are given in Appendix~\ref{app:gaussian}.

\subsection{Infall asymmetry from the $\delta V$ analysis}
\label{sec:deltaV_results}

To compare with previous statistical infall surveys, we also calculated the normalised velocity difference, $\delta V$, following the classical method of \citet{1997mardones} and later studies \citep[e.g.][]{2005fuller,2022tatematsu}. This quantity compares the velocity of an optically thick line with the velocity traced by an optically thinner reference line, normalised by the linewidth of the optically thinner line,
\begin{equation} 
\hspace*{\fill} 
\delta V =
\frac{V_{\rm thick}-V_{\rm thin}}{\Delta V_{\rm thin}}
, \hspace*{\fill} \label{eq:dV} 
\end{equation}
where $V_{\rm thick}$ is the adopted velocity of the optically thick transition, and $V_{\rm thin}$ and $\Delta V_{\rm thin}$ are the centroid velocity and FWHM linewidth of the corresponding optically thinner reference line. For the optically thick tracers, $V_{\rm thick}$ was measured directly from the observed line shapes as the velocity of the relevant emission peak. For blue-asymmetric spectra this corresponds to $V_{\rm blue}$, while for red-asymmetric spectra it corresponds to $V_{\rm red}$. H$^{13}$CO$^+$ is used as the optically thinner reference for HCO$^+$, and HN$^{13}$C for HNC. For H$_2$CO, no dedicated optically thin isotopologue is available in the present data, and H$^{13}$CO$^+$ is therefore adopted as a proxy velocity reference.

\begin{figure}
\centering
\includegraphics[width=0.9\columnwidth]{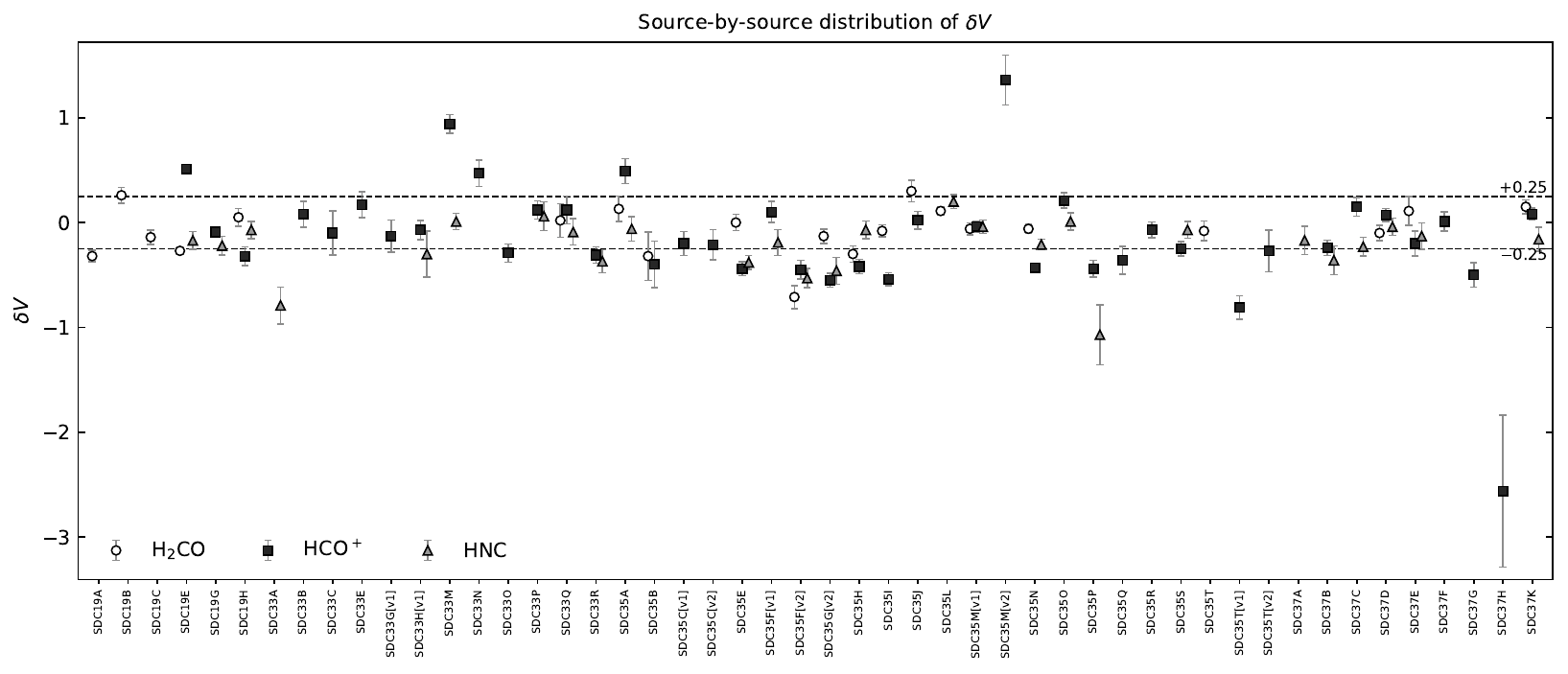}
\caption{Source-by-source distribution of the normalised velocity difference, $\delta V$, for H$_2$CO, HCO$^+$, and HNC. The vertical error bars show the propagated $1\sigma$ uncertainties in $\delta V$. The horizontal dashed lines mark the adopted classification thresholds, $\delta V=-0.25$ and $\delta V=+0.25$. Points with central values below the lower threshold are classified as blue, points above the upper threshold are classified as red, and points between the two thresholds are classified as neutral. The statistics are calculated only for sources with velocity-matched optically thinner reference components, as listed in Table~\ref{tab:deltaV_by_source_revised}.
}
\label{fig:deltav_distribution}
\end{figure}

The uncertainty in the normalised velocity difference was propagated as
\begin{equation}
\sigma_{\delta V}^{2}
=
\frac{
\sigma_{V_{\rm thick}}^{2}
+
\sigma_{V_{\rm thin}}^{2}
}{
\Delta V_{\rm thin}^{2}
}
+
\delta V^{2}
\left(
\frac{\sigma_{\Delta V_{\rm thin}}}
{\Delta V_{\rm thin}}
\right)^{2}.
\label{eq:deltaV_uncertainty}
\end{equation}
Here, $\sigma_{V_{\rm thin}}$ and $\sigma_{\Delta V_{\rm thin}}$ are the Gaussian-fit uncertainties in the centroid velocity and FWHM linewidth of the corresponding optically thinner reference transition. For the optically thick tracers, $V_{\rm thick}$ was taken as the velocity of the channel containing the relevant local emission maximum. Since no parametric fit was applied to these asymmetric profiles, we assign half of one smoothed spectral channel as the uncertainty associated with the spectral sampling of the peak position. This uncertainty was propagated together with the Gaussian-fit uncertainties in $V_{\rm thin}$ and $\Delta V_{\rm thin}$ using Eq. \ref{eq:deltaV_uncertainty}. The resulting $1\sigma$ uncertainties are listed in Table~\ref{tab:deltaV_by_source_revised} and displayed in Fig.~\ref{fig:deltav_distribution}.

For the 31 velocity components reliably detected in both H$^{13}$CO$^+$ and HN$^{13}$C, the median signed centroid-velocity difference, $V_{\rm LSR}({\rm H^{13}CO^+})- V_{\rm LSR}({\rm HN^{13}C})$, is $-0.15$~km~s$^{-1}$, and the median absolute difference is $0.15$~km~s$^{-1}$. In total, 22 of the 31 components differ by no more than $0.26$~km~s$^{-1}$, approximately one velocity-resolution element. The reference-velocity differences are therefore generally small compared with the typical fitted linewidths of approximately $2$~km~s$^{-1}$, although they may affect the $\delta V$ classification of individual profiles close to the adopted boundaries.

Profiles with $\delta V<-0.25$, $\delta V>+0.25$, and $-0.25\leq\delta V\leq+0.25$ were classified as blue, red, and neutral, respectively \citep{1997mardones}. The blue excess was then calculated as
\begin{equation} 
\hspace*{\fill}
E =
\frac{N_{\rm blue}-N_{\rm red}}{N_{\rm total}}, \hspace*{\fill} \label{eq:E}
\end{equation}
where $N_{\rm blue}$, $N_{\rm red}$, and $N_{\rm total}$ are the numbers of blue, red, and total classified profiles for a given tracer. We also calculated a one-sided binomial probability, $p_{\rm bin}$, for obtaining at least the observed number of blue spectra among the non-neutral asymmetric spectra, assuming equal probabilities of blue and red asymmetries. Representative blue, red, and neutral $\delta V$ classifications are shown in Fig.~\ref{fig:examples}.

The resulting census is listed in Table~\ref{tab:blue_excess}. HNC shows the largest blue excess, with 8 blue profiles and no red profiles among 27 velocity-matched sources, giving $E=0.30$ and $p_{\rm bin}=0.003$. HCO$^+$ also shows a significant blue excess, with 16 blue and 5 red among 43 velocity-matched sources, giving $E=0.26$ and $p_{\rm bin}=0.01$. H$_2$CO shows a weaker excess, with 5 blue and 2 red among 21 velocity-matched sources, giving $E=0.14$ and $p_{\rm bin}=0.23$. The classifications and blue-excess statistics are based on the nominal $\delta V$ values calculated from the measured $V_{\rm thick}$ and the best-fitting $V_{\rm thin}$ and $\Delta V_{\rm thin}$. The propagated uncertainties indicate that the classifications of some profiles close to the adopted $\delta V=\pm0.25$ boundaries are uncertainty-sensitive.

\begin{table}
\centering
\caption{Census of $\delta V$ classifications and blue-excess statistics.}
\label{tab:blue_excess}
\begingroup
\scriptsize
\setlength{\tabcolsep}{2.2pt}
\renewcommand{\arraystretch}{1.05}
\begin{tabular*}{\columnwidth}{@{\extracolsep{\fill}}llcccccc@{}}
\hline\hline
Tracer & Ref. line & $N_{\rm tot}$ & $N_{\rm blue}$ & $N_{\rm red}$ & $N_{\rm neu}$ & $E$ & $p_{\rm bin}$ \\
\hline
H$_2$CO & H$^{13}$CO$^+$ (proxy) & 21 & 5  & 2 & 14 & 0.14 & 0.23 \\
HCO$^+$ & H$^{13}$CO$^+$         & 43 & 16 & 5 & 22 & 0.26 & 0.01 \\
HNC      & HN$^{13}$C             & 27 & 8  & 0 & 19 & 0.30 & 0.003 \\
\hline
\end{tabular*}
\vspace{1mm}
\begin{minipage}{\columnwidth}
\scriptsize
Notes. Blue and red profiles are defined using $\delta V < -0.25$ and $\delta V > +0.25$, respectively. $N_{\rm neu}$ is the number of profiles with $-0.25 \leq \delta V \leq +0.25$. The blue excess is $E=(N_{\rm blue}-N_{\rm red})/N_{\rm tot}$, where $N_{\rm tot}$ includes blue, red, and neutral classifications. The probability $p_{\rm bin}$ is the one-sided binomial probability of obtaining at least the observed number of blue profiles among the asymmetric spectra, $N_{\rm blue}+N_{\rm red}$, assuming equal probabilities of blue and red asymmetries. Neutral profiles are excluded only in the calculation of $p_{\rm bin}$. Only sources with a velocity-matched optically thinner reference component are included.
\end{minipage}
\endgroup
\end{table}

The source-by-source $\delta V$ values are shown in Fig.~\ref{fig:deltav_distribution}, with the full classifications listed in Table~\ref{tab:deltaV_by_source_revised}. HCO$^+$ and HNC both show an excess of blue over red classifications, consistent with the blue-excess statistics in Table~\ref{tab:blue_excess}. H$_2$CO also shows a positive excess, but its interpretation is limited by the smaller matched sample and the use of a proxy velocity reference. To assess this, we shifted the H$^{13}$CO$^+$ reference velocity for H$_2$CO by $\pm0.1~{\rm km~s^{-1}}$, approximately 0.4 of the H$^{13}$CO$^+$ velocity-resolution element. This illustrative sub-channel perturbation changes the H$_2$CO blue excess from $E=0.10$ to $E=0.19$. The blue excess remains positive but weak, and we therefore treat the H$_2$CO result as a secondary indication of blue asymmetry.

\section{Discussion}
\label{sec: Discussion}

\subsection{Tracer-dependent infall signatures in IRDCs}
\label{sec:discussion_tracers}

The ALOHA IRDCs exhibit tracer-dependent infall signatures in both Fuller's measurements and the $\delta V$ analysis. Among the clearly self-absorbed spectra, HCO$^+$ shows the strongest median blue-asymmetric morphology, with $P_{\rm r}=1.69$ and $D_{\rm r}=0.35$, followed by HNC ($P_{\rm r}=1.57$ and $D_{\rm r}=0.50$), while H$_2$CO shows the weakest contrast ($P_{\rm r}=1.43$ and $D_{\rm r}=0.57$). This ordering is consistent with HCO$^+$ being an effective optically thick tracer of clump-scale self-absorption in massive star-forming regions \citep{2005fuller,2013smith,2018yoo,2021xieraa}, probably reflecting its high opacity and sensitivity to extended dense gas \citep{2026jackson}.

The tracers also differ in linewidth, with median $w_{1/2}=5.25~{\rm km~s^{-1}}$ for H$_2$CO, compared with 3.86 and 3.27~${\rm km~s^{-1}}$ for HCO$^+$ and HNC, respectively. Given the larger beam size of the H$_2$CO transition ($24.2$ arcsec versus $19.7$--$19.4$ arcsec), together with differences in excitation and opacity, the broader H$_2$CO profiles likely include a larger contribution from extended envelope gas rather than tracing intrinsically more disturbed motions.

The $\delta V$ classifications show tracer-dependent source-by-source behaviour (Table~\ref{tab:tracer_agreement}). HCO$^+$ and HNC show the strongest overlap in blue classifications, whereas H$_2$CO shows less overlap with either tracer. Most clumps in the ALOHA IRDCs sample are either neutral or blue in only one tracer. Only SDC35F[v2] in SDC35F is blue in all three tracers, while SDC19E is the only clear blue/red disagreement between HCO$^+$ and H$_2$CO. Thus, inward-motion signatures are common in the sample, but they are not uniformly recovered by all molecular lines. This is expected because line asymmetry depends on opacity, excitation gradient, abundance structure, beam filling and the relation between the optically thick line and the adopted systemic-velocity tracer \citep[e.g.][]{1997mardones,2008tsamis,2012smith,2013smith}.

A further tracer-dependent difference is that HCO$^+$ shows five red classifications among 43 velocity-matched sources, whereas no red profiles are identified among the 27 velocity-matched HNC sources. Given the unequal sample sizes, however, the absence of red HNC profiles should not be overinterpreted. Moreover, a red asymmetry does not have a unique physical interpretation and may arise from outward motions, foreground absorption, unresolved velocity components, rotation, or differences in the excitation-temperature structure \citep[e.g.][]{2023pillai}. Similar tracer-dependent behaviour was reported by \citet{2018traficantea}, which found that HCO$^+$~$(1-0)$ was generally more asymmetric and broader than HNC~$(1-0)$ in massive clumps. This difference may partly reflect the generally higher optical depth of HCO$^+$, which increases its sensitivity to excitation and velocity gradients along the line of sight \citep{2019jackson,2026jackson}. The lower effective excitation density of HCO$^+$~$(1-0)$ may make the line more sensitive to lower-density, extended gas \citep{2015shirley}, while emission associated with outflows and outflow--envelope interfaces may also affect the line profile and complicate the interpretation of any infall signature \citep{2004rawlings,2023pillai,2026morii}. Differences in opacity, excitation, abundance structure, and spatial weighting may therefore cause HCO$^+$ and HNC to respond differently to the same underlying velocity field. The broader environmental dependence of tracer weighting is also illustrated by mapping of the $\rho$ Ophiuchus cloud, where [C\,{\sc i}] and CO abundances respond differently to H$_2$ column density and the UV radiation field \citep{2026xiajifeng}. Spatially resolved multi-tracer observations and radiative-transfer modelling would be required to determine the origin of the individual red HCO$^+$ profiles.

\begin{table}
\centering
\caption{Source-by-source comparison of $\delta V$ classifications between tracers.}
\label{tab:tracer_agreement}
\begingroup
\scriptsize
\setlength{\tabcolsep}{2.4pt}
\renewcommand{\arraystretch}{1.05}
\begin{tabular*}{\columnwidth}{@{\extracolsep{\fill}}lccccc@{}}
\hline\hline
Tracer pair & $N_{\rm match}$ & Same & Blue--blue
& $P({\rm B}_2|{\rm B}_1)$ & Opp. B/R \\
\hline
HCO$^+$--HNC      & 24 & 16 & 5 & 5/8 & 0 \\
HCO$^+$--H$_2$CO  & 16 &  8 & 3 & 3/8 & 1 \\
HNC--H$_2$CO      & 14 & 10 & 1 & 1/3 & 0 \\
\hline
\end{tabular*}
\vspace{1mm}
\begin{minipage}{\columnwidth}
\scriptsize
Notes. Classifications are based on the $\delta V$ criterion. $N_{\rm match}$ is the number of velocity-matched entries classified in both tracers. ``Same'' means that both tracers give the same blue/red/neutral classification. ``Blue--blue'' means that both tracers are classified as blue. $P({\rm B}_2\mid{\rm B}_1)=\frac{N(B_1\cap B_2)}{N(B_1)}$ gives the fraction of entries classified as blue in the first-listed tracer that are also classified as blue in the second-listed tracer. It is therefore a directional quantity and depends on the tracer order. ``Opp. B/R'' counts cases where one tracer is blue and the other is red. The only opposite blue/red case is SDC19E between HCO$^+$ and H$_2$CO.
\end{minipage}
\endgroup
\end{table}

\subsection{Comparison among infall surveys}
\label{sec:discussion_comparisons}

The blue-excess statistics of the ALOHA IRDCs are broadly consistent with previous statistical infall surveys, while showing important differences linked to tracer and sample selection (Table~\ref{tab:previous_comparison}). For HCO$^+$ J$\,=1-0$, the ALOHA value, $E=0.26$, is comparable to the blue excesses measured toward high-mass protostellar objects (HMPOs) by \citet{2007wu} and \citet{2018yoo}, and higher than the value reported by \citet{2005fuller}. It is also qualitatively consistent with Millimetre Astronomy Legacy Team 90 GHz (MALT90)-based analyses that found a preference for blue-asymmetric HCO$^+$ spectra using integrated-flux asymmetry parameters rather than the $\delta V$ blue-excess statistic \citep{2019jackson,2026jackson}. \citet{2018traficantea} similarly identified clear double-peaked blue HCO$^+$ J$\,=1-0$ profiles among Hi-GAL/MALT90 massive clumps, although their visually selected sample is more directly comparable to our morphological profile classification than to the full $\delta V$ census.

The ALOHA HCO$^+$ blue excess is much larger than that found for Bolocam Galactic Plane Survey (BGPS) massive starless clump candidates, for which \citet{2018calahan} reported only a small HCO$^+$ J$\,=1-0$ blue excess. This contrast is significant because both surveys use the same transition as a clump-scale optically thick tracer. The difference is therefore more likely related to target selection than to tracer choice. The BGPS sample was selected to identify starless clump candidates across the Galactic plane, whereas the ALOHA sample selects nearby, high-density IRDC structures. The stronger blue excess in ALOHA suggests that this selection efficiently isolates dense clumps in which inward motions are already prominent, while still probing an early phase before bright HII-region feedback dominates the gas kinematics.

HNC shows a similarly strong blue excess in ALOHA, $E=0.30$, in good agreement with the IRDC value reported by \citet{2016jin}. The same study found much weaker or absent HNC blue excess in HMPOs and ultracompact H\,{\sc ii} (UCH\,{\sc ii}) regions, suggesting that HNC is most effective as an infall-asymmetry tracer in cold IRDC gas. This behaviour is chemically plausible because HNC is favoured in cold dense environments, while its abundance, excitation and optical depth may change in warmer, more evolved regions. H$_2$CO gives a weaker but still positive blue excess, $E=0.14$, comparable to values reported for HMPOs by \citet{2005fuller} and \citet{2018yoo}. Its lower statistical weight in the present work is likely due to the smaller matched sample, excitation sensitivity and use of a proxy systemic velocity.

Large MALT90 studies provide a consistent broader context. \citet{2015he} and \citet{2016he} found that infall candidates are common among massive clumps and tend to have higher column and volume densities than non-infall sources, with the infall candidate fraction decreasing from early to later evolutionary stages. The ALOHA results are consistent with this picture of dense IRDC clumps showing signatures of ongoing mass accumulation. The prominence of HCO$^+$ and HNC blue excess therefore supports the use of nearby, high-density IRDCs as efficient targets for studying the early mass assembly of massive star and cluster formation.

The comparison with previous work is further complicated by the fact that different surveys have targeted different transitions, angular resolutions, and source classes. In addition to the statistical surveys listed in Table~\ref{tab:previous_comparison}, infall signatures have been searched for using HCN J$\,=3-2$, HCO$^+$ J$\,=3-2$ and $4-3$, HCN J$\,=4-3$, CS, H$_2$CO, and other optically thick dense-gas tracers toward massive cores, HMPOs, IRDCs, extended green objects (EGOs), high-extinction clumps, and more evolved massive star-forming regions \citep[e.g.][]{1998zhang,2003wu,2006purcell,2008velusamy,2010barnes,2011reiter,2013rygl,2015he,2016qin,2018contreras,2021xieraa,2023xu}. These studies should not be interpreted as forming a simple sequence in which higher transitions are necessarily better infall tracers. Rather, different transitions sample different combinations of excitation, optical depth, beam filling factor, and physical scale \citep{2012li}. For a given telescope, higher-frequency transitions are observed with smaller beams and may be more sensitive to warmer or denser inner gas, but their detectability as infall tracers also depends on whether sufficient optical depth is maintained. For example, multi-transition HCO$^+$ observations of IRDCs found broadly comparable infall properties from J$\,=1-0$ and J$\,=3-2$, whereas J$\,=4-3$ showed infall signatures only toward a smaller subset of massive clumps, probably because of lower opacity \citep{2021xiescpma,2021xieraa}. A systematic, mapped multi-transition survey of the same IRDC sample, ideally with matched angular resolution, optically thin isotopologues, and radiative-transfer modelling, would therefore be needed to determine how the observed asymmetries vary from extended clump envelopes to denser inner regions, and on which physical scales inward motions are most prominent.

\begin{table*}
\centering
\caption{Comparison with previous statistical infall and blue-asymmetry studies.}
\label{tab:previous_comparison}
\scriptsize
\setlength{\tabcolsep}{2.2pt}
\renewcommand{\arraystretch}{1.15}
\begin{tabular}{llrrrr rrrr rrrr}
\hline\hline
\multirow{2}{*}{Sample} & \multirow{2}{*}{Reference}
& \multicolumn{4}{c}{H$_2$CO}
& \multicolumn{4}{c}{HCO$^+$}
& \multicolumn{4}{c}{HNC} \\
\cline{3-14}
&
& $N_{\rm blue}$ & $N_{\rm red}$ & $N_{\rm total}$ & $E$
& $N_{\rm blue}$ & $N_{\rm red}$ & $N_{\rm total}$ & $E$
& $N_{\rm blue}$ & $N_{\rm red}$ & $N_{\rm total}$ & $E$ \\
\hline
ALOHA IRDCs
& This work
& 5 & 2 & 21 & 0.14
& 16 & 5 & 43 & 0.26
& 8 & 0 & 27 & 0.30 \\

IRDCs
& \citet{2016jin}
& -- & -- & -- & --
& -- & -- & -- & --
& 8 & 1 & 25 & 0.28 \\

HMPOs
& \citet{2005fuller}
& 17 & 5 & 64 & 0.19
& 21 & 11 & 68 & 0.15
& -- & -- & -- & -- \\

HMPOs
& \citet{2007wu}
& -- & -- & -- & --
& 16 & 4 & 41 & 0.29
& -- & -- & -- & -- \\

HMPOs
& \citet{2016jin}
& -- & -- & -- & --
& -- & -- & -- & --
& 2 & 4 & 28 & $-0.07$ \\

HMPOs
& \citet{2018yoo}
& 18 & 9 & 71 & 0.13
& 29 & 14 & 74 & 0.20
& -- & -- & -- & -- \\

UCH\,{\sc ii} regions
& \citet{2016jin}
& -- & -- & -- & --
& -- & -- & -- & --
& 3 & 3 & 23 & 0.00 \\
\hline
\end{tabular}

\begin{minipage}{0.98\textwidth}
\footnotesize
Notes. $E=(N_{\rm blue}-N_{\rm red})/N_{\rm total}$ is the blue excess based on blue/red counts. Dashes indicate that the corresponding tracer was not reported or was not directly used in that study. For this work, H$_2$CO, HCO$^+$, and HNC are compared using the velocity-matched $\delta V$ classifications listed in Table~\ref{tab:deltaV_by_source_revised}. The HCO$^+$ and HNC comparisons refer to the J$\,=1-0$ transitions where applicable, while the H$_2$CO comparisons in \citet{2005fuller} and \citet{2018yoo} use the $2_{12}-1_{11}$ transition. Values should not be interpreted as one-to-one comparisons because the studies differ in sample selection, beam size, transition, tracer, and classification method.
\end{minipage}
\end{table*}

\subsection{Physical interpretation and limitations}
\label{sec:discussion_others}

Fuller's measurements and the $\delta V$ census are complementary tools. The former describe the morphology of individual self-absorbed candidate infall profiles, whereas the latter provides a population-level measure of the prevalence of candidate inward motions across the targets and enables comparison with previous infall surveys. The Fuller's measurement parameters are measured directly from self-absorbed, double-peaked blue-asymmetric profiles, and therefore remain closest to the observed infall morphology. Large $P_{\rm r}$, low $D_{\rm r}$, and a well-defined $v_{\rm sep}$ identify spectra with strong blue-to-red peak contrast and deep self-absorption. In contrast, $\delta V$ and the blue excess, $E$, describe population-level asymmetry, rather than the detailed morphology of each source.

Both diagnostics consistently indicate tracer-dependent infall signatures. HCO$^+$ most effectively isolates strong individual self-absorbed profiles, while HNC provides the most statistically robust blue-excess signal in the velocity-matched sample. H$_2$CO is less stable in both diagnostics due to beam effects, excitation differences, and the use of a proxy systemic velocity.

To explore possible evolutionary connections, we compare the $\delta V$ classifications with NH$_2$D detections and H$_2$O maser associations from \citet{2026aloha}. Each velocity-matched entry inherits the parent clump’s NH$_2$D and maser status, ensuring consistent classification across multiple velocity components. These quantities are used only for contextual comparison and are not used to define infall candidates.

\begin{table*}
\centering
\caption{Infall statistics grouped by NH$_2$D detection status and H$_2$O maser association.}
\label{tab:nh2d_h2o_census}
\renewcommand{\arraystretch}{1.25}
\begin{tabular}{llccc}
\hline\hline
\multirow{2}{*}{Indicator} & \multirow{2}{*}{Detection}
& \multicolumn{3}{c}{Infall tracer} \\
\cline{3-5}
&  & H$_2$CO & HCO$^+$ & HNC \\
\hline
NH$_2$D
& Detected
& $10;\ 3/1/6;\ 0.20$
& $17;\ 6/3/8;\ 0.18$
& $15;\ 5/0/10;\ 0.33$ \\
& Not detected
& $11;\ 2/1/8;\ 0.09$
& $26;\ 10/2/14;\ 0.31$
& $12;\ 3/0/9;\ 0.25$ \\
\hline
H$_2$O maser
& Detected
& $11;\ 3/1/7;\ 0.18$
& $18;\ 7/4/7;\ 0.17$
& $12;\ 3/0/9;\ 0.25$ \\
& Not detected
& $10;\ 2/1/7;\ 0.10$
& $25;\ 9/2/14;\ 0.28$
& $15;\ 5/0/10;\ 0.33$ \\
\hline
\end{tabular}

\begin{minipage}{0.95\textwidth}
\footnotesize
Notes. Each tracer entry is given as
$N_{\rm total};\ N_{\rm blue}/N_{\rm red}/N_{\rm neutral};\ E$, where $E=(N_{\rm blue}-N_{\rm red})/N_{\rm total}$.
The NH$_2$D detection status and H$_2$O maser association are taken from \citet{2026aloha} and assigned to each $\delta V$ entry according to the corresponding ALOHA clump. For clumps with multiple velocity components, the same NH$_2$D and H$_2$O maser information is used for all associated components. Only entries included in the velocity-matched $\delta V$ census are considered here. The source-by-source $\delta V$ classifications used for this comparison are listed in Table~\ref{tab:deltaV_by_source_revised}. The H$_2$O-maser-associated group includes the newly identified water maser toward SDC33O.
\end{minipage}
\end{table*}

The grouped statistics in Table~\ref{tab:nh2d_h2o_census} show no monotonic dependence of blue excess on either NH$_2$D detection or H$_2$O maser association. NH$_2$D-detected sources show a larger HNC blue excess than NH$_2$D non-detections, but the opposite trend is seen for HCO$^+$. Similarly, H$_2$O-maser-associated sources do not show systematically stronger blue excess than non-maser sources. Differences between tracers are larger than those associated with evolutionary indicators, suggesting that tracer selection has a stronger influence than chemical or activity-based grouping in the present sample. This behaviour is consistent with previous maser-selected and EGO studies. \citet{2010chen} found that the HCO$^+$ blue excess in EGOs varies with IRDC, 6.7 GHz methanol maser, and UCH\,{\sc ii} region association, indicating that outflows, thermalisation, and evolutionary state can all affect the observed asymmetry. \citet{2013chen} further showed that GLIMPSE II EGOs frequently exhibit CS and HNC line wings and strong class I methanol maser associations, confirming that maser-rich samples can be dominated by outflow activity. Similarly, \citet{2006purcell} found approximately equal numbers of red- and blue-skewed HCO$^+$ profiles in a methanol-maser-selected sample, although the mid-infrared dark-cloud subsample preferentially showed blue-skewed profiles.

Maser association is interpreted as a signpost of embedded activity rather than as a direct predictor of infall asymmetry. The absence of a simple H$_2$O-maser dependence in the ALOHA sample does not weaken the infall interpretation. Instead, it suggests that the measured blue excess is controlled more strongly by tracer choice, opacity, excitation gradient, and beam-averaged velocity structure than by the presence or absence of a maser alone. In evolutionary terms, the ALOHA IRDCs appear to occupy an early active stage: they retain cold dense gas, include clumps with embedded activity, and show statistically significant inward motion signatures.

More generally, clump--filament studies have reported blue-asymmetric HCO$^+$ J$\,=1-0$ and HNC J$\,=1-0$ profiles in dynamically active, relatively high-surface-density regions, consistent with an increasing contribution from gravity to the observed motions \citep{2020traficante}. Magnetic fields and filamentary environment may also affect the observed line widths and velocity structure by regulating turbulent motions, collapse and gas flows \citep[e.g.][]{2015pillai}. Recent observations of sub-parsec supercritical filaments have further revealed local velocity gradients with no preferred alignment relative to either the filament axes or the local gravitational fields, highlighting the complexity of their internal gas motions \citep{2026zhang}. The present spectra do not provide the polarization or spatially resolved kinematic information required to isolate these effects. Whether the strongest multi-tracer infall candidates are preferentially associated with SCUBA-2 filament junctions or hub-like structures will be examined in the companion ALOHA continuum/filament analysis.

Several observational limitations should also be considered. The analysis is based on single-pointing observations and cannot distinguish global collapse from localised motions or resolve internal spatial structure \citep[cf.][]{2013li}. Beam size differences and distance variations introduce additional scale-dependent effects, while unresolved line-of-sight velocity components may be blended into a single, apparently broader spectral feature, particularly towards kinematically crowded inner-Galaxy sightlines \citep{2026xiaruilin}.

Finally, the parameters derived from Fuller's measurements, $P_{\rm r}$, $D_{\rm r}$, $v_{\rm sep}$, and $w_{1/2}$, together with the $\delta V$ statistic, do not directly measure infall velocities or mass accretion rates. Such quantities require radiative transfer modelling with assumptions about geometry, excitation, optical depth, density structure, and velocity fields. Previous studies have applied two-layer or hill models to selected clear profiles to estimate infall velocities and mass accretion rates \citep[e.g.][]{1996myers,2005hill,2018traficantea,2021xieraa}, while synthetic observations show that line asymmetries also depend on chemistry, excitation, and viewing geometry \citep{2008tsamis,2012smith,2013smith}. We do not adopt such modelling here in order to maintain a uniform, radiative-transfer-model-free, multi-tracer census across the full ALOHA IRDCs sample.

Despite these limitations, the agreement between Fuller's measurements and the $\delta V$ classifications for many sources provides statistical evidence that inward-motion signatures are present in a substantial fraction of the ALOHA molecular-line subsample. The results support a picture in which many nearby, high-density clumps in the ALOHA IRDCs sample are dynamically active infall candidates. The prominent HCO$^+$ and HNC blue excesses support the view that the ALOHA selection is well suited for studying the early mass-accumulation phase of massive star formation.

\section{Summary}
\label{sec:summary}

We have analysed infall signatures towards 56 clumps in the ALOHA IRDCs sample using Yebes W-band molecular line observations. We used two empirical methods: direct Fuller's measurements for clearly self-absorbed blue-asymmetric morphologies, and the normalised velocity difference, $\delta V$, together with the blue excess, $E$, for the velocity-matched sample. The main results are as follows.

\begin{enumerate}
\item We identified 37 clumps as infall candidates in at least one tracer. In total, 49 self-absorbed blue-asymmetric profiles were measured, including 6 in H$_2$CO, 22 in HCO$^+$, and 21 in HNC. For these profiles, $P_{\rm r}=T_{\rm blue}/T_{\rm red}$, $v_{\rm sep}=V_{\rm red}-V_{\rm blue}$, $w_{1/2}$, and $D_{\rm r}=T_{\rm dip}/T_{\rm blue}$ provide an empirical description of the observed line morphology.

\item HCO$^+$ shows the strongest median self-absorbed blue morphology among the three main tracers. It has the largest median blue-to-red peak contrast, $P_{\rm r}=1.69$, and the lowest median relative dip depth, $D_{\rm r}=0.35$, indicating strong blue-asymmetric profiles with relatively deep absorption dips. HNC shows slightly weaker but still clear self-absorbed blue profiles, while the smaller H$_2$CO sample shows weaker self-absorption and broader empirical line shape widths.

\item The $\delta V$ analysis finds a blue excess in each of the three tracers, with weaker statistical support for H$_2$CO. HNC has the largest blue-excess value. HCO$^+$ has the most profiles with matched optically thinner reference velocities, with 43 compared with 27 for HNC and 21 for H$_2$CO.

\item The source-by-source comparison shows tracer-dependent behaviour. Among 24 entries classified in both HCO$^+$ and HNC, 16 have the same blue/red/neutral class, and five are blue in both species, giving the strongest overlap in blue classifications among the three tracer pairs. One velocity component, SDC35F[v2] in source SDC35F, is classified as blue in all three tracers.

\item Comparison with previous infall surveys shows that the ALOHA HCO$^+$ blue excess is comparable to values found in HMPO samples and higher than that reported for BGPS massive starless clump candidates. The HNC result agrees with previous IRDC studies and is stronger than values found towards more evolved HMPOs and UCH\,{\sc ii} regions. These comparisons suggest that the ALOHA selection of nearby, high-density IRDCs efficiently identifies dense clumps in which inward motions are already prominent.

\item The comparison with NH$_2$D detections and H$_2$O maser associations shows no simple monotonic dependence of blue excess on either chemical or activity-related source properties. Blue-asymmetric profiles are found in clumps with different NH$_2$D and maser properties, suggesting that infall signatures are shaped not only by evolutionary or activity indicators, but also by opacity, excitation, feedback, unresolved velocity structure, and beam-averaged source structure. 

\end{enumerate}

In summary, the ALOHA IRDCs sample shows widespread but tracer-dependent blue asymmetry. HCO$^+$ provides the strongest median self-absorbed profiles, HNC yields the strongest statistical blue-excess, and H$_2$CO provides weaker constraints. These results support a picture in which many nearby, high-density ALOHA IRDCs clumps are dynamically active infall candidates, making the sample well suited for studying the early mass accumulation of massive star formation.

\section*{Acknowledgements}

We thank the anonymous referee for many constructive comments and suggestions, which helped to improve the quality of this manuscript. This work is supported by  the National Natural Science Foundation of China grant No. 12588202, National Key R\&D Program of China under grant Nos. (2022YFA1603103 and 2023YFA1608002), Natural Science Foundation of Xinjiang Uyghur Autonomous Region No. 2025D01B173, and Tianshan Talent Training Program 2024TSYCTD0013. It was also partially funded by the NSFC under grants 12373029, 12103082, 12563004 and 12403033, the XBZG-ZDSYS-202212, the Tianchi Talent Project of Xinjiang Uygur Autonomous Region, the Central Guidance on Local Science and Technology Development Fund (ZYYD2025ZY23).

G.A.F acknowledges support by the Deutsche Forschungsgemeinschaft (DFG, German Research Foundation) under Germany's Excellence Strategy via EXC 3037, the Cluster of Excellence "Our Dynamic Universe", no. 533607693, funding via the DFG Collaborative Research Center 1601 (SFB 1601, sub-project B1) and from the University of Cologne and its Global Faculty programme.  

X.-P.C. acknowledges support from the National Science Foundation of China (12041305), the Tianchi Talent Program of Xinjiang Uygur Autonomous Region, and the Tianshan Talent Training Program (2024TSYCTD0013). 

C.W.L is supported by the Basic Science Research Program through the NRF funded by the Ministry of Education, Science and Technology (grant No. NRF-2019R1A2C1010851) and by the Korea Astronomy and Space Science Institute grant funded by the Korea government (MSIT; project No. 2025-1-841-02).

W.Y. acknowledges the support from the National Natural Science Foundation of China (12403027), China Postdoctoral Science Foundation (2024M751376), and Jiangsu Funding Program for Excellent Postdoctoral Talent (2024ZB347). 

C.H. has been funded by Chinese Academy of Sciences President's International Fellowship Initiative grant No. 2025PVA0048. 


Y.T. Yan is supported by JSPS KAKENHI grant No. 26H02071.

A.M. Sobolev has been funded by Chinese Academy of Sciences President's International Fellowship Initiative grant No. 2024VMA0002. 
Yebes Observatory thanks the ERC for funding support under grant ERC-2013-Syg-610256-NANOCOSMOS.

D.A.L. was supported by the Ministry of Science and Higher Education of the Russian Federation (theme No. FEUZ-2025-0003). 

Based on observations carried out with the Yebes 40 m telescope 22A020. The 40 m radio telescope at Yebes Observatory is operated by the Spanish Geographic Institute (IGN; Ministerio de Transportes y Movilidad Sostenible). Maria Jesus Jimenez Donaire has carried out the observations and the first inspection of the data quality.

\section*{Data Availability}

The reduced Yebes 40-m spectra and the derived measurements
underlying this article will be shared on reasonable request to the corresponding author. The Gaussian fit parameters, the parameters derived from Fuller's measurements, and $\delta V$ classifications used in the analysis are provided in the appendices and tables of this article. H$_2$O maser information used for comparison is from the companion Effelsberg study and will be made available with that work.



\bibliographystyle{mnras}
\bibliography{references} 
\par\bigskip
{\footnotesize
\noindent$^{1}$State Key Laboratory of Radio Astronomy and Technology, Xinjiang Astronomical Observatory, Chinese Academy of Sciences, 150 Science 1-Street, \\Urumqi, Xinjiang 830011, China\\
$^{2}$Xinjiang Key Laboratory of Radio Astrophysics, Urumqi 830011, China\\
$^{3}$University of the Chinese Academy of Sciences, Beijing 100080, China \\
$^{4}$Purple Mountain Observatory, Chinese Academy of Sciences, No.10 Yuanhua Road, Qixia District, Nanjing, Jiangsu 210023, China\\
$^{5}$School of Astronomy and Space Science, University of Science and Technology of China, 96 Jinzhai Road, Hefei 230026, China\\
$^{6}$Space Engineering University, Beijing 101416, China\\
$^{7}$Jodrell Bank Centre for Astrophysics, Department of Physics and Astronomy, University of Manchester, Oxford Road, Manchester M13 9PL, UK\\
$^{8}$Cluster of Excellence `Our Dynamic Universe' (DYNAVERSE), I. Physikalisches Institut, University of Cologne, Z\"ulpicher Straße 77, 50937 K\"oln, Germany\\
$^{9}$School of Physics \& Astronomy, Cardiff University, Queen\textquotesingle s Building, The Parade, Cardiff, CF24 3AA, UK\\
$^{10}$School of Physical Sciences, The Open University, Walton Hall, Milton Keynes, MK7 6AA, UK\\
$^{11}$RAL Space, STFC Rutherford Appleton Laboratory, Chilton, Didcot, Oxfordshire, OX11 0QX, UK\\
$^{12}$Korea Astronomy and Space Science Institute, 776 Daedeokdae-ro, Yuseong-gu, Daejeon 34055, Republic of Korea\\
$^{13}$University of Science and Technology, Korea (UST), 217 Gajeong-ro, Yuseong-gu, Daejeon 34113, Republic of Korea\\
$^{14}$School of Astronomy and Space Science, Nanjing University, 163 Xianlin Avenue, Nanjing 210023, China\\
$^{15}$ Max-Plank-Institut f\"ur Radioastronomie, Auf dem H\"ugel 69, 53121 Bonn, Germany\\
$^{16}$National Astronomical Observatories, Chinese Academy of Sciences, Beijing 100101, China\\
$^{17}$Department of Physics, National Sun Yat-Sen University, No. 70, Lien-Hai Road, Kaohsiung City 80424, Taiwan\\
$^{18}$Center of Astronomy and Gravitation, National Taiwan Normal University, Taipei 116, Taiwan\\
$^{19}$School of Physics and Electronic Science, Guizhou Normal University, Guiyang 550025, China\\
$^{20}$Netherlands Institute for Radio Astronomy ASTRON, NL-7991 PD Dwingeloo, The Netherlands\\
$^{21}$Global Banking School, Universal Square, Devonshire Street North, Manchester, M12 6JH, UK\\
$^{22}$Armagh Observatory and Planetarium, College Hill, Armagh, BT61 9DB, UK\\
$^{23}$National Astronomical Observatory of Japan, National Institutes of Natural Sciences, 2-21-1 Osawa, Mitaka, Tokyo 181-8588, Japan\\
$^{24}$Astronomical Science Program, The Graduate University for Advanced Studies, SOKENDAI, 2-21-1 Osawa, Mitaka, Tokyo 181-8588, Japan\\
$^{25}$\label{inst.dos.naoj}Division of Science, National Astronomical Observatory of Japan, 2-21-1 Osawa, Mitaka, Tokyo 181-8588, Japan\\
$^{26}$Department of Astronomy, School of Science, The University of Tokyo, 7-3-1 Hongo, Bunkyo, Tokyo 113-0033, Japan\\
$^{27}$Institute of Astronomy, Department of Physics, National Tsing Hua University, Hsinchu, Taiwan\\
$^{28}$SARAO, Hartebeesthoek Radio Astronomy Observatory, Krugers-
dorp, South Africa\\
$^{29}$Ural Federal University, 51 Lenin Str., 620051 Ekaterinburg, Russia\\
$^{30}$AURA for the European Space Agency (ESA), ESA Office, Space Telescope Science Institute, 3700 San Martin Drive, Baltimore, MD 21218, USA\\
$^{31}$Observatorio de Yebes, IGN, Cerro de la Palera s/n, 19141 Yebes, Guadalajara, Spain\\
$^{32}$New Cornerstone Science Laboratory, Department of Astronomy, Tsinghua University, Beijing 100084, China
\par}



\clearpage
\onecolumn
\appendix

\section{Spectra of the molecular tracers used in the infall analysis}\label{app:spectra}

 \begin{figure}
  \centering
\includegraphics[width=0.32\textwidth,height=0.39\textwidth, trim={0 0  0 0}, clip]{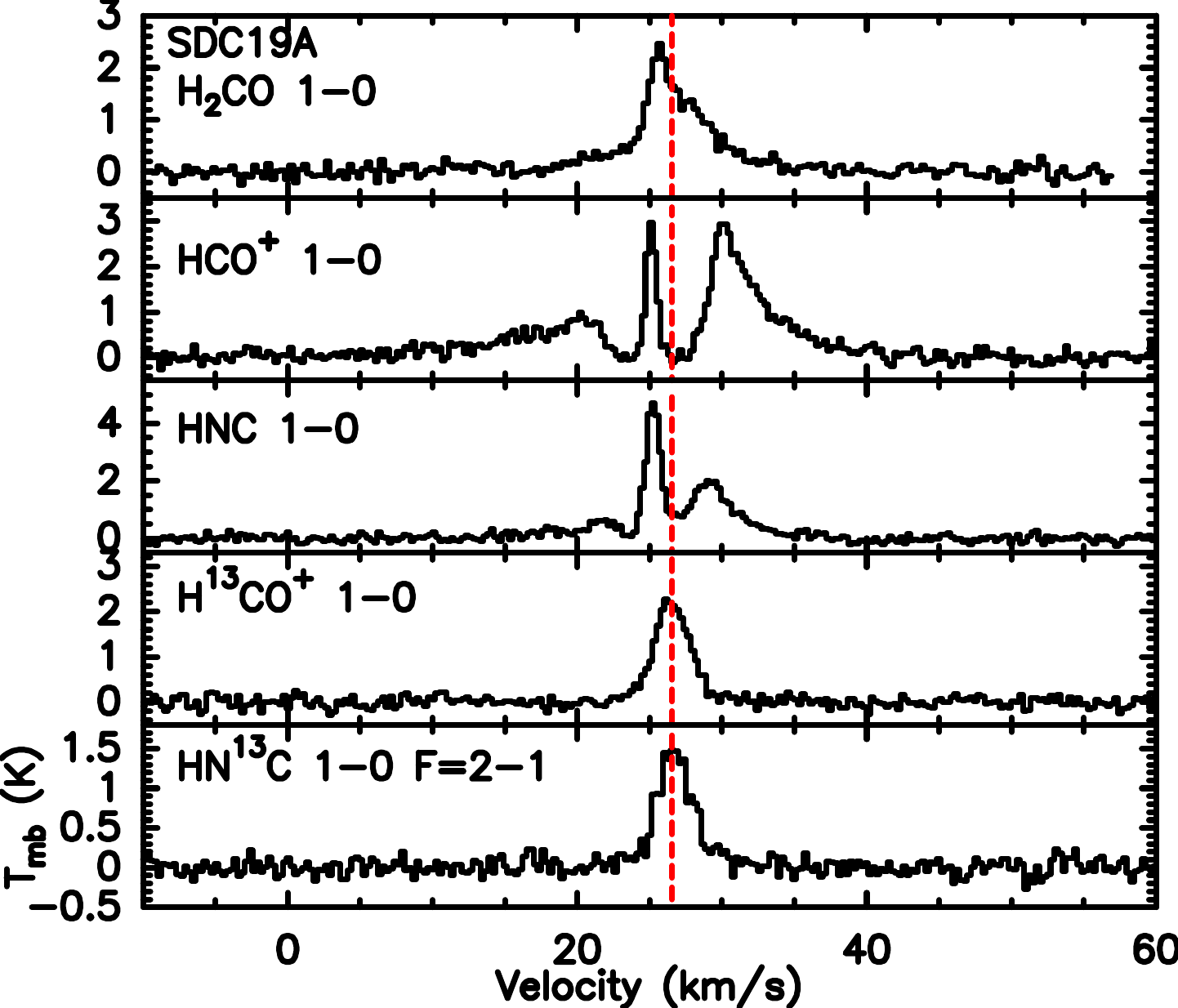}
\includegraphics[width=0.32\textwidth,height=0.39\textwidth,trim={0 0  0 0}, clip]{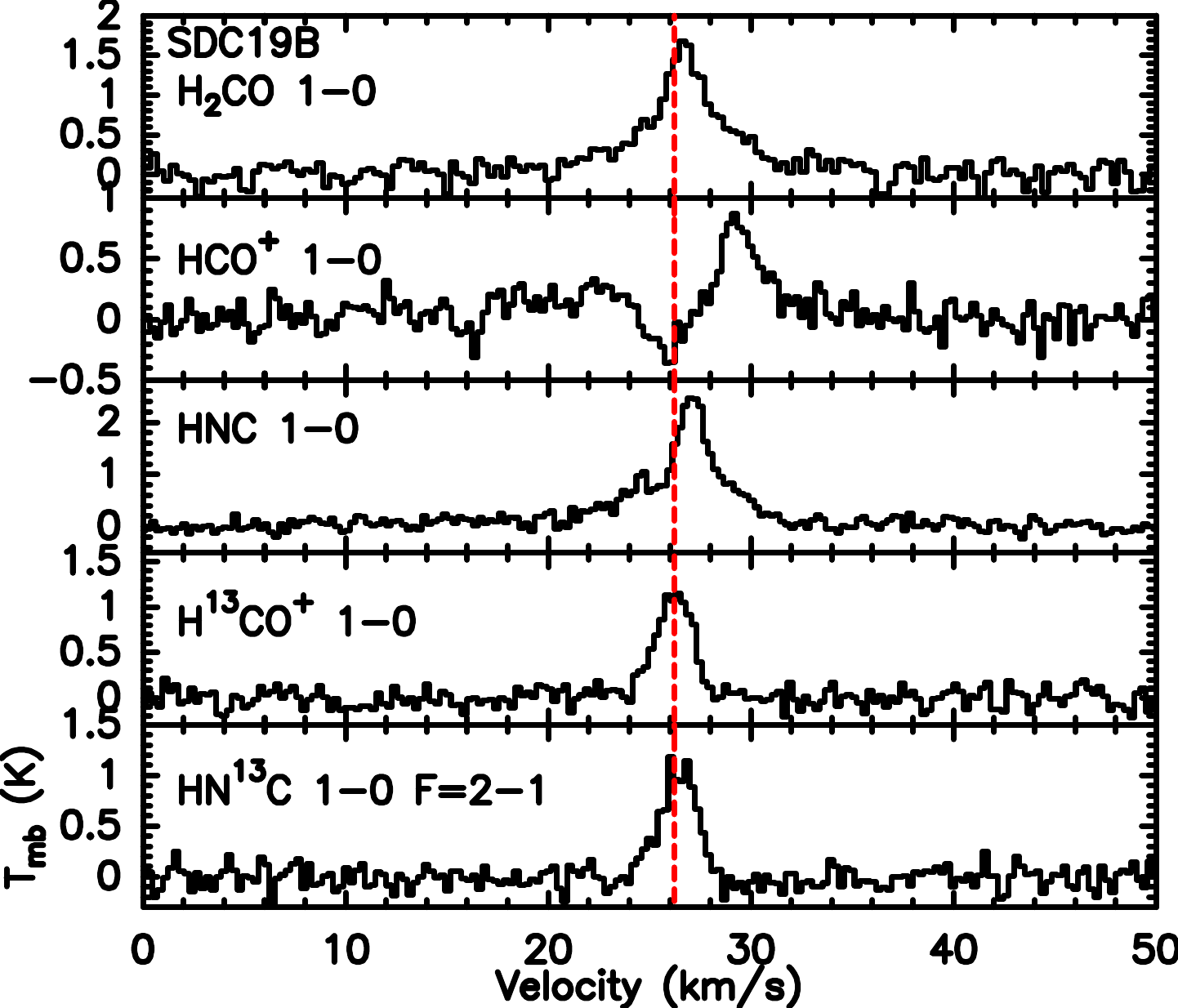}
\includegraphics[width=0.32\textwidth,height=0.39\textwidth,trim={0 0  0 0}, clip]{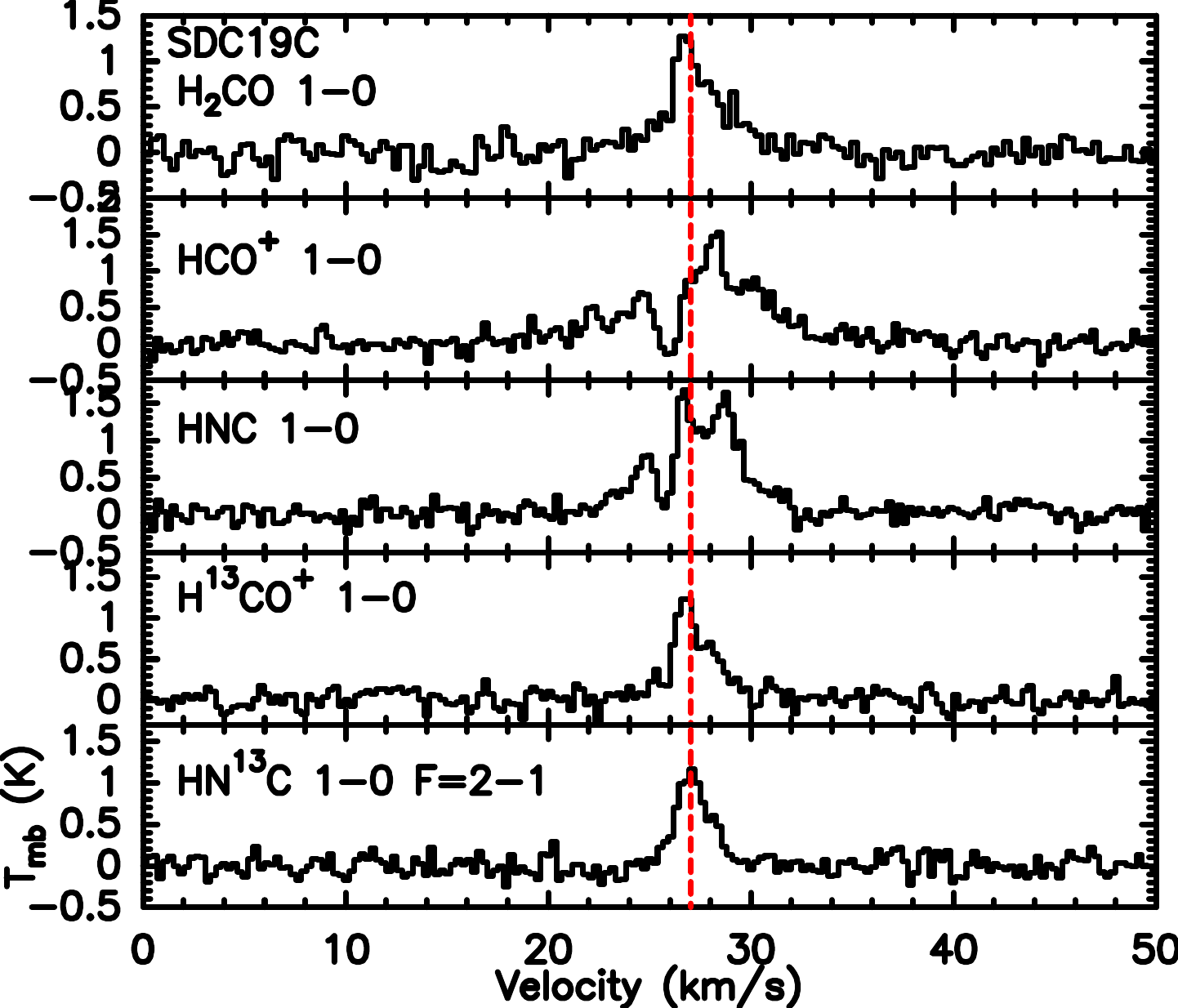}\\[1mm]

\includegraphics[width=0.32\textwidth,height=0.39\textwidth,trim={0 0  0 0}, clip]{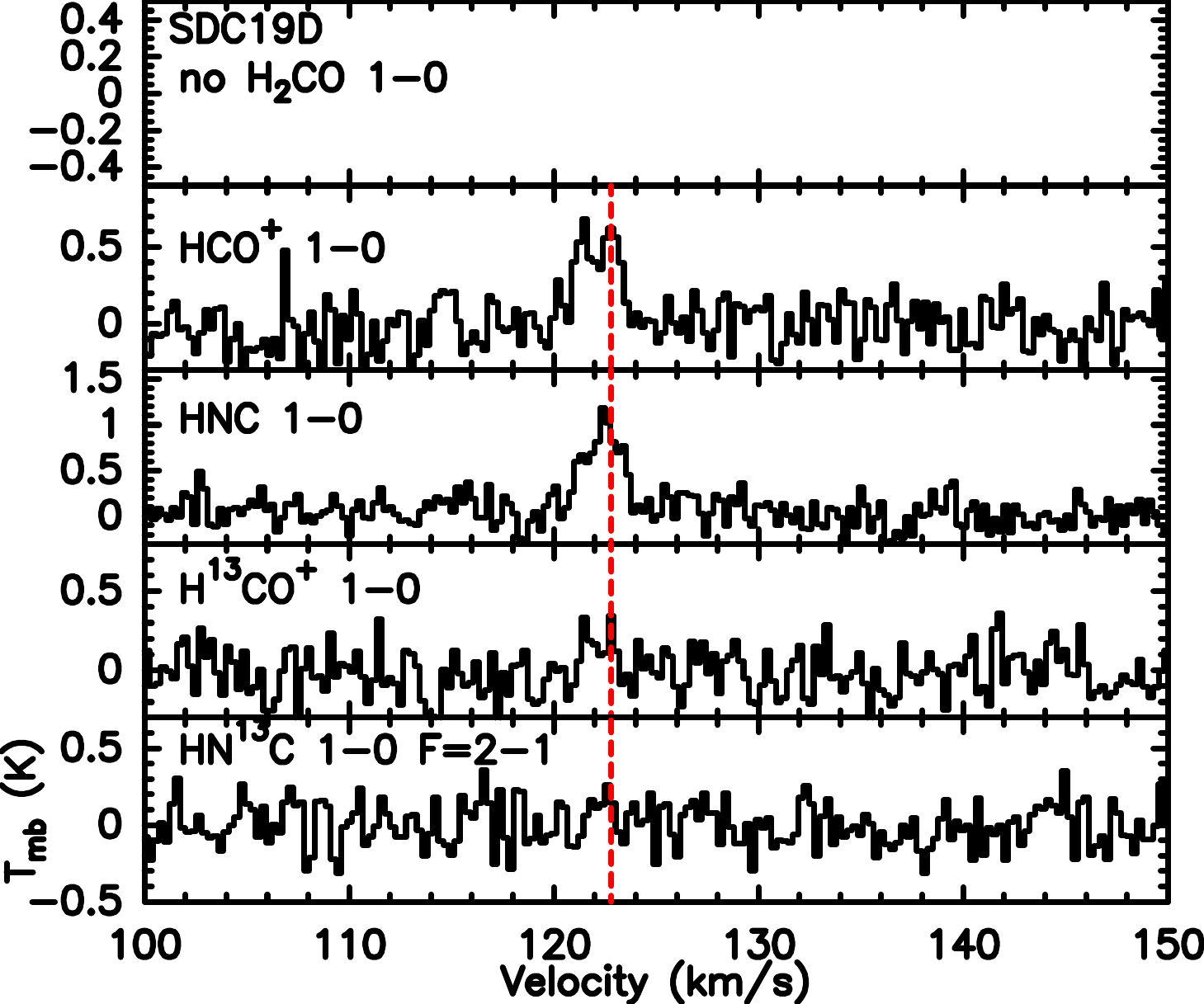}
\includegraphics[width=0.32\textwidth,height=0.39\textwidth,trim={0 0  0 0}, clip]{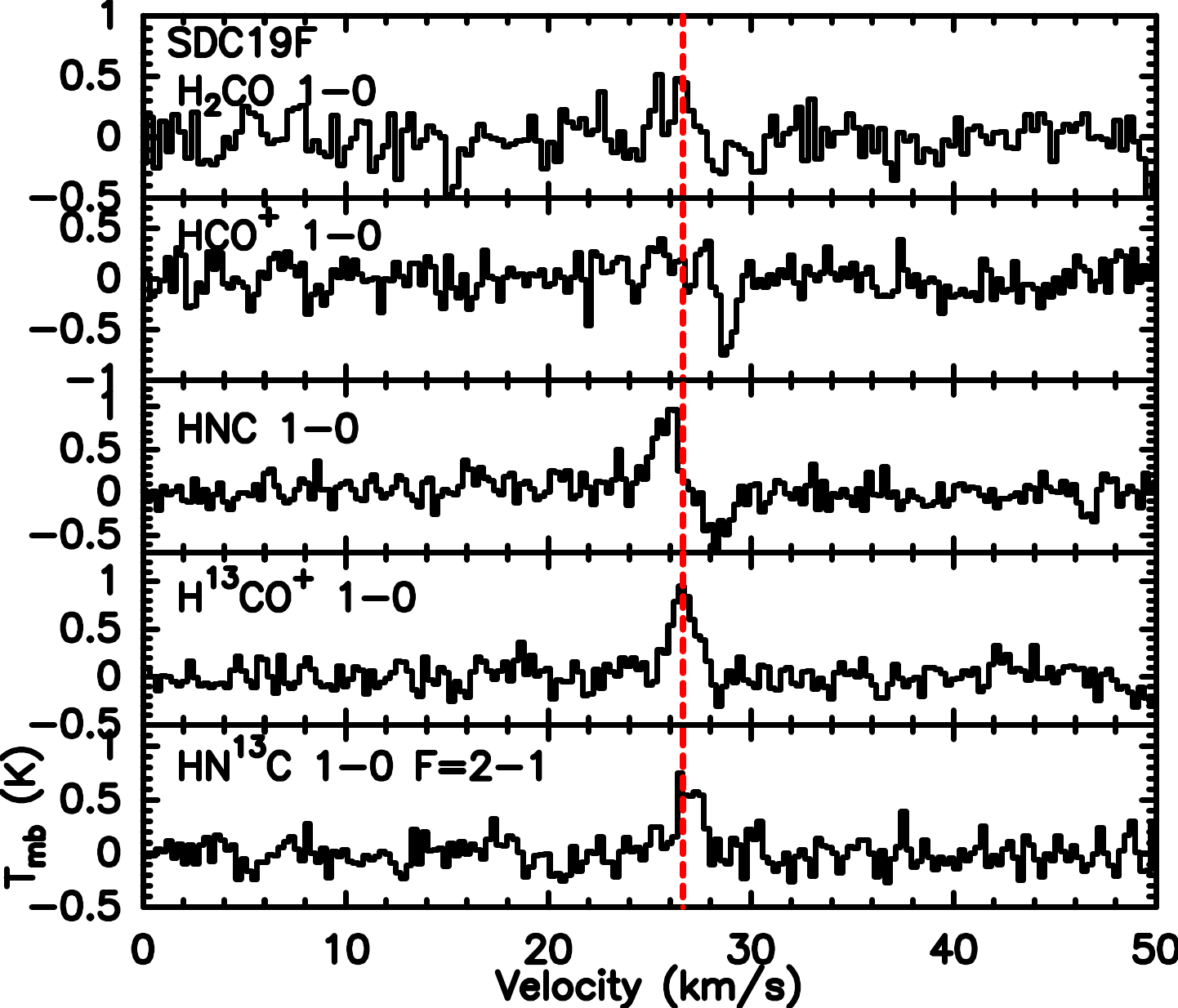}
\includegraphics[width=0.32\textwidth,height=0.39\textwidth,trim={0 0  0 0}, clip]{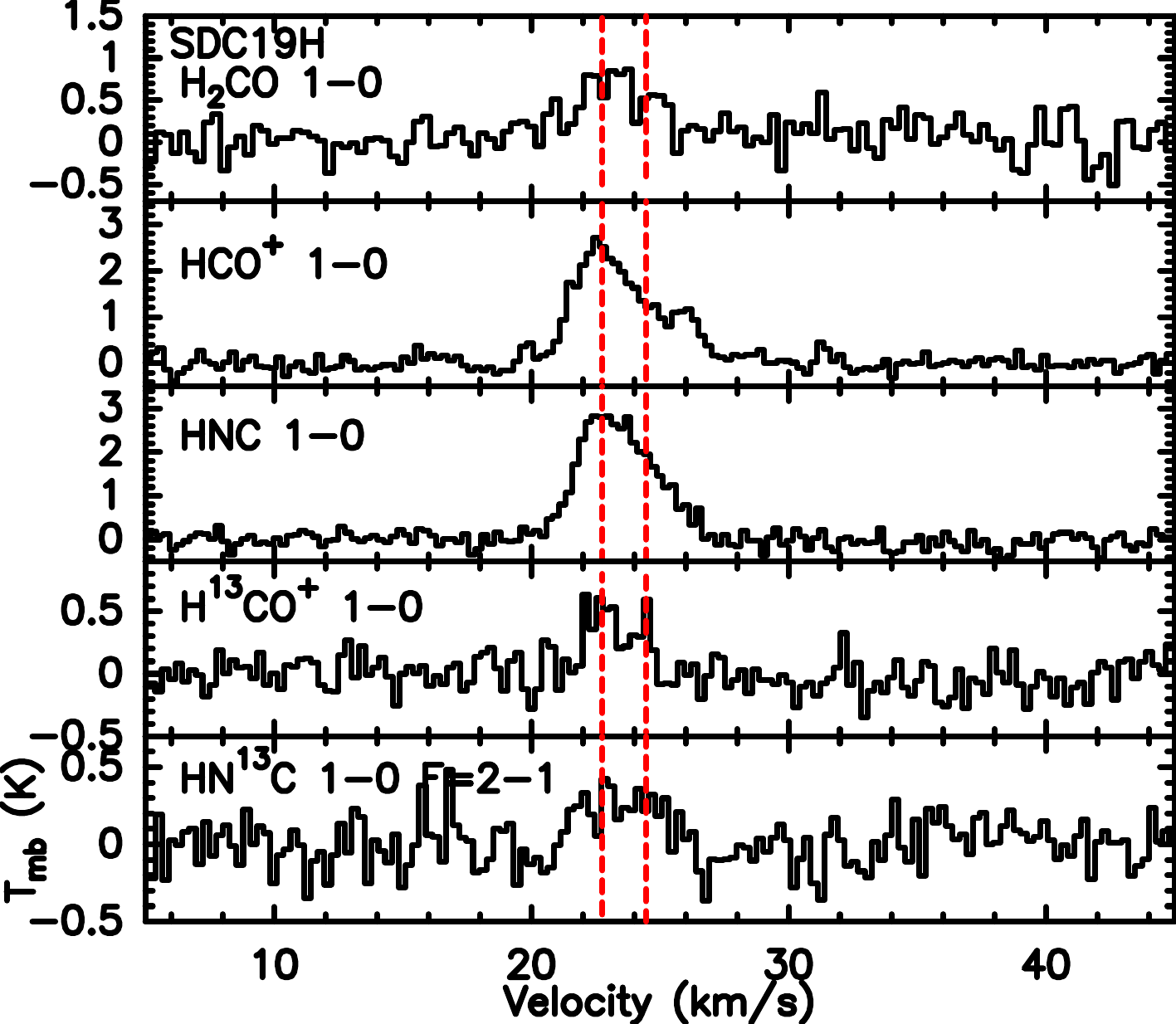}\\[1mm]

\includegraphics[width=0.32\textwidth,height=0.39\textwidth,trim={0 0  0 0}, clip]{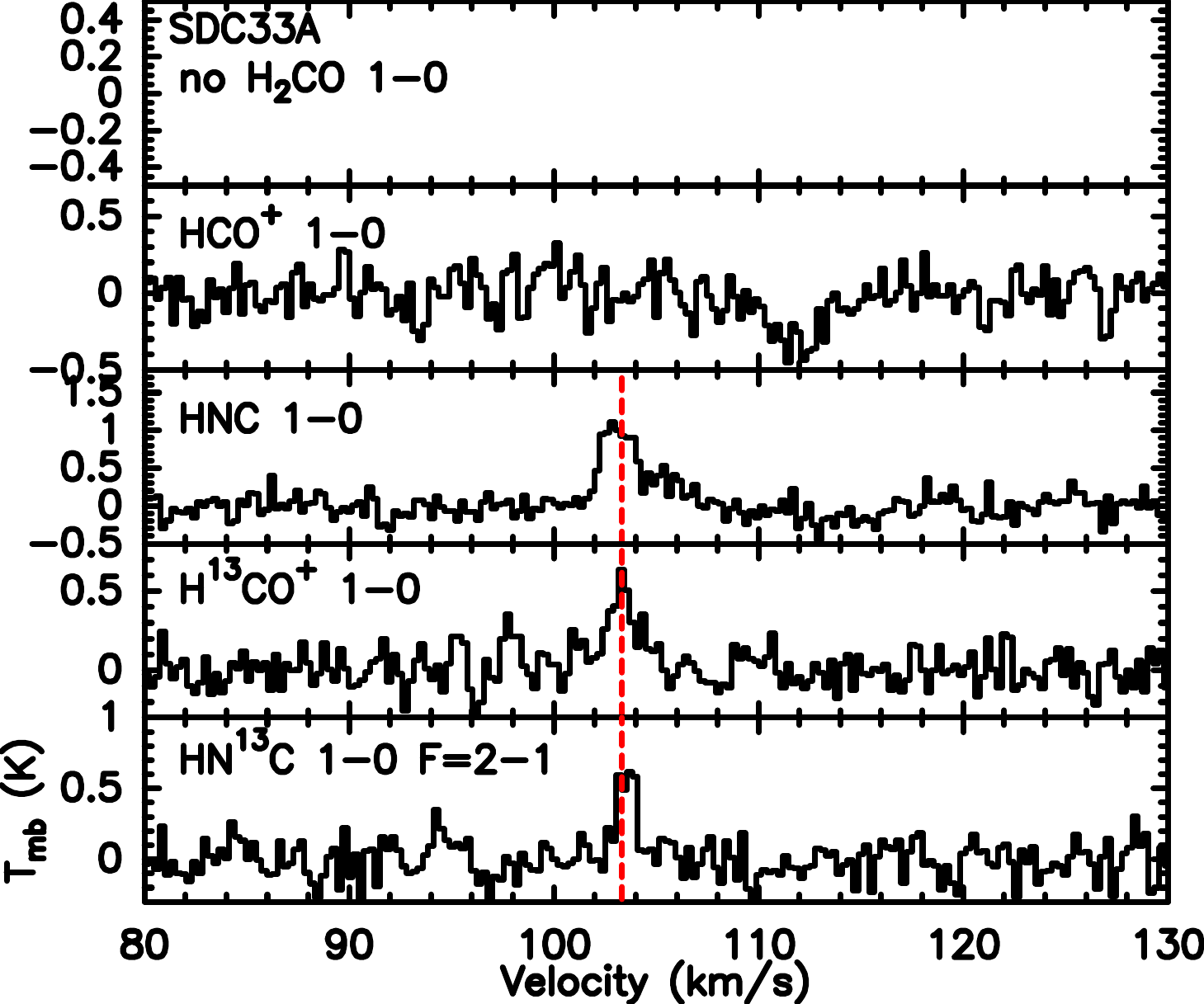}
\includegraphics[width=0.32\textwidth,height=0.39\textwidth,trim={0 0  0 0}, clip]{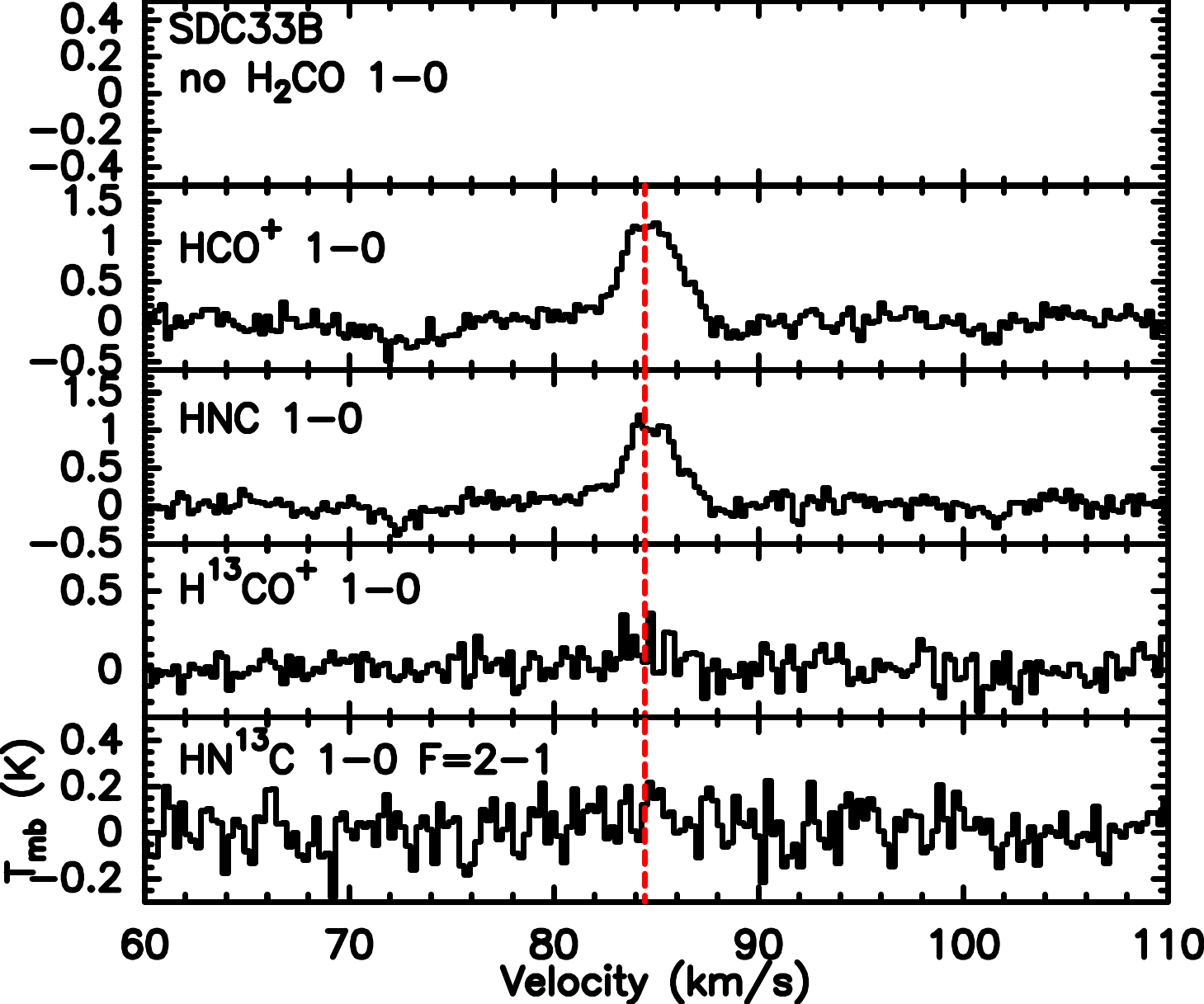}
\includegraphics[width=0.32\textwidth,height=0.39\textwidth,trim={0 0  0 0}, clip]{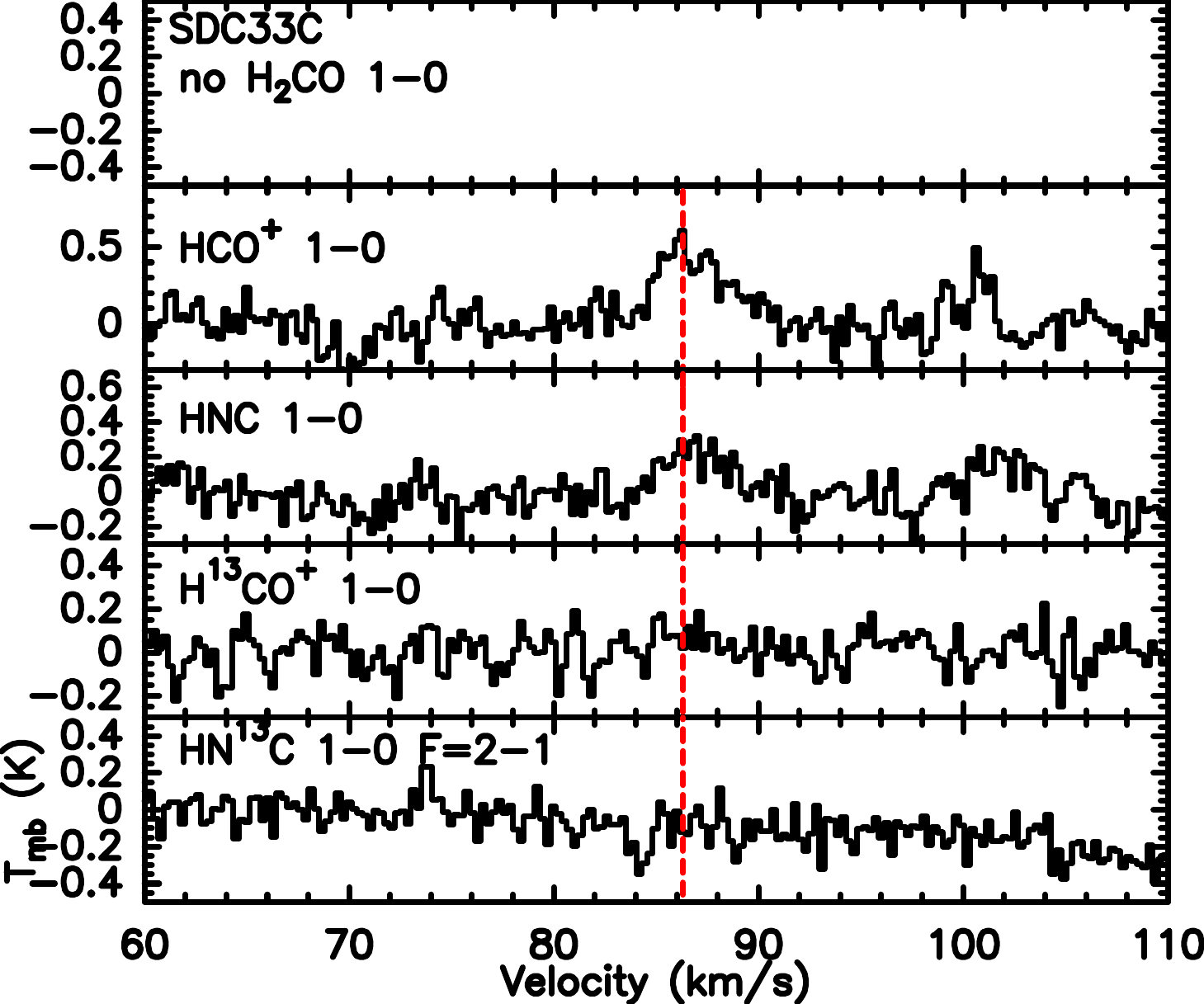}
\caption{Molecular-line spectra toward the remaining ALOHA IRDCs clumps in the sample. The displayed transitions comprise H$_2$CO, HCO$^+$, HNC, H$^{13}$CO$^+$, and HN$^{13}$C, where available. The source name is given above each group of panels, and the molecular transition is indicated in each panel. The spectra are shown on the main-beam brightness-temperature scale as a function of LSR velocity. The complete figure is continued on the following pages.
}
\label{fig:appendix_spectra_1}
\end{figure}

 \begin{figure}
  \centering
\includegraphics[width=0.32\textwidth,height=0.39\textwidth, trim={0 0  0 0}, clip]{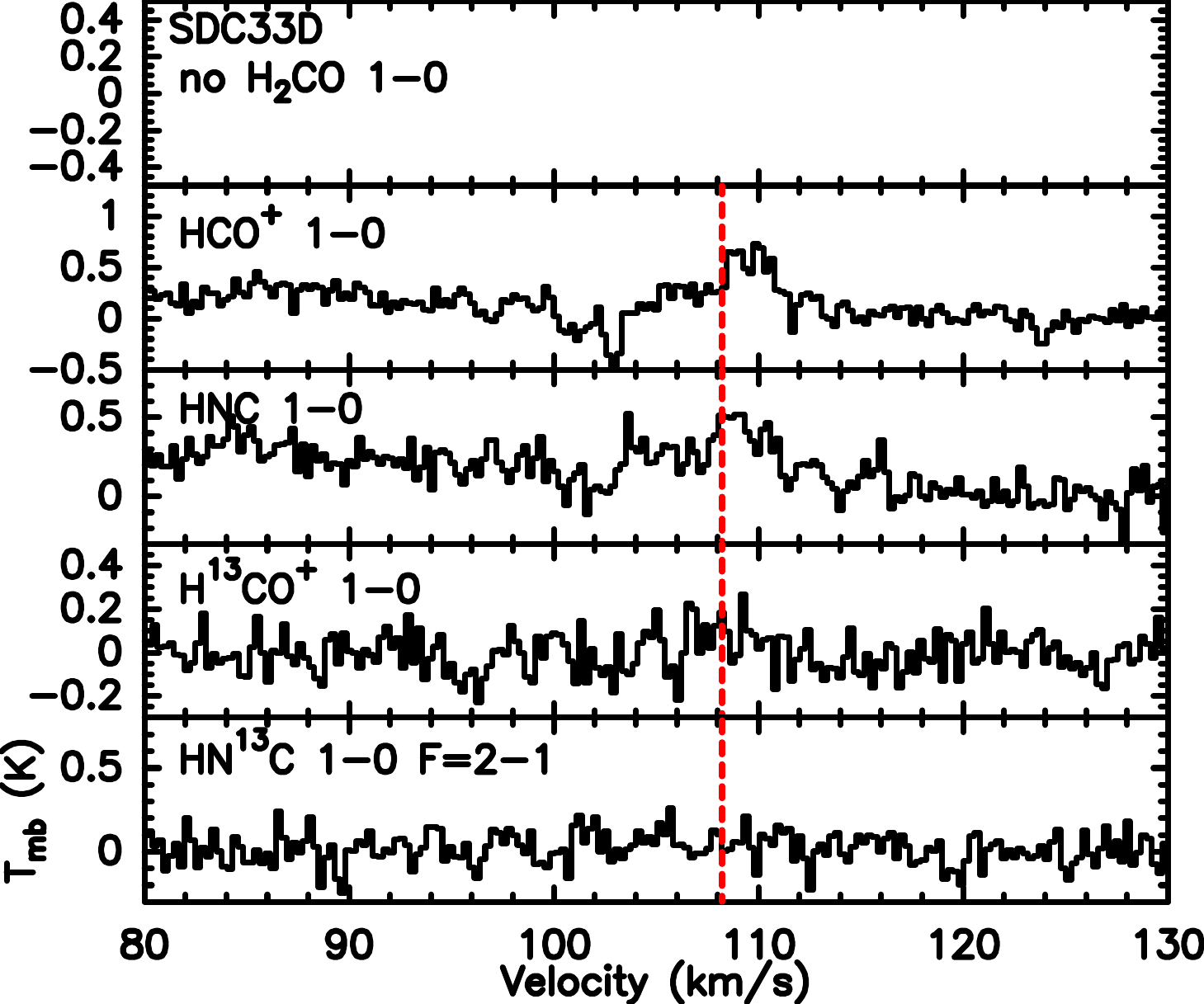}
\includegraphics[width=0.32\textwidth,height=0.39\textwidth,trim={0 0  0 0}, clip]{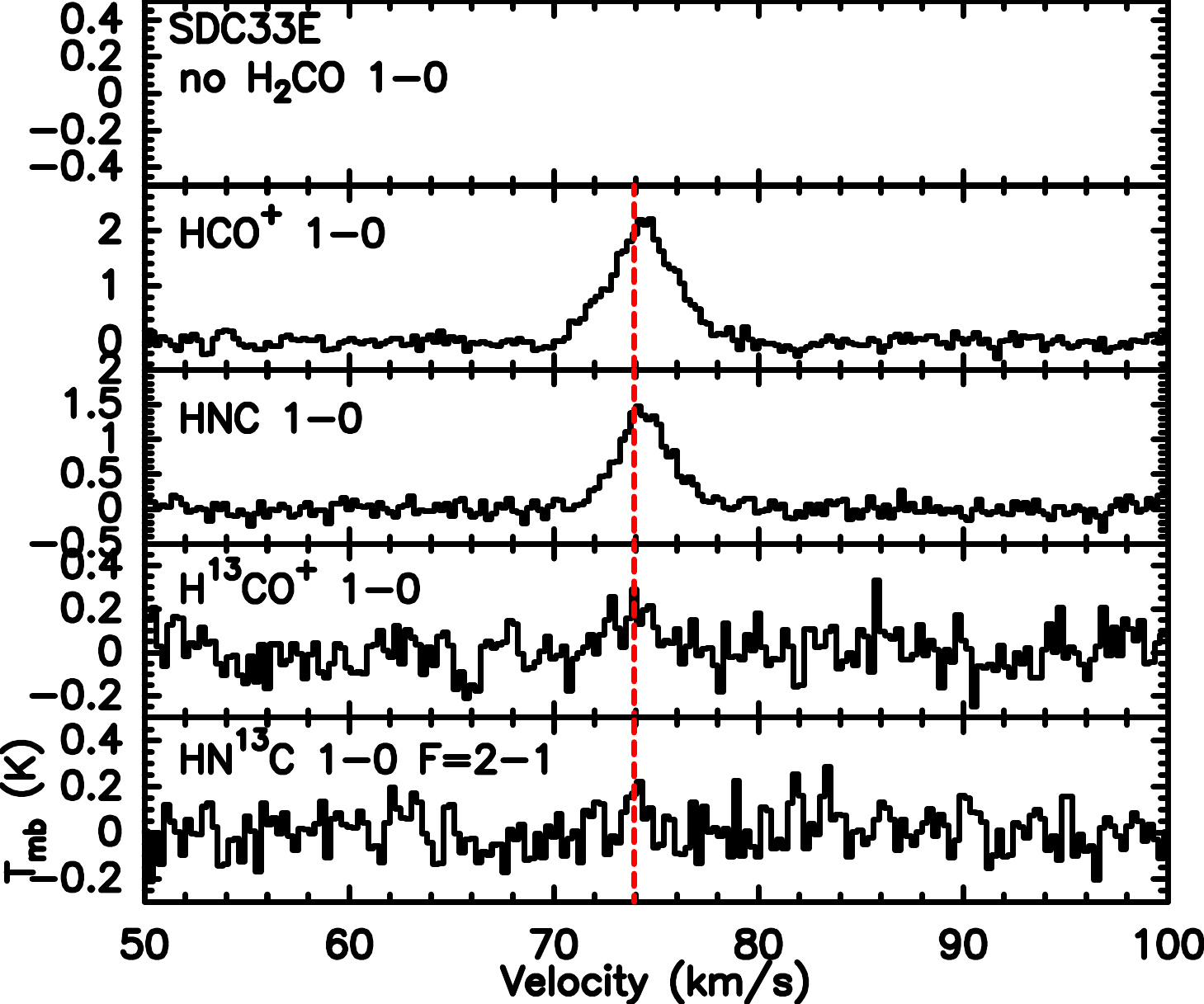}
\includegraphics[width=0.32\textwidth,height=0.39\textwidth,trim={0 0  0 0}, clip]{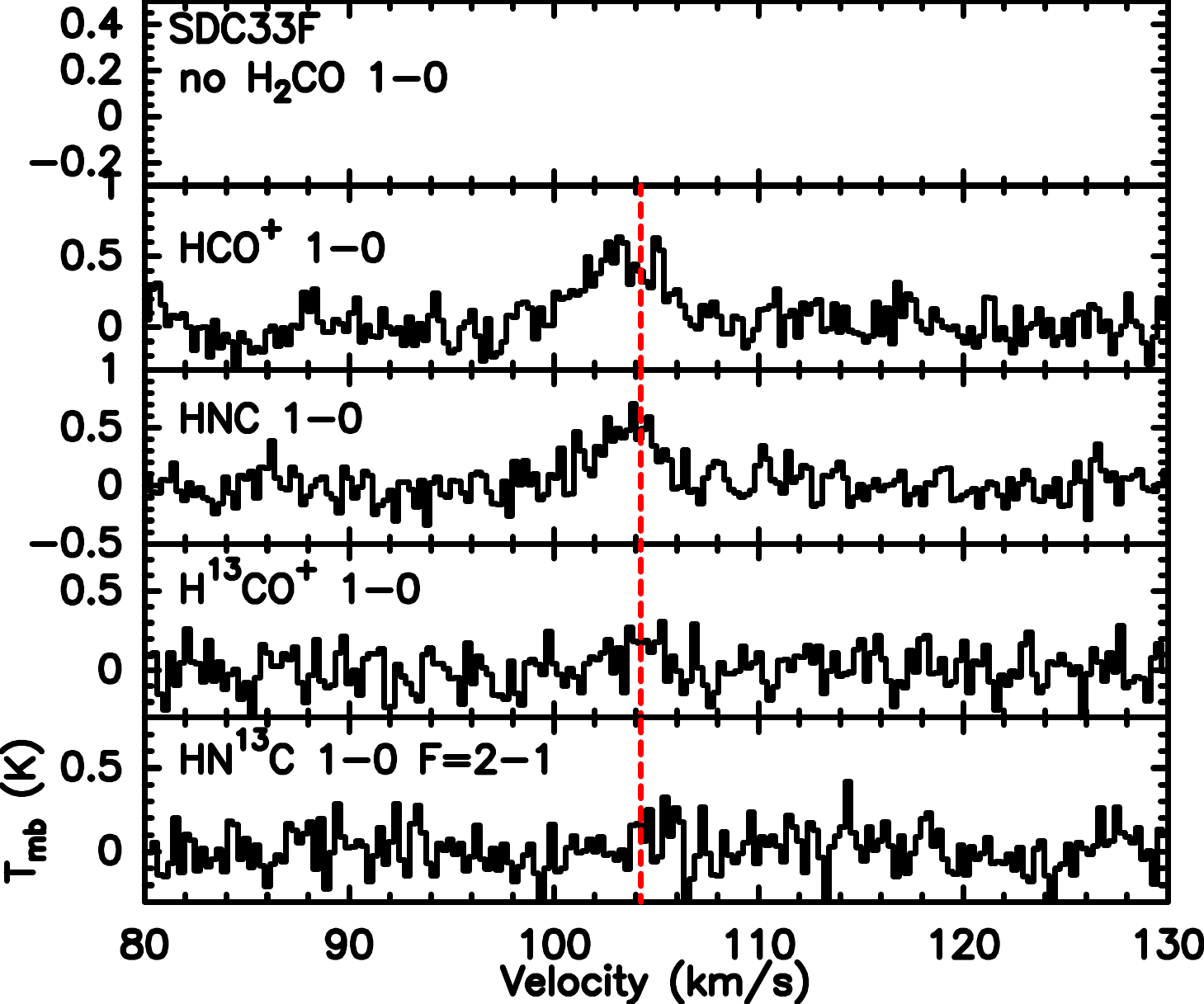}\\[1mm]

\includegraphics[width=0.32\textwidth,height=0.39\textwidth,trim={0 0  0 0}, clip]{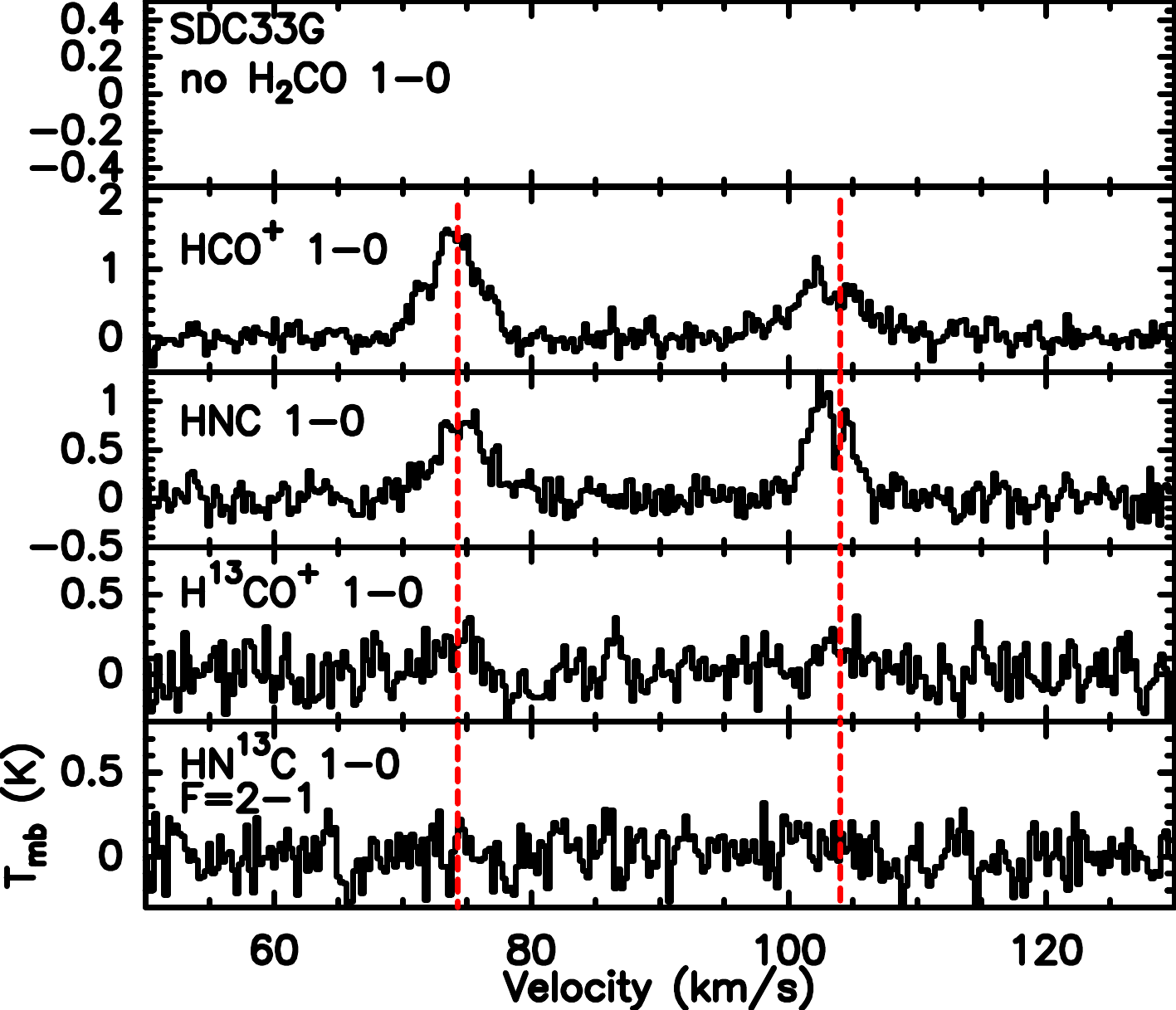}
\includegraphics[width=0.32\textwidth,height=0.39\textwidth,trim={0 0  0 0}, clip]{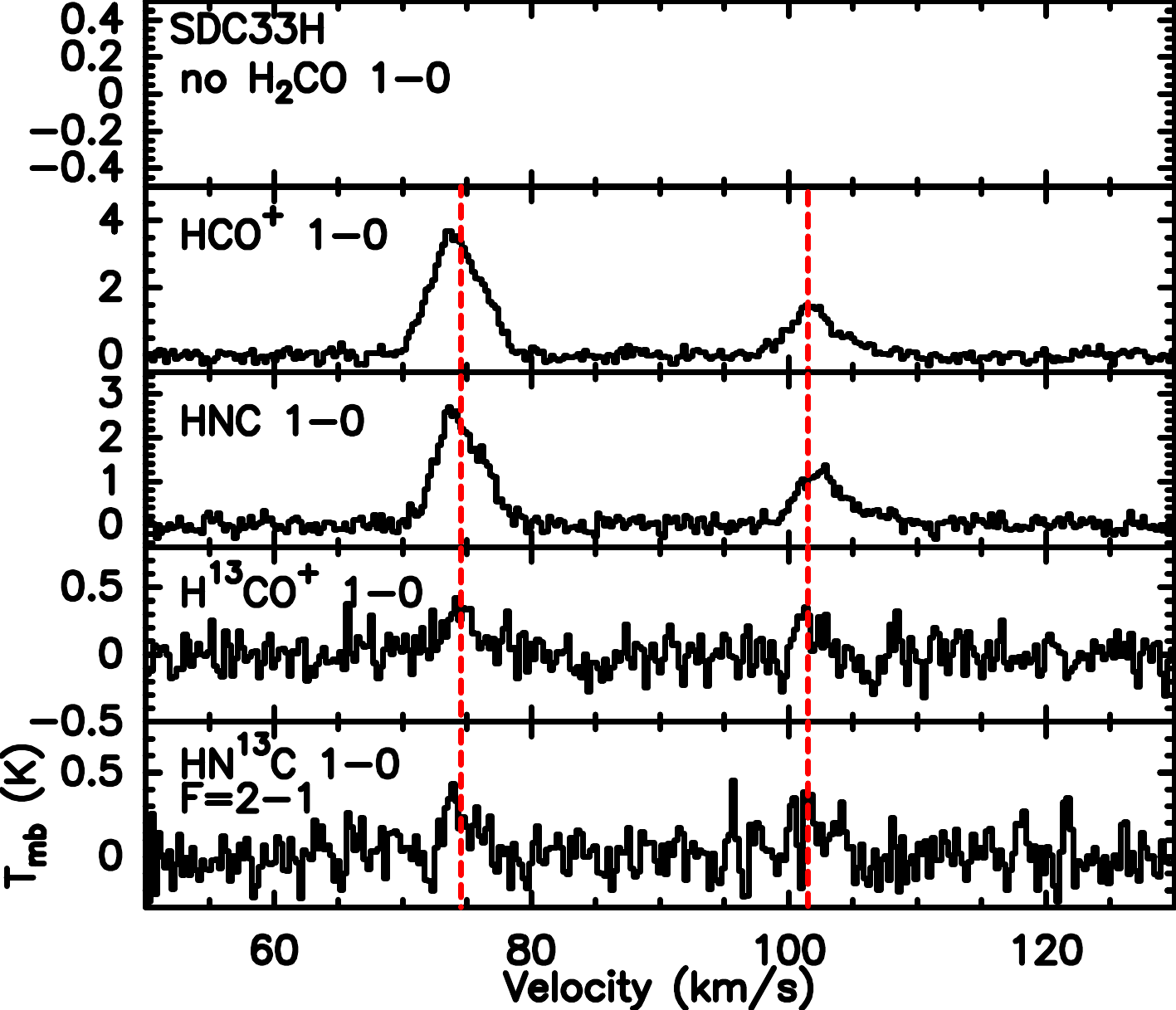}
\includegraphics[width=0.32\textwidth,height=0.39\textwidth,trim={0 0  0 0}, clip]{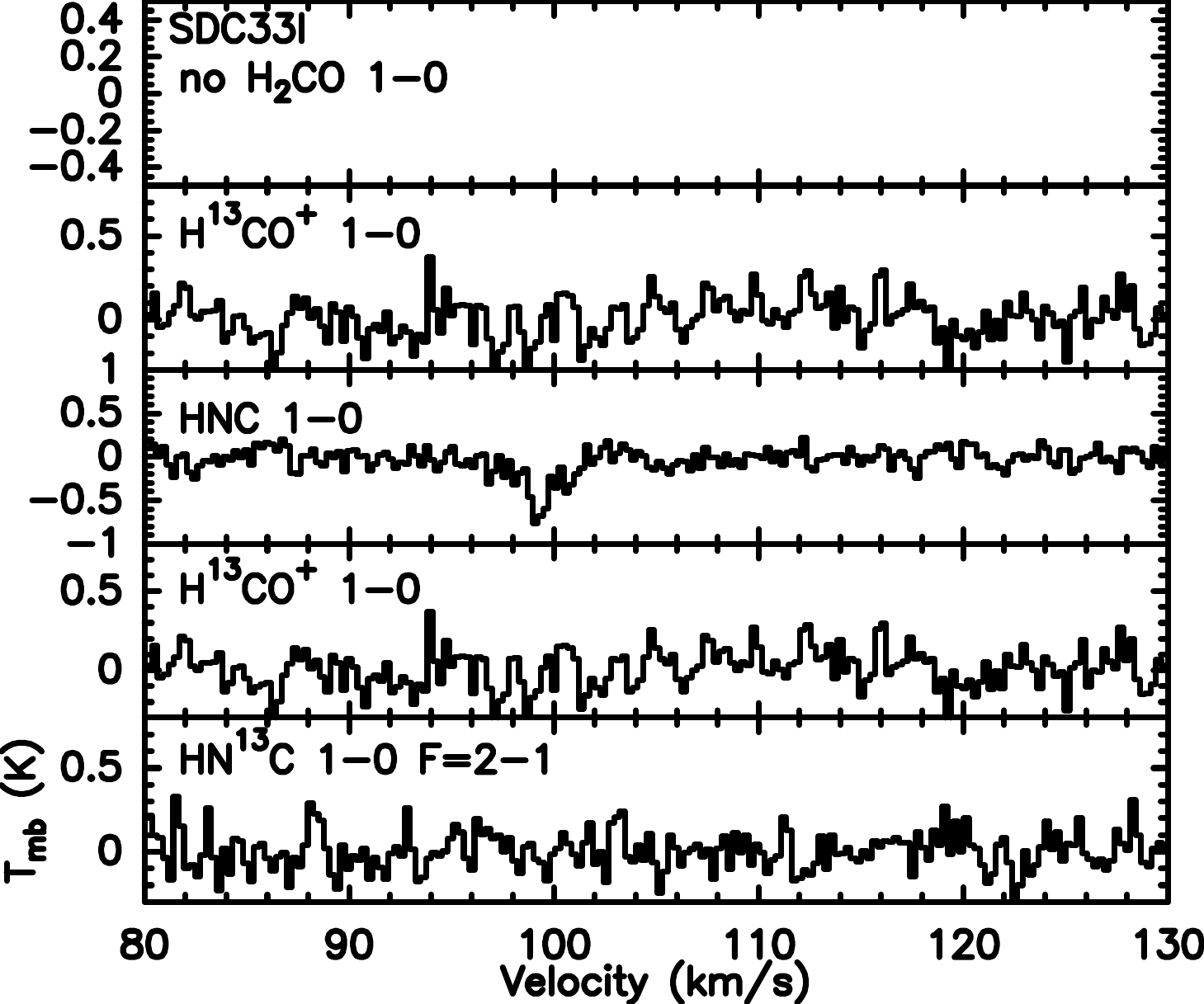}\\[1mm]

\includegraphics[width=0.32\textwidth,height=0.39\textwidth,trim={0 0  0 0}, clip]{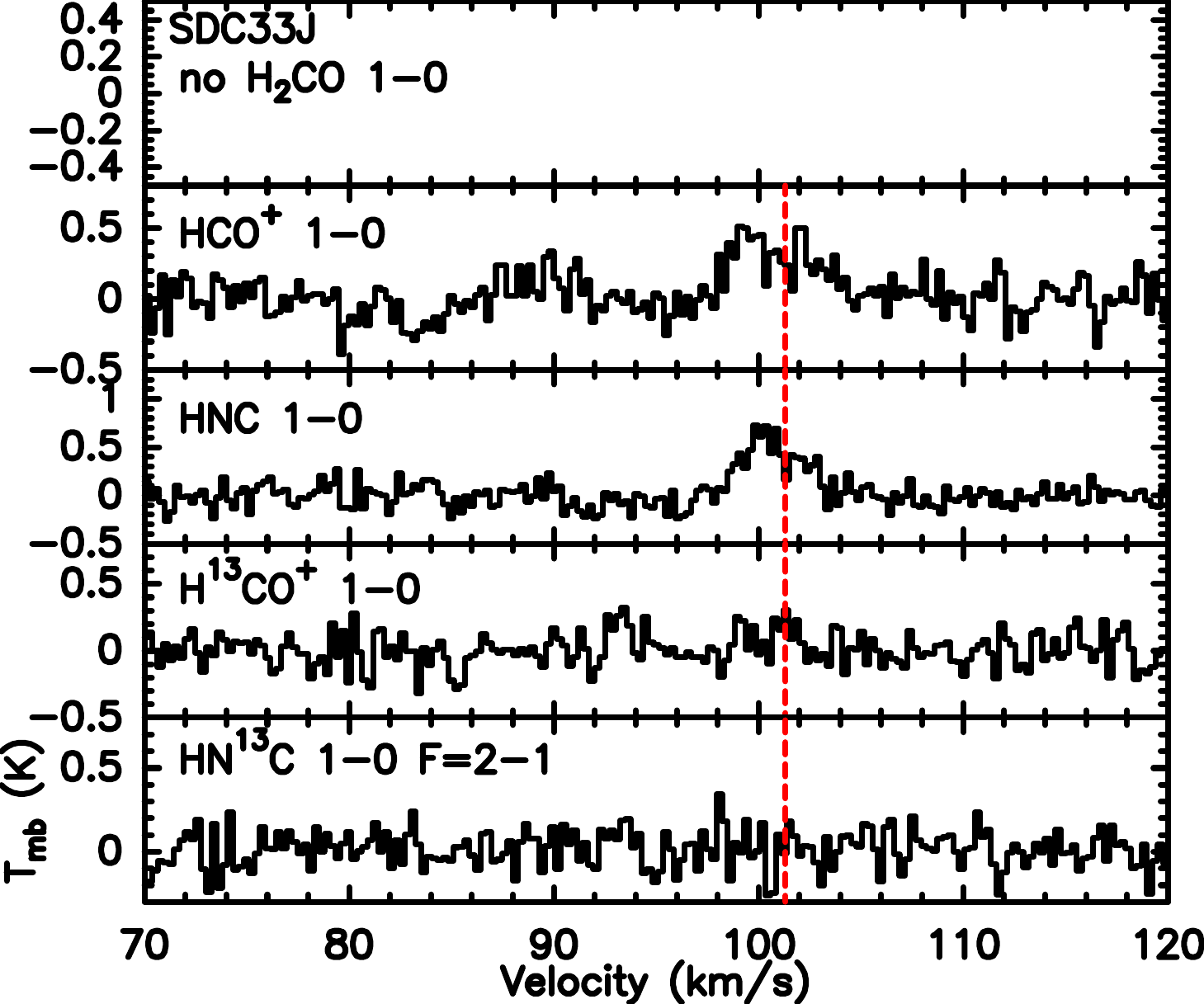}
\includegraphics[width=0.32\textwidth,height=0.39\textwidth,trim={0 0  0 0}, clip]{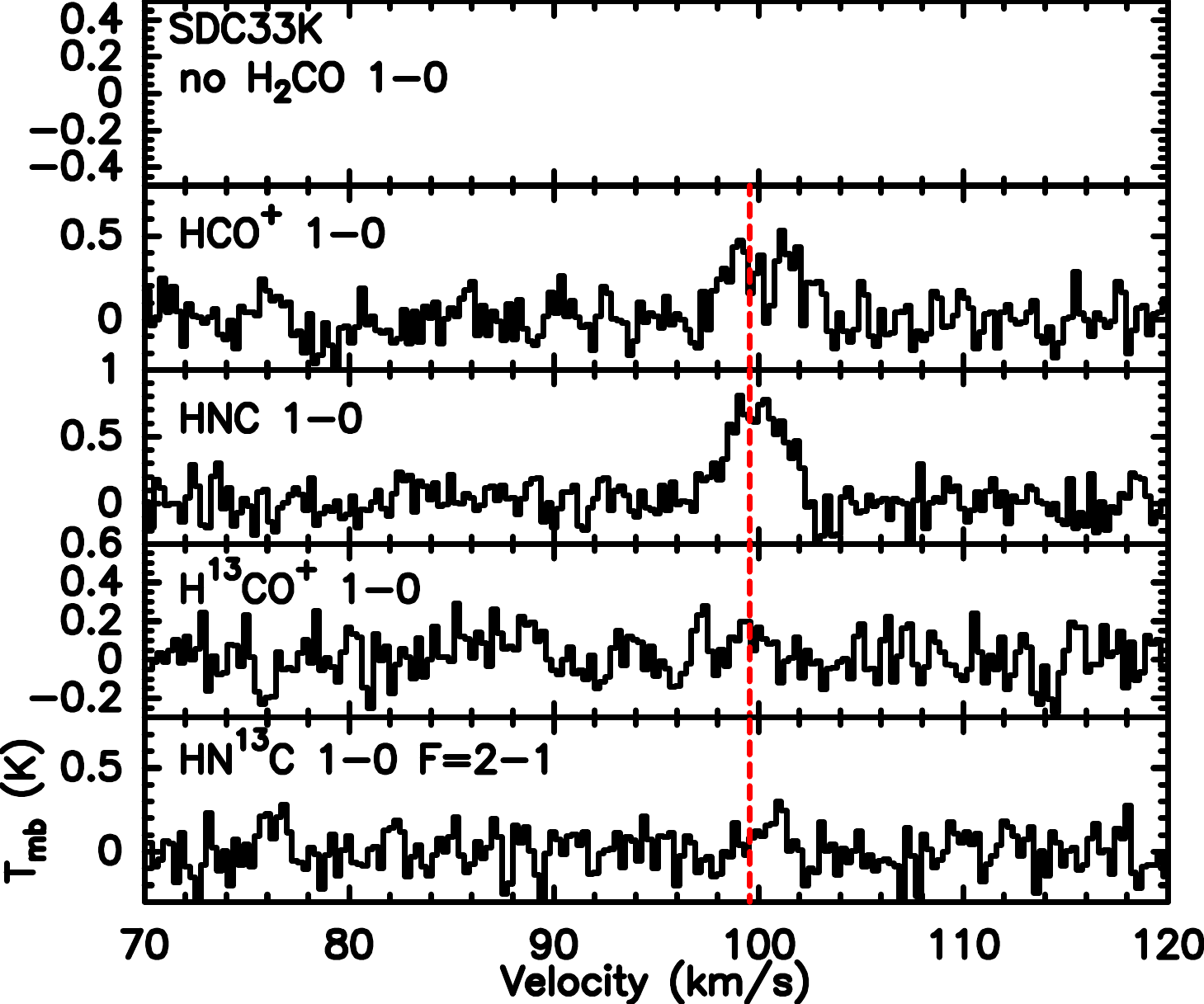}
\includegraphics[width=0.32\textwidth,height=0.39\textwidth,trim={0 0  0 0}, clip]{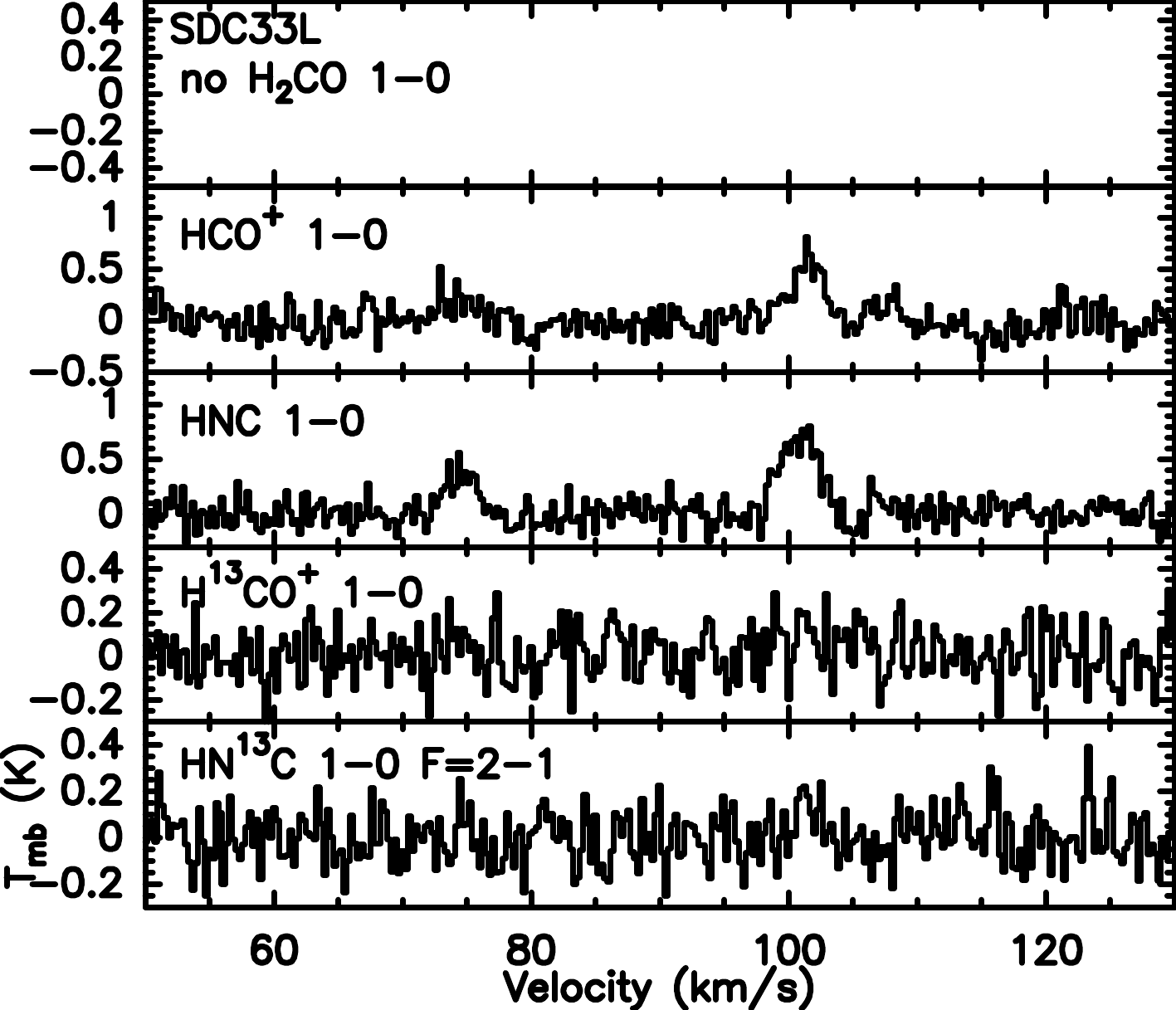}
\caption{Continued. Molecular-line spectra toward the ALOHA IRDCs clumps.
}
\label{fig:appendix_spectra_2}
\end{figure}

 \begin{figure}
  \centering
\includegraphics[width=0.32\textwidth,height=0.39\textwidth, trim={0 0  0 0}, clip]{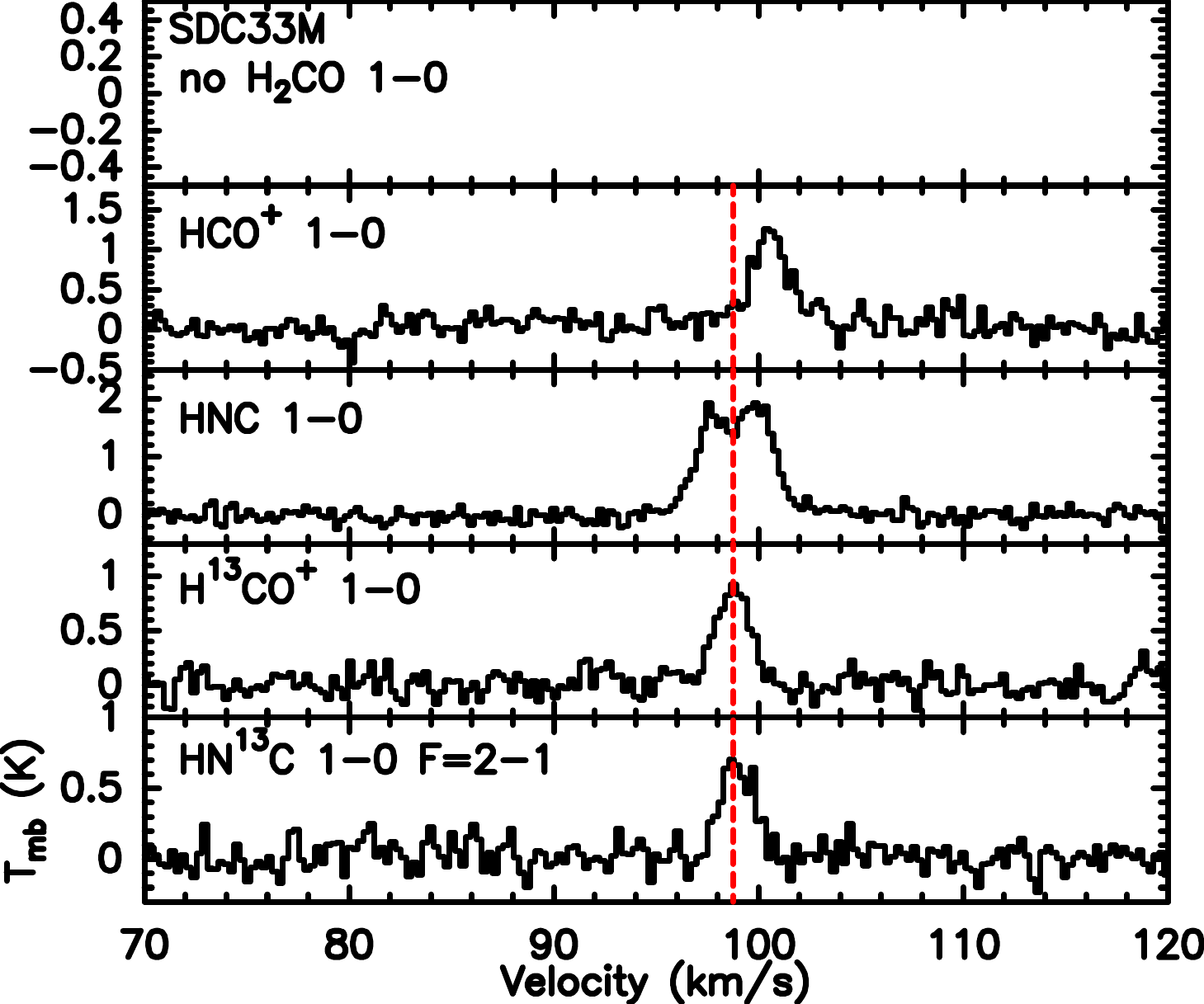}
\includegraphics[width=0.32\textwidth,height=0.39\textwidth,trim={0 0  0 0}, clip]{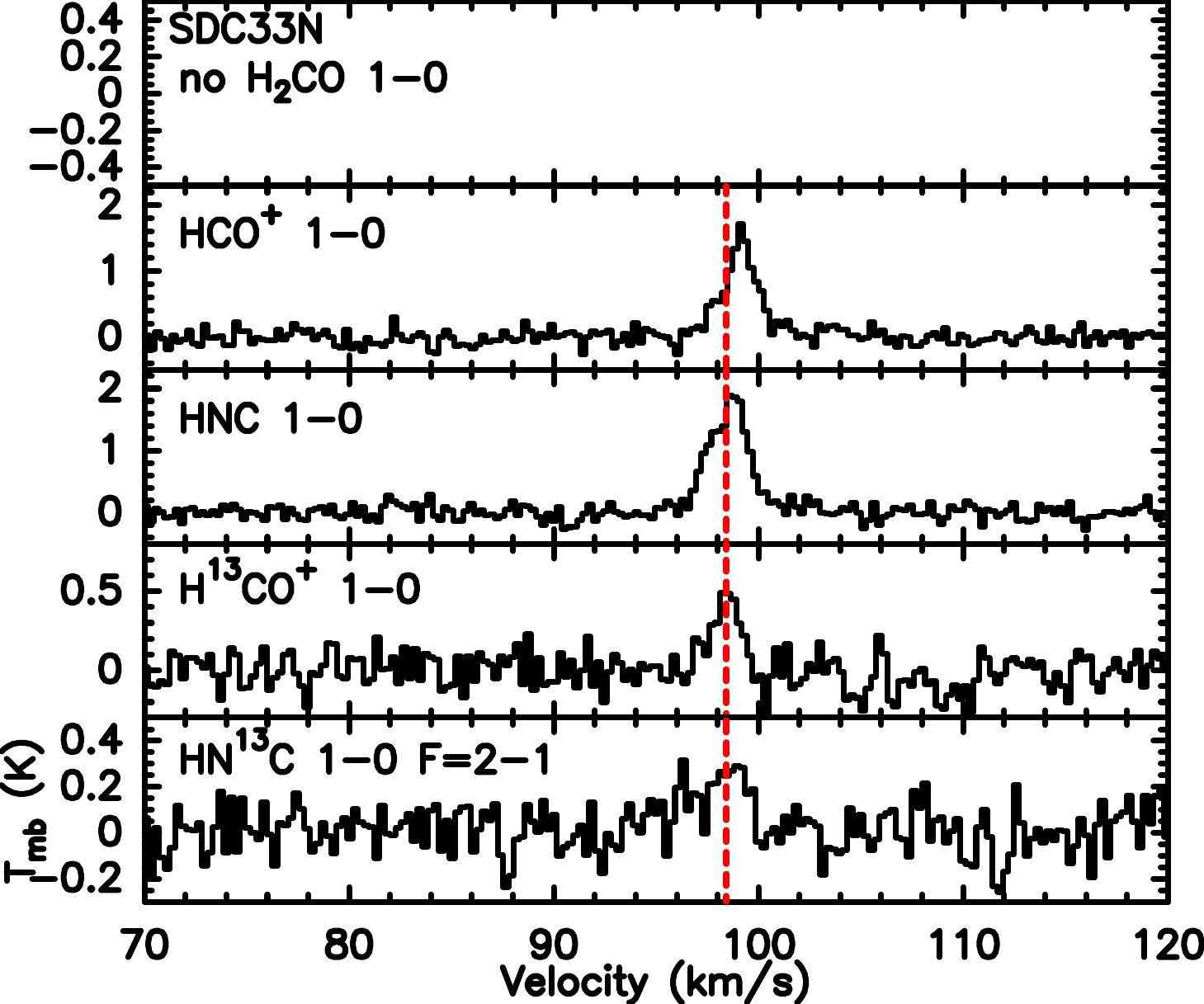}
\includegraphics[width=0.32\textwidth,height=0.39\textwidth,trim={0 0  0 0}, clip]{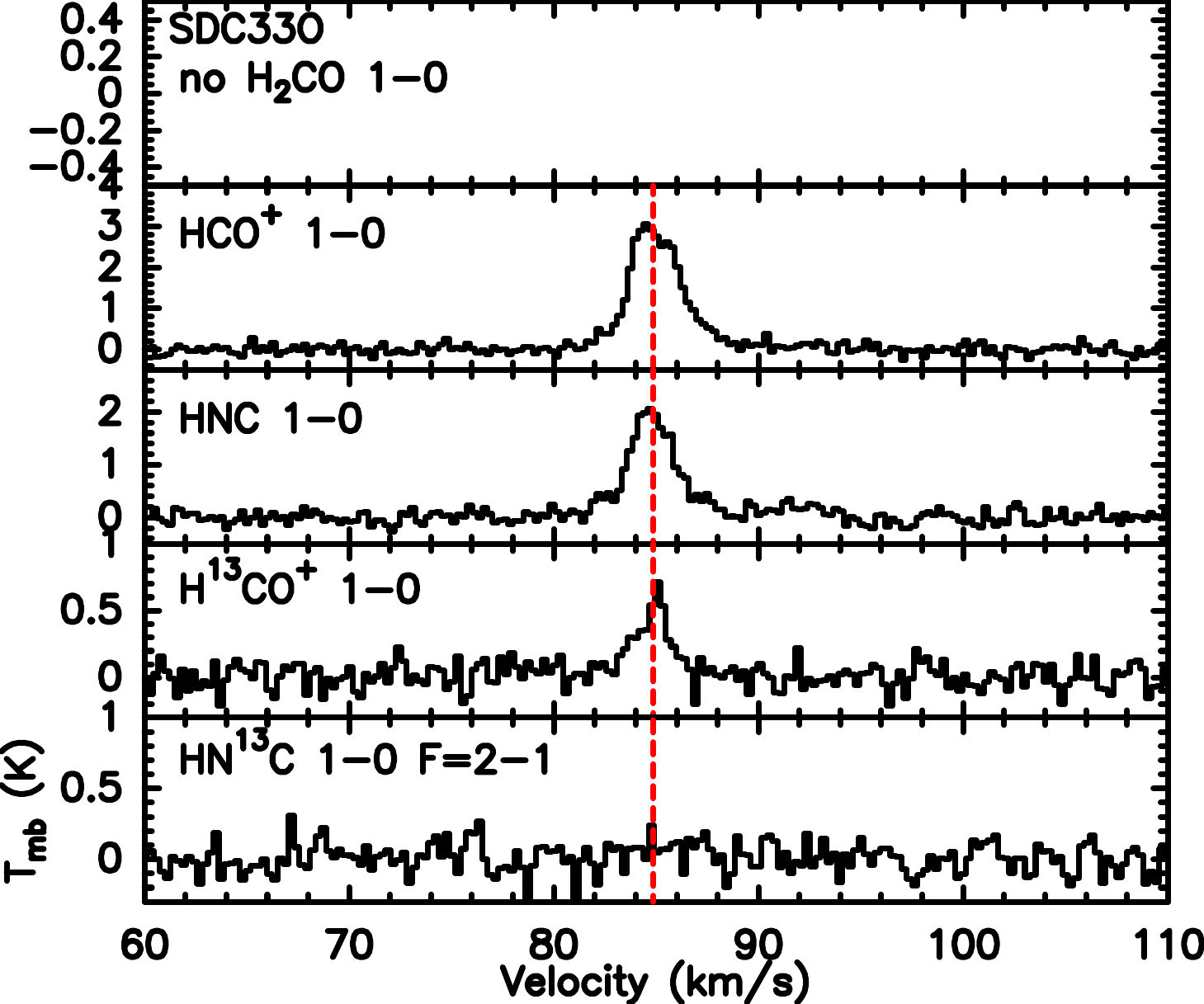}\\[1mm]

\includegraphics[width=0.32\textwidth,height=0.39\textwidth,trim={0 0  0 0}, clip]{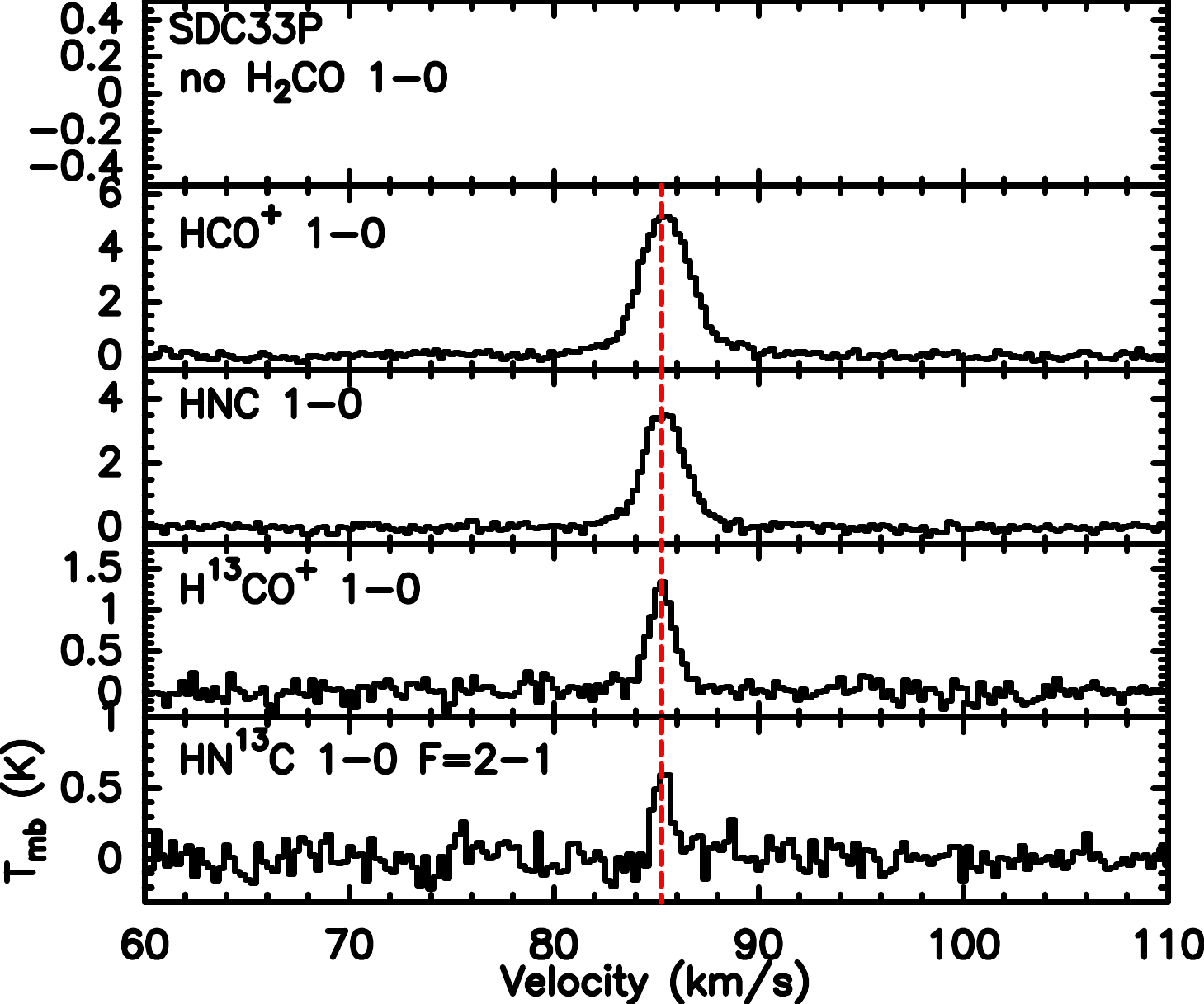}
\includegraphics[width=0.32\textwidth,height=0.39\textwidth,trim={0 0  0 0}, clip]{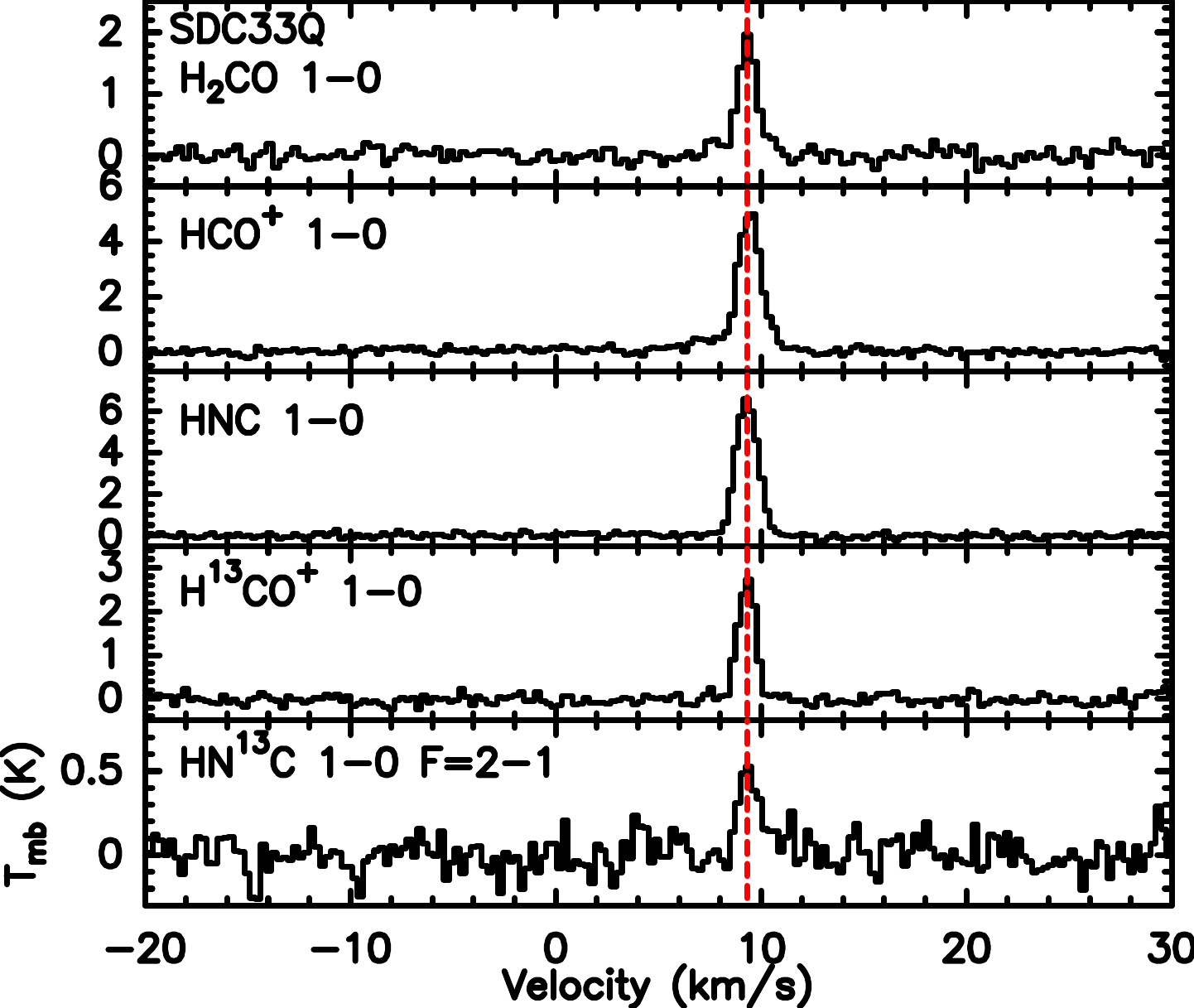}
\includegraphics[width=0.32\textwidth,height=0.39\textwidth,trim={0 0  0 0}, clip]{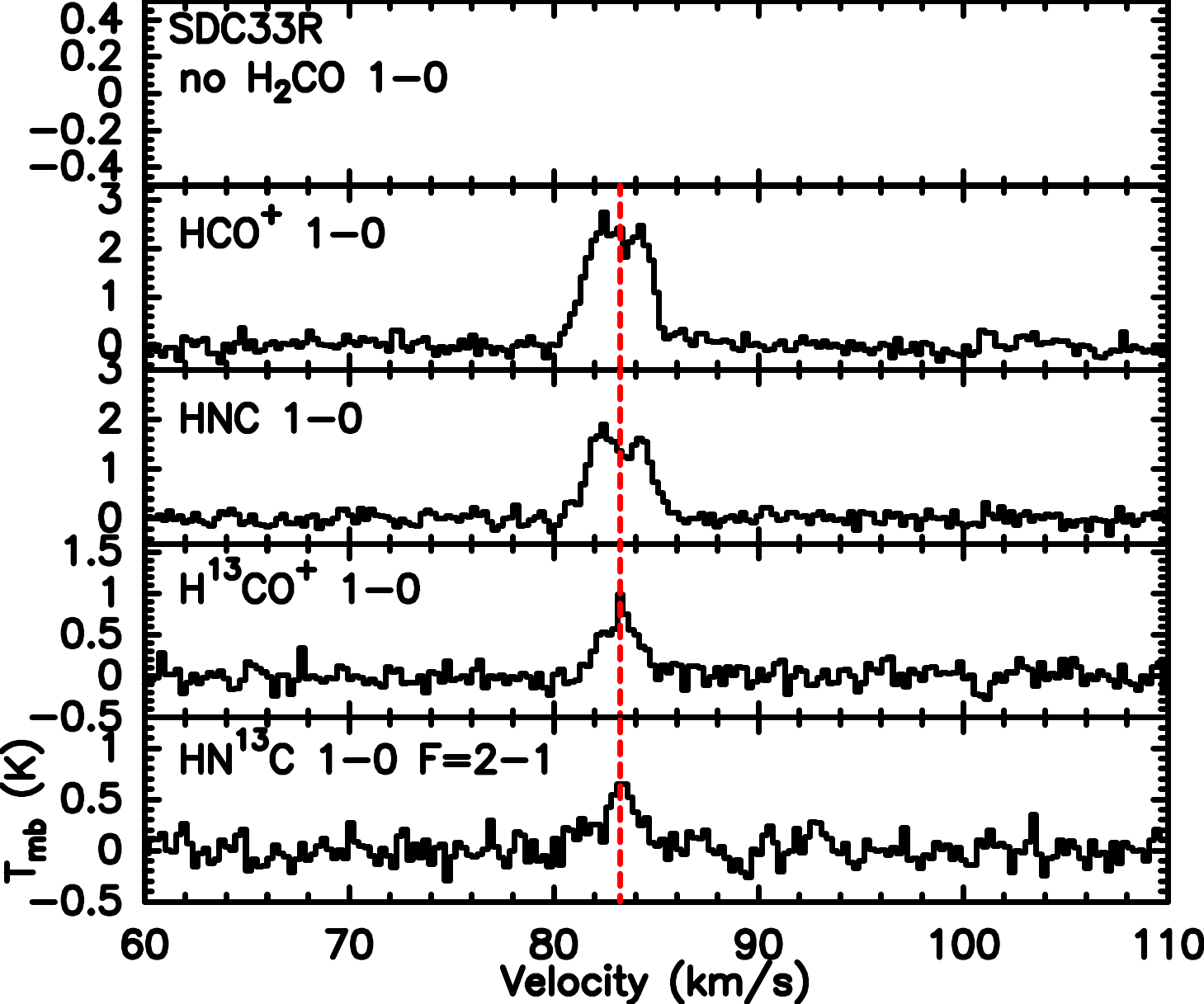}\\[1mm]

\includegraphics[width=0.32\textwidth,height=0.39\textwidth,trim={0 0  0 0}, clip]{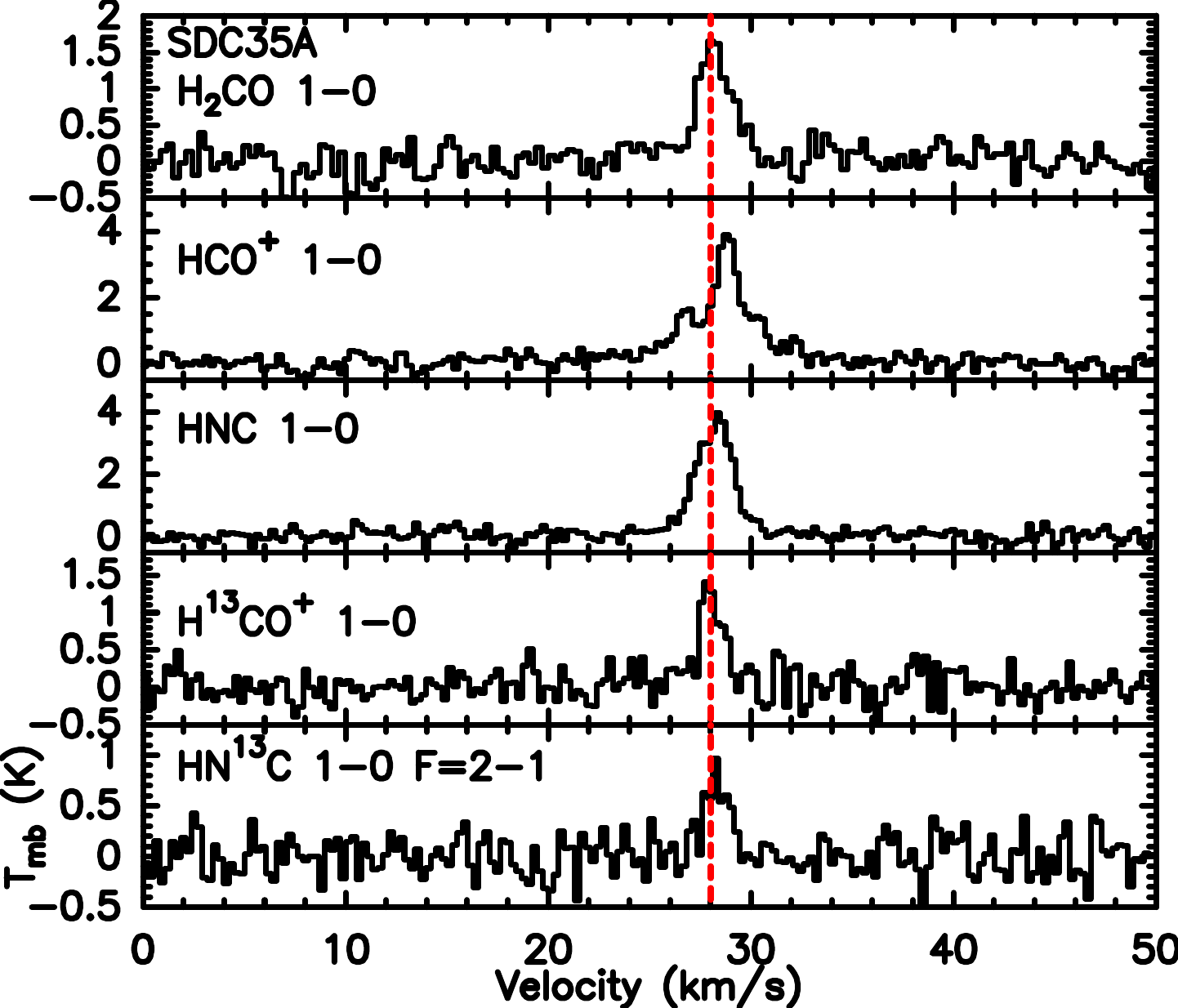}
\includegraphics[width=0.32\textwidth,height=0.39\textwidth,trim={0 0  0 0}, clip]{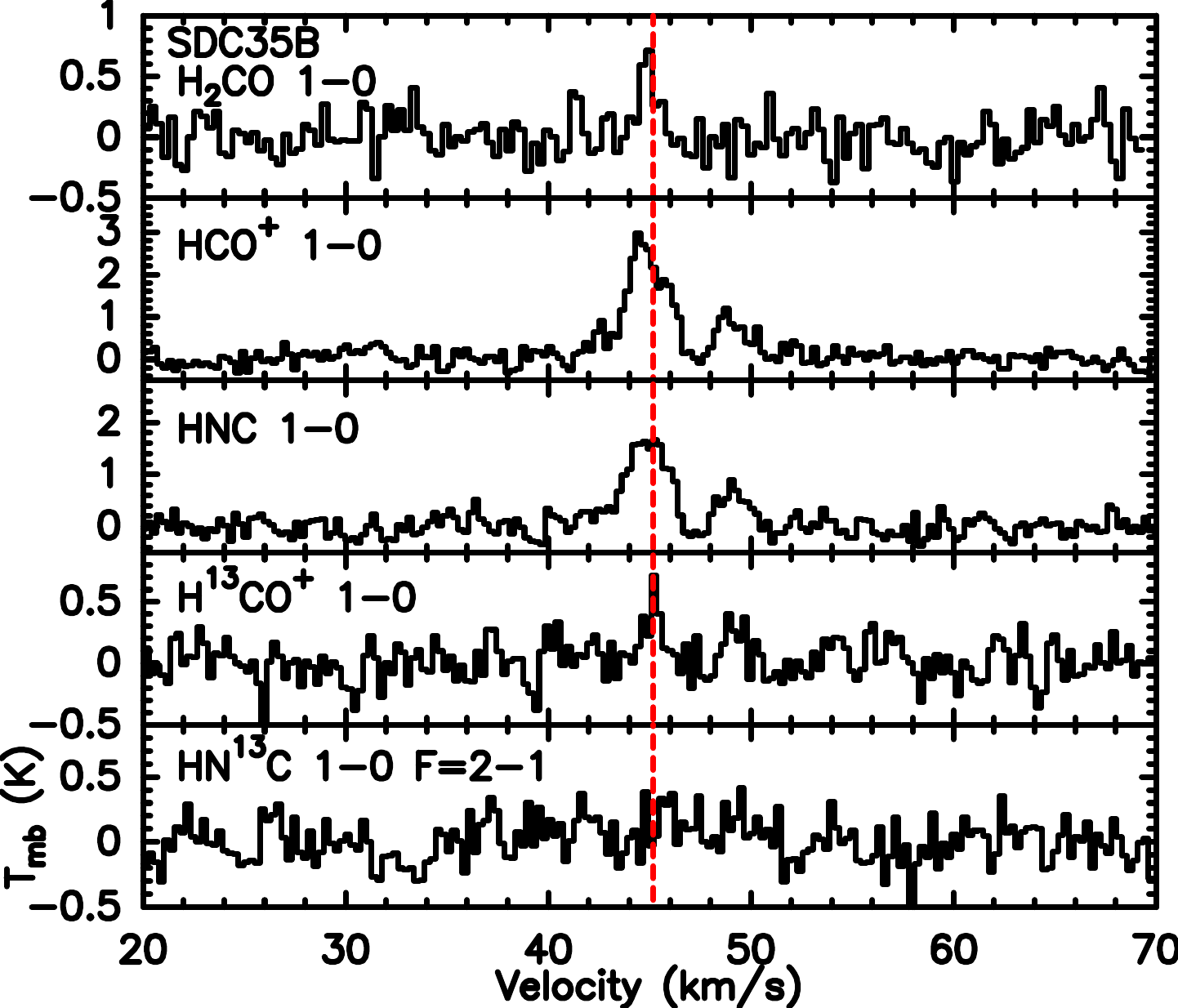}
\includegraphics[width=0.32\textwidth,height=0.39\textwidth,trim={0 0  0 0}, clip]{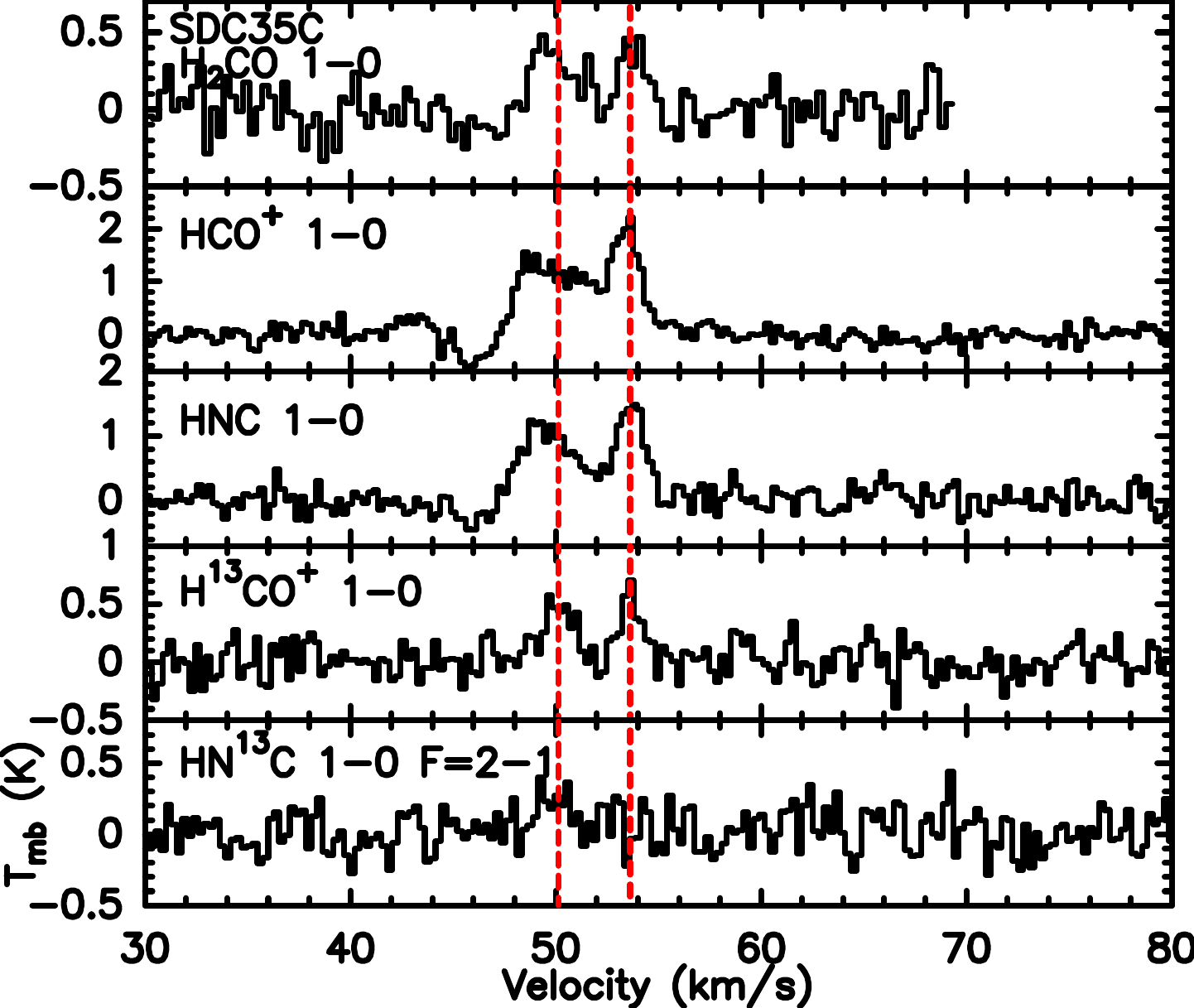}
\caption{Continued. Molecular-line spectra toward the ALOHA IRDCs clumps.
}
\label{fig:appendix_spectra_3}
\end{figure}

 \begin{figure}
  \centering
\includegraphics[width=0.32\textwidth,height=0.39\textwidth, trim={0 0  0 0}, clip]{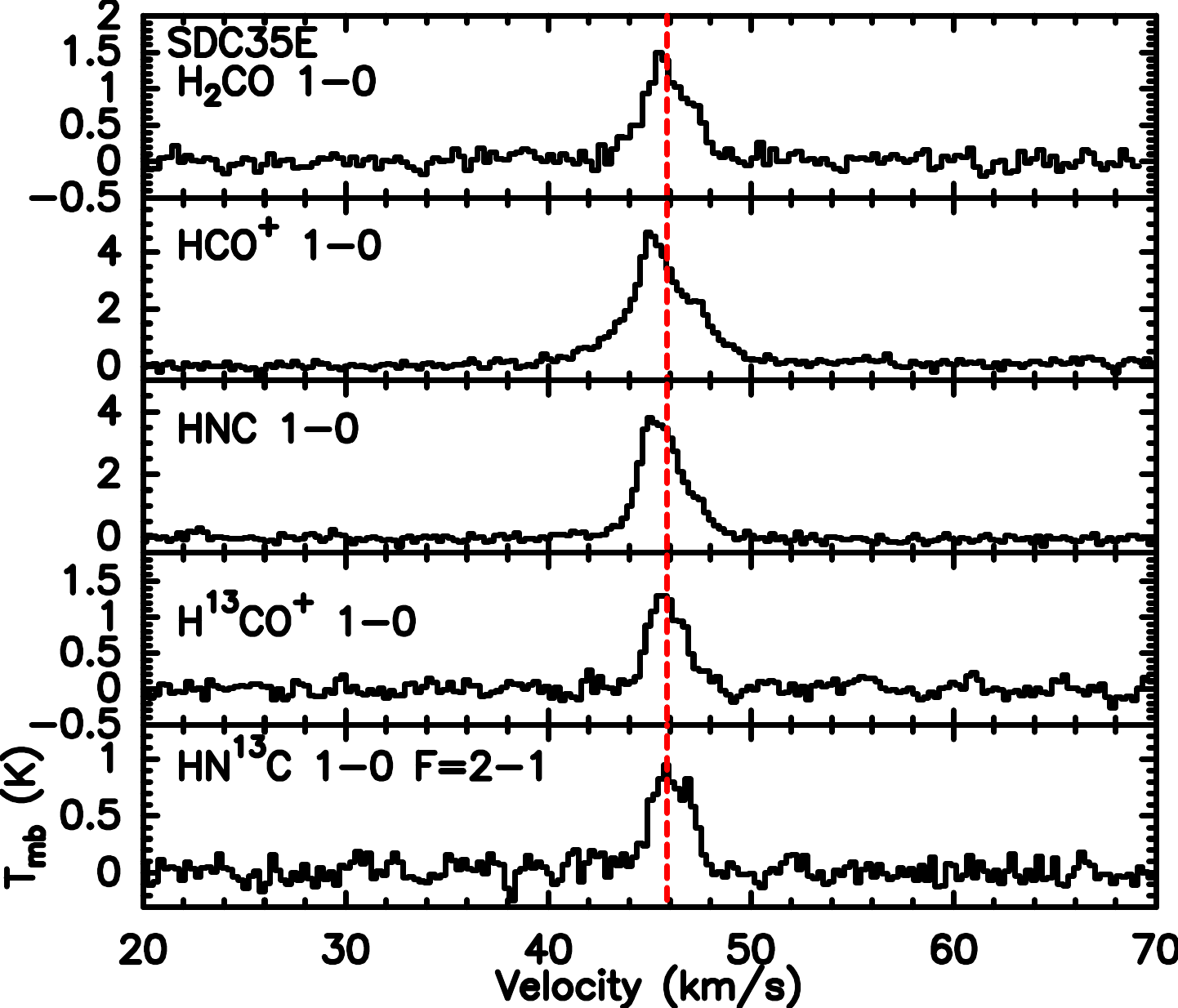}
\includegraphics[width=0.32\textwidth,height=0.39\textwidth,trim={0 0  0 0}, clip]{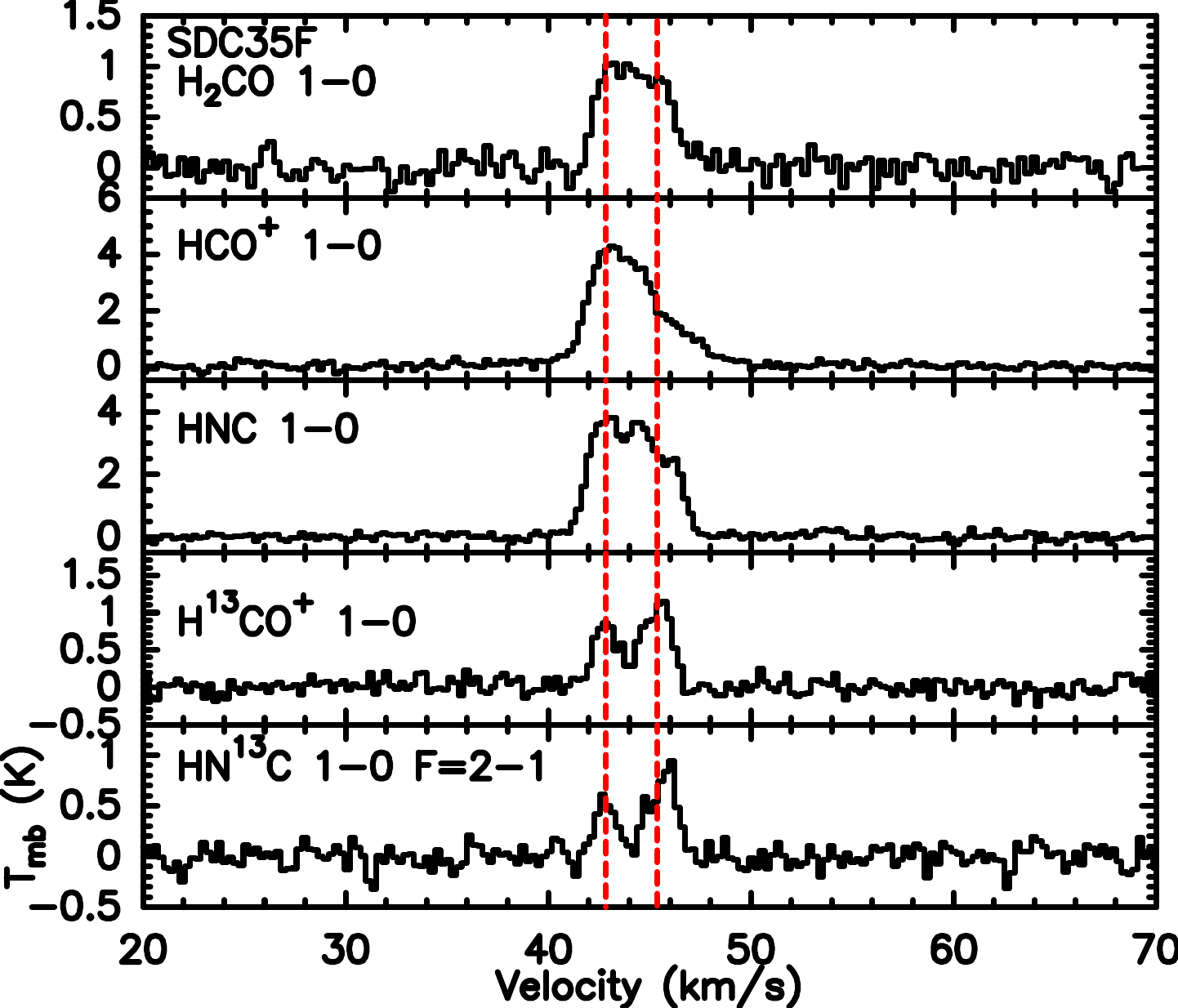}
\includegraphics[width=0.32\textwidth,height=0.39\textwidth,trim={0 0  0 0}, clip]{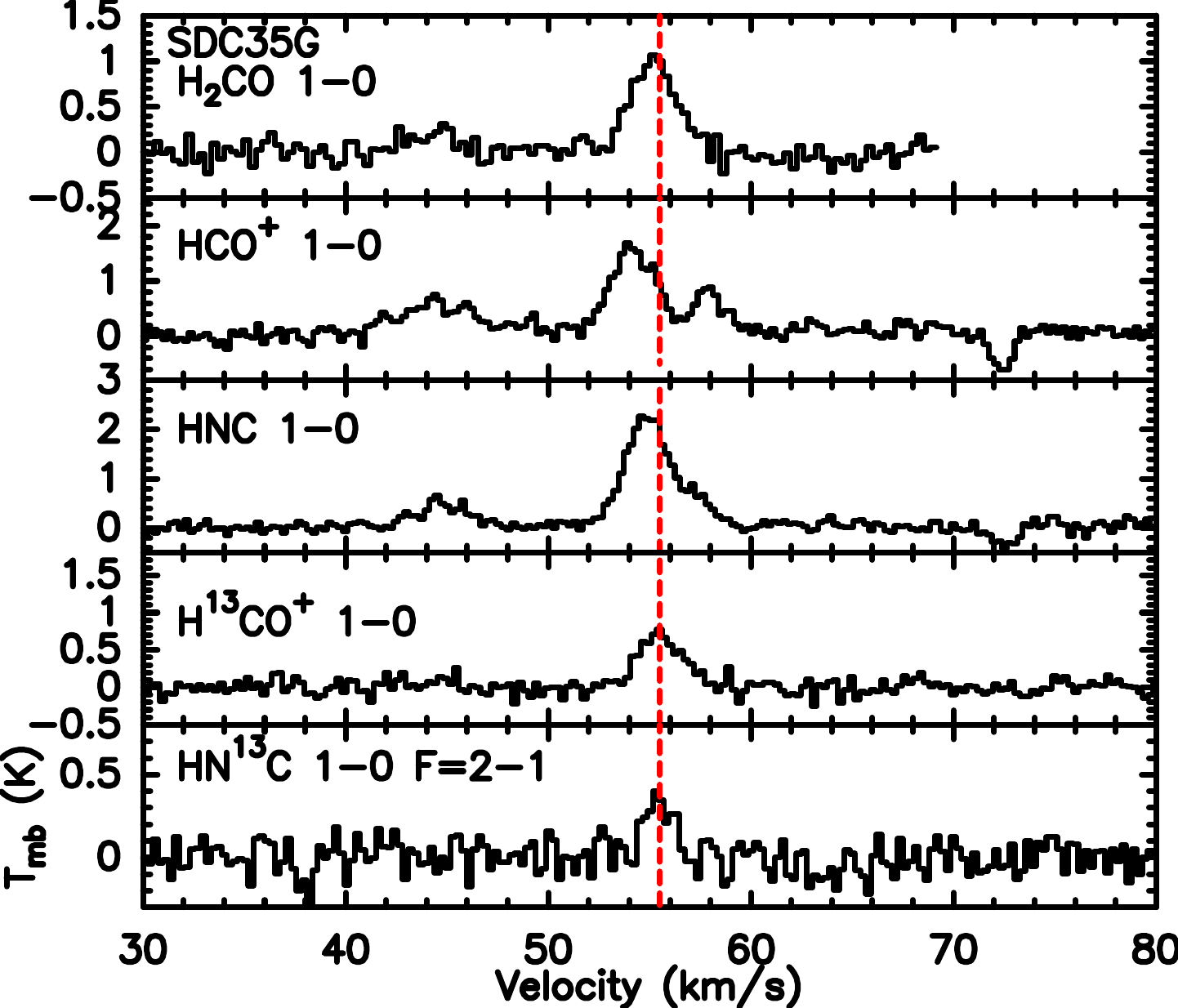}\\[1mm]

\includegraphics[width=0.32\textwidth,height=0.39\textwidth,trim={0 0  0 0}, clip]{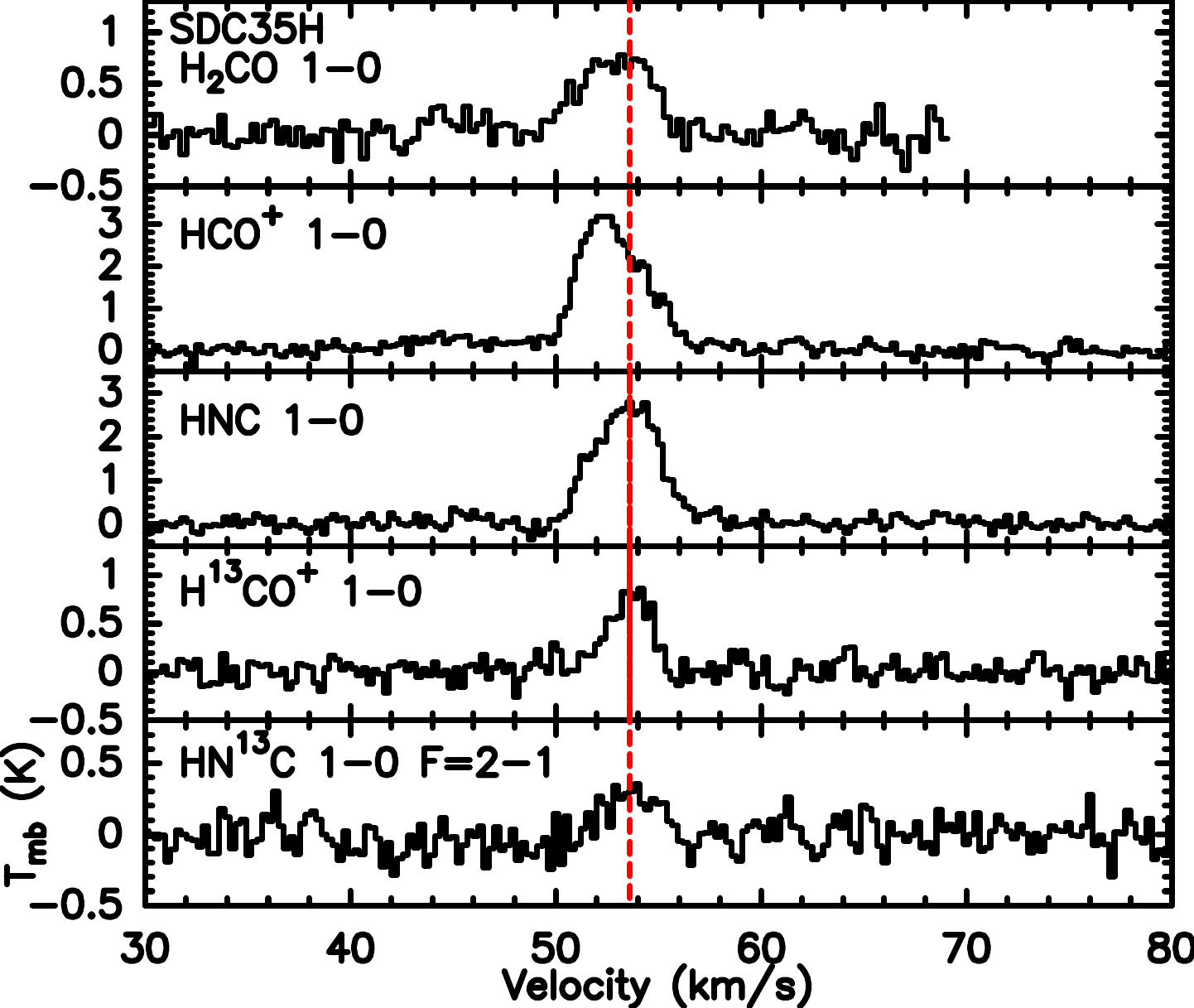}
\includegraphics[width=0.32\textwidth,height=0.39\textwidth,trim={0 0  0 0}, clip]{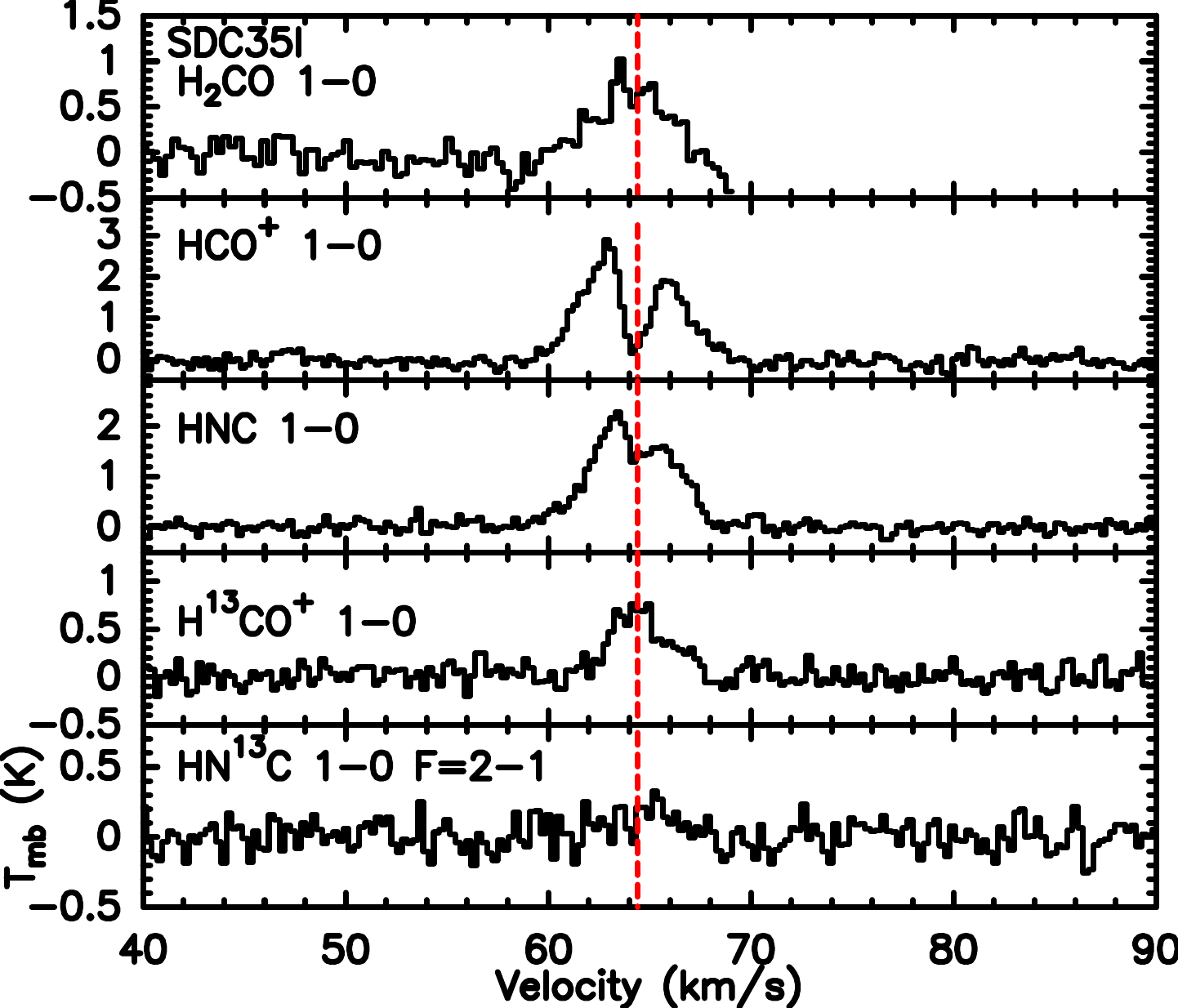}
\includegraphics[width=0.32\textwidth,height=0.39\textwidth,trim={0 0  0 0}, clip]{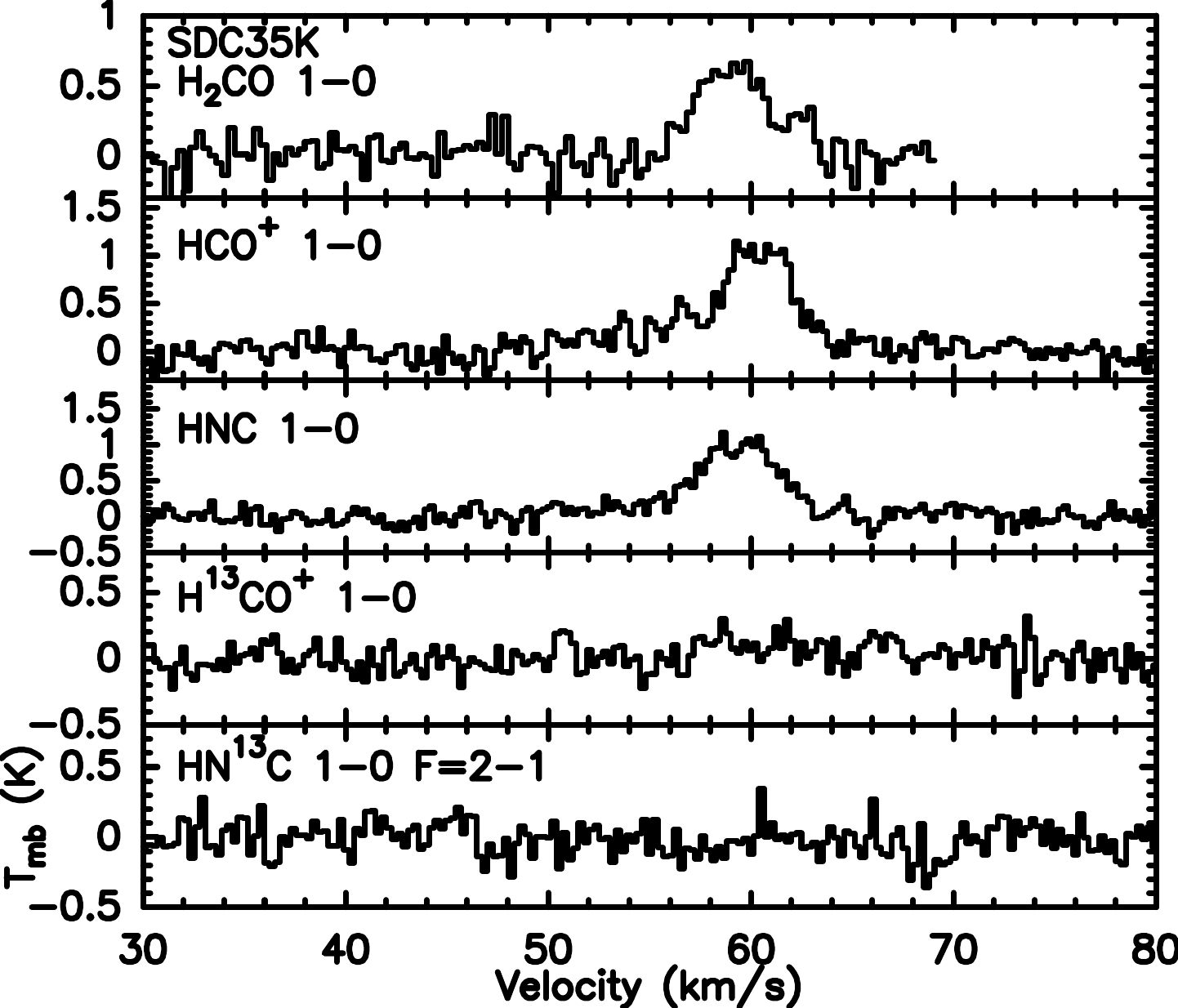}\\[1mm]

\includegraphics[width=0.32\textwidth,height=0.39\textwidth,trim={0 0  0 0}, clip]{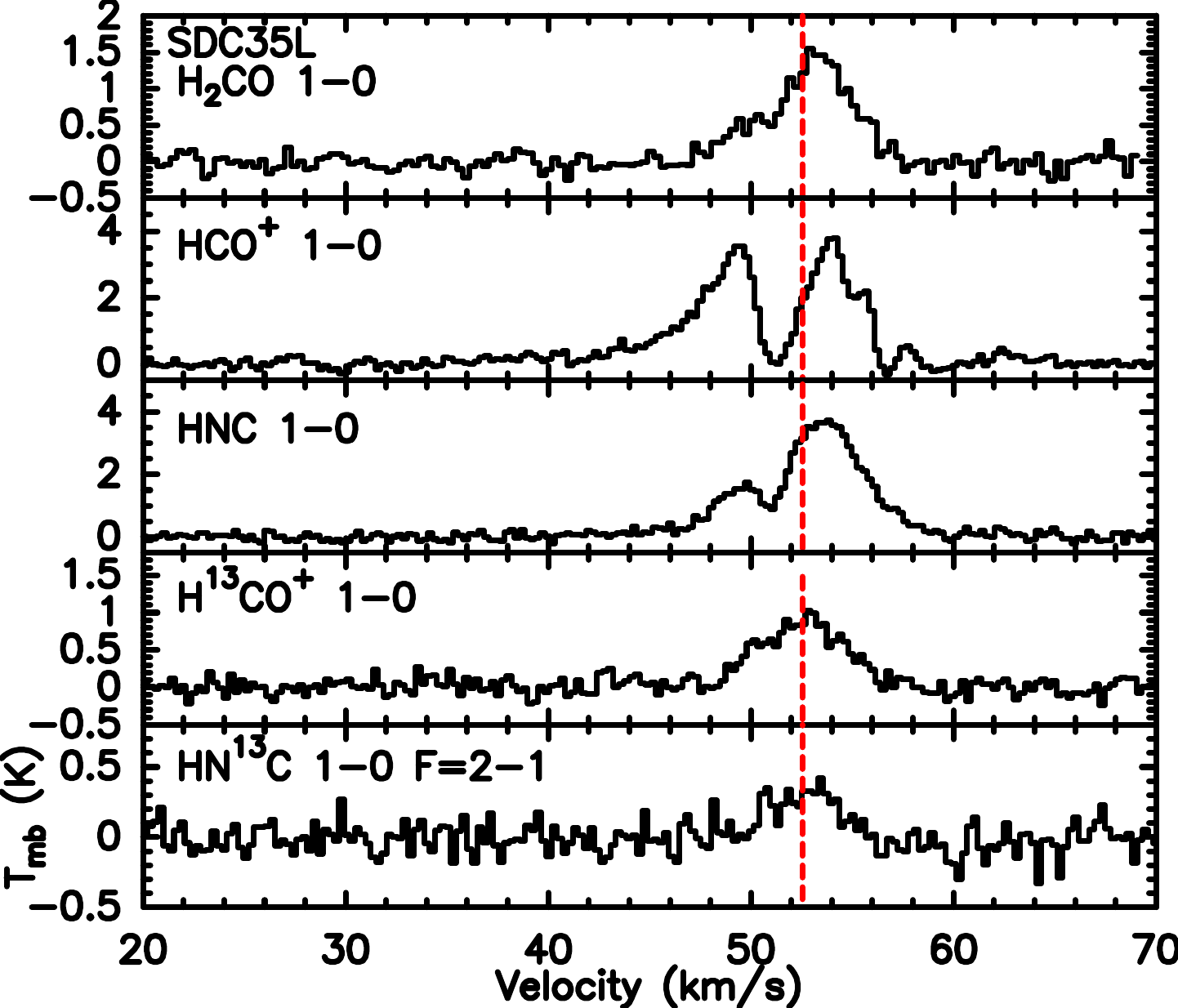}
\includegraphics[width=0.32\textwidth,height=0.39\textwidth,trim={0 0  0 0}, clip]{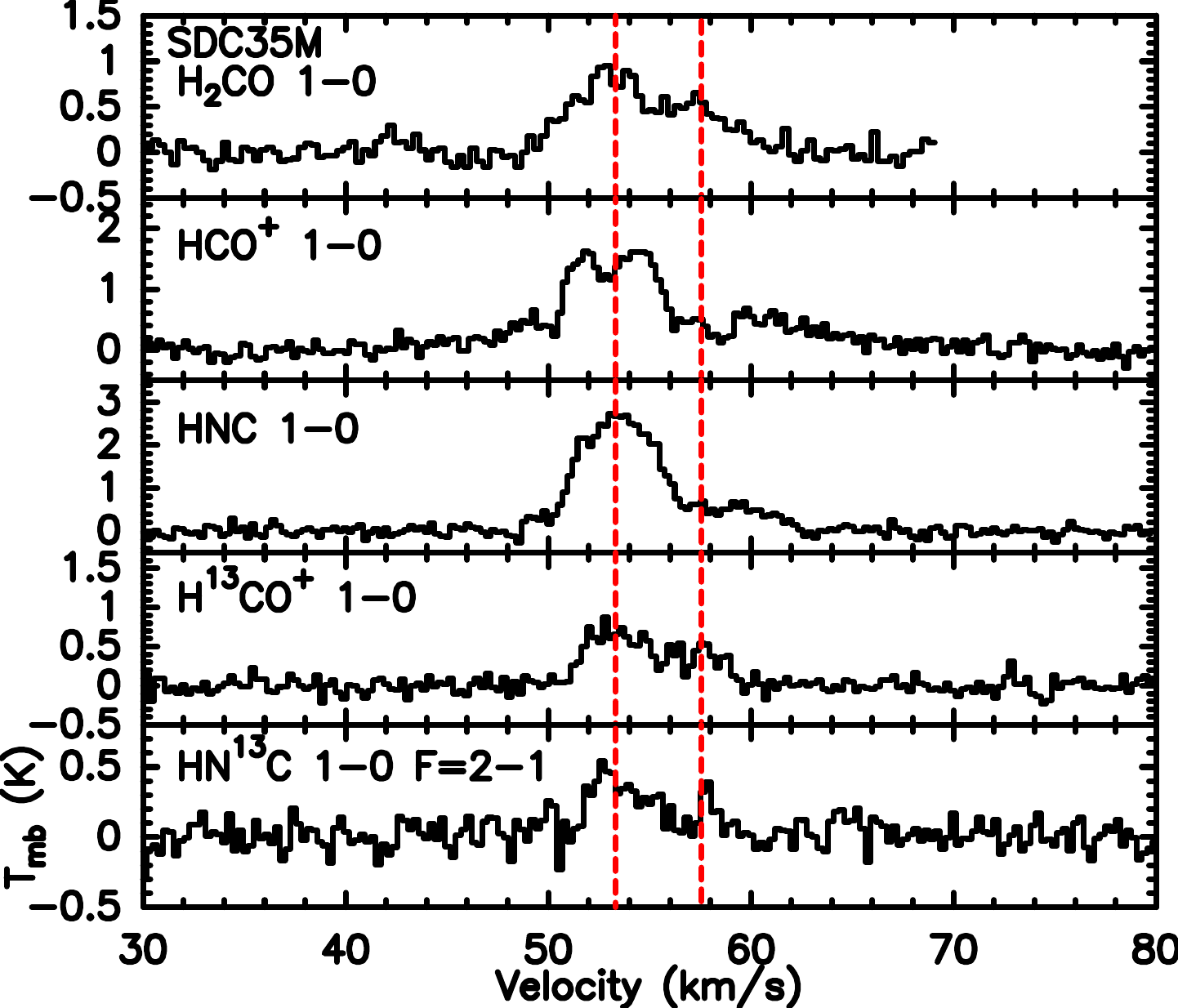}
\includegraphics[width=0.32\textwidth,height=0.39\textwidth,trim={0 0  0 0}, clip]{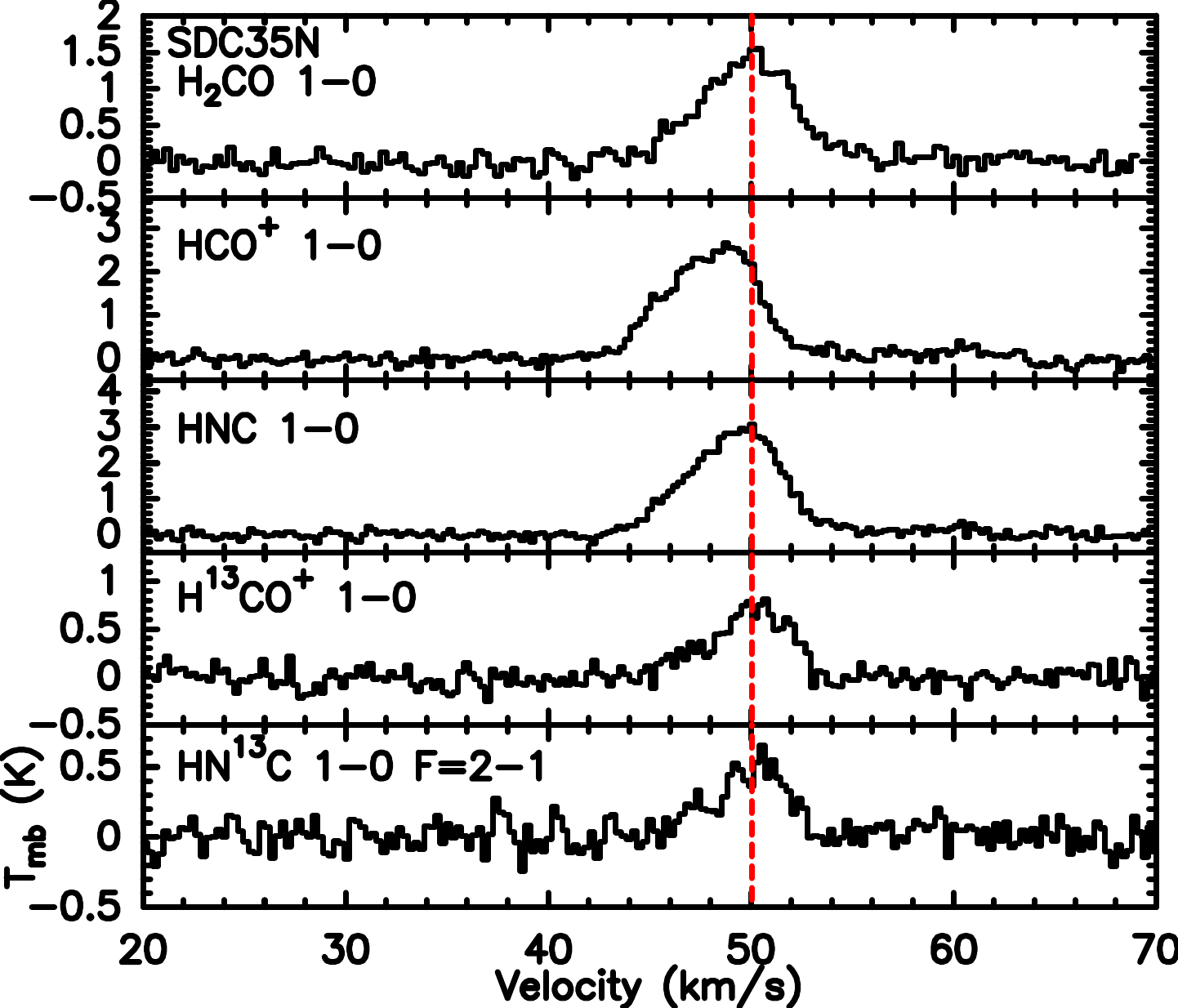}
\caption{Continued. Molecular-line spectra toward the ALOHA IRDCs clumps.
}
\label{fig:appendix_spectra_4}
\end{figure}

 \begin{figure}
  \centering
\includegraphics[width=0.32\textwidth,height=0.39\textwidth, trim={0 0  0 0}, clip]{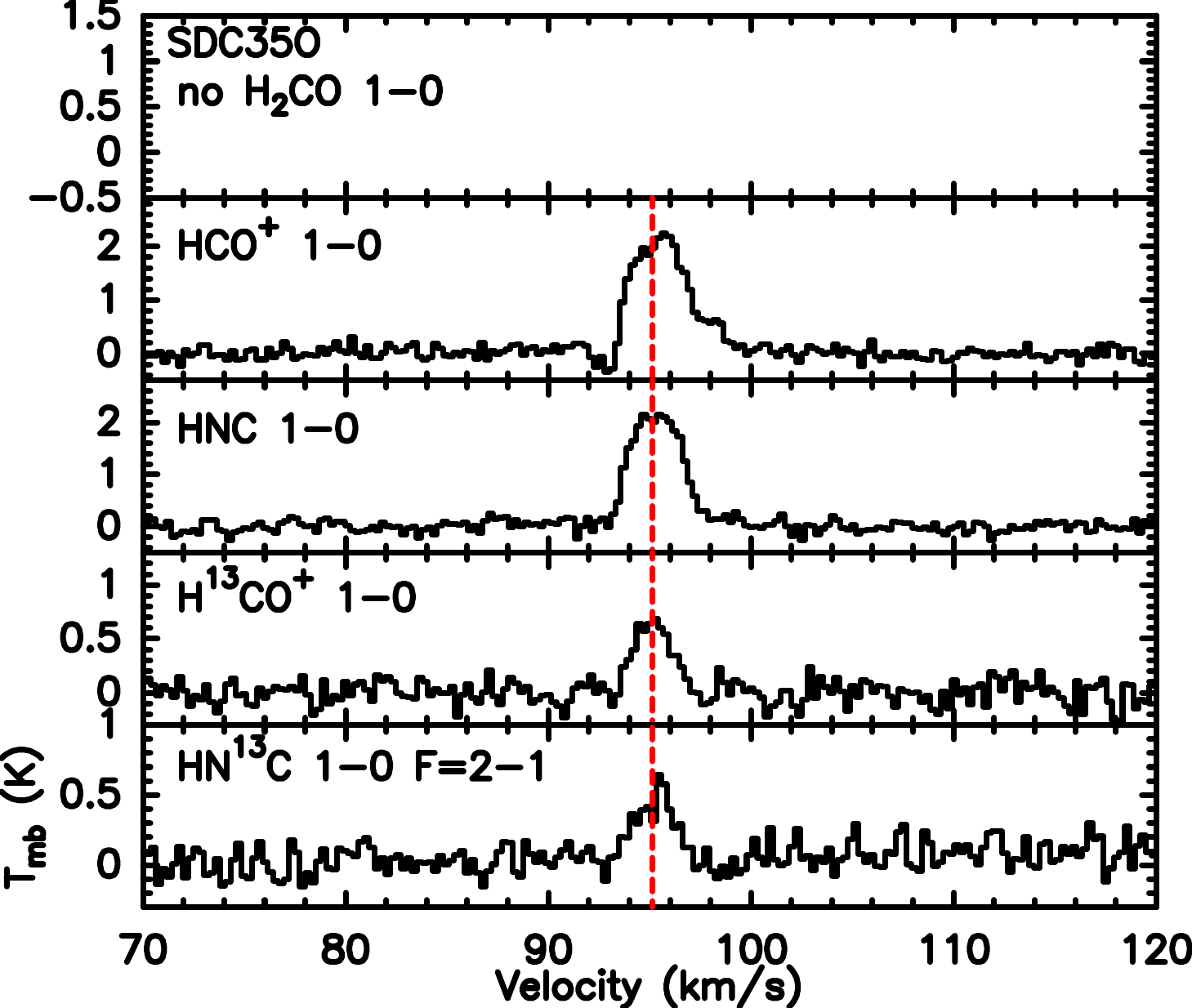}
\includegraphics[width=0.32\textwidth,height=0.39\textwidth,trim={0 0  0 0}, clip]{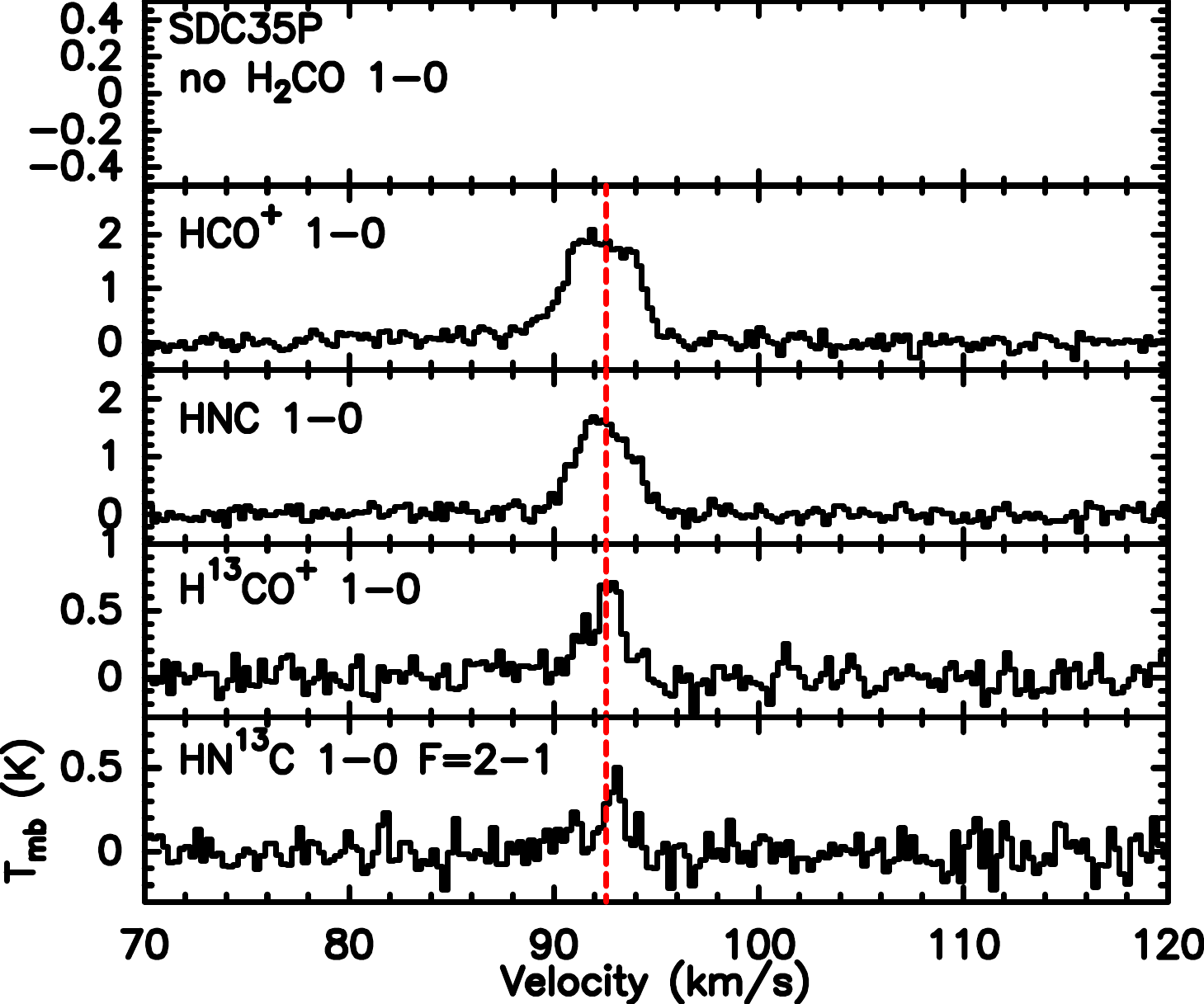}
\includegraphics[width=0.32\textwidth,height=0.39\textwidth,trim={0 0  0 0}, clip]{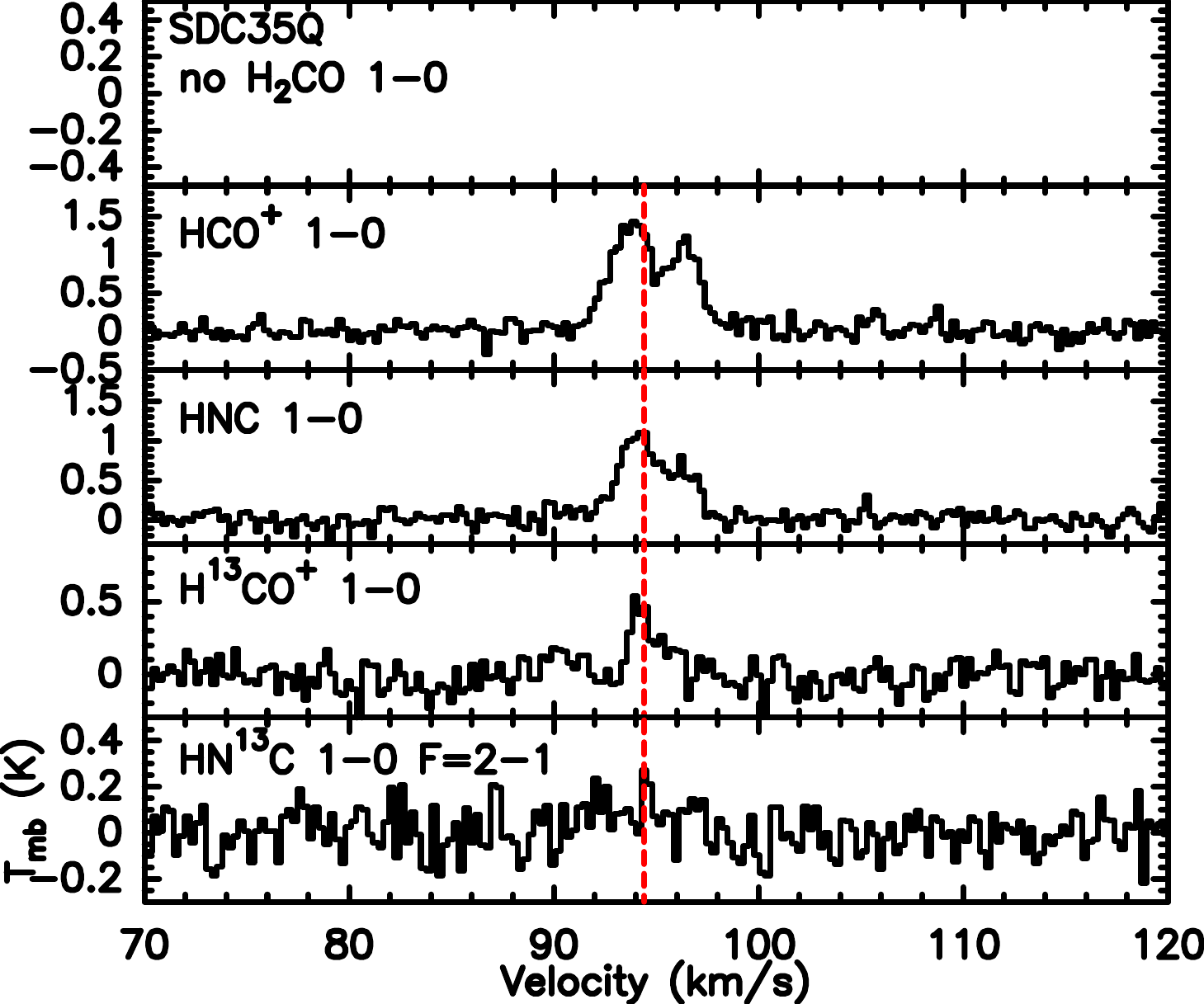}\\[1mm]

\includegraphics[width=0.32\textwidth,height=0.39\textwidth,trim={0 0  0 0}, clip]{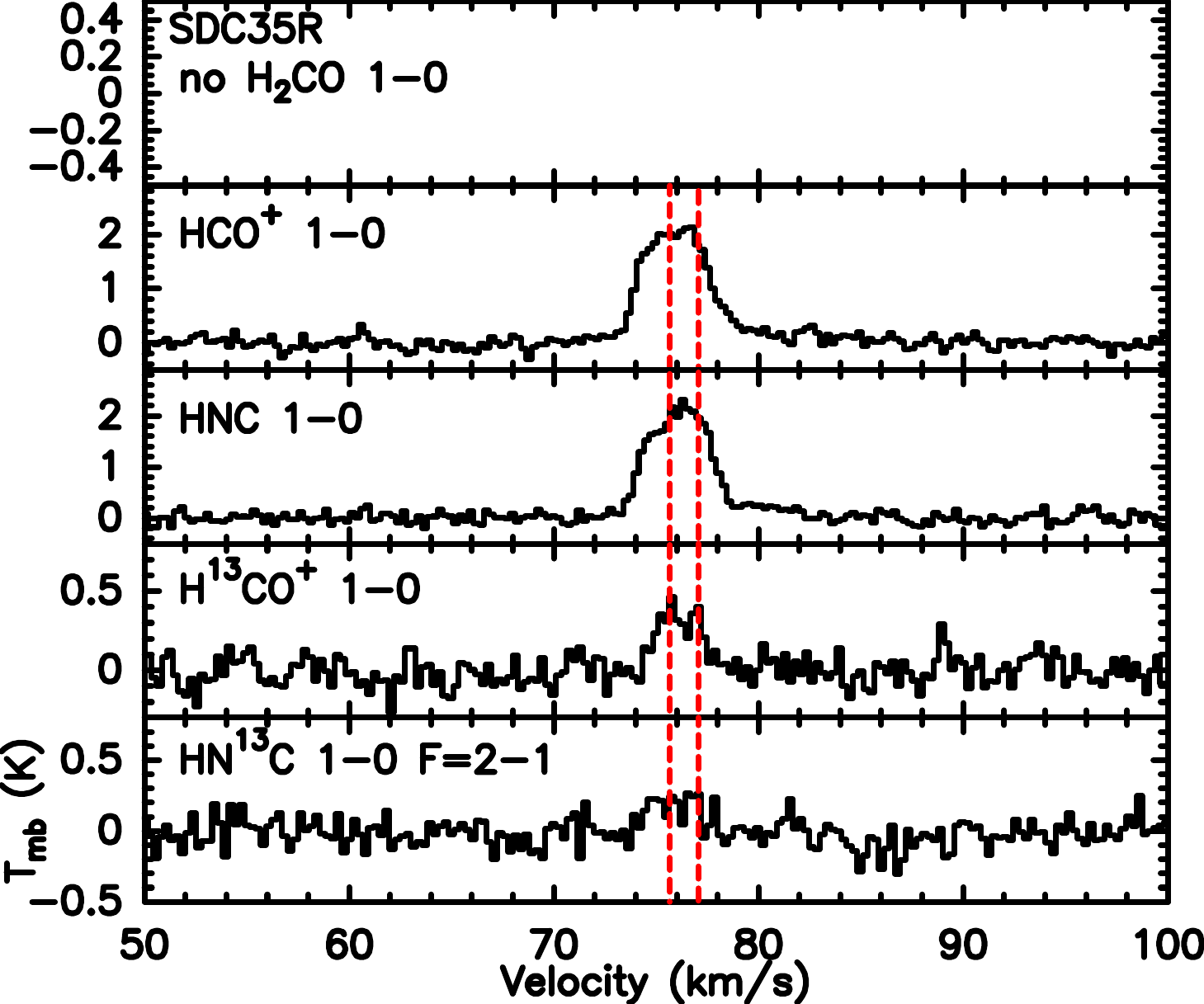}
\includegraphics[width=0.32\textwidth,height=0.39\textwidth,trim={0 0  0 0}, clip]{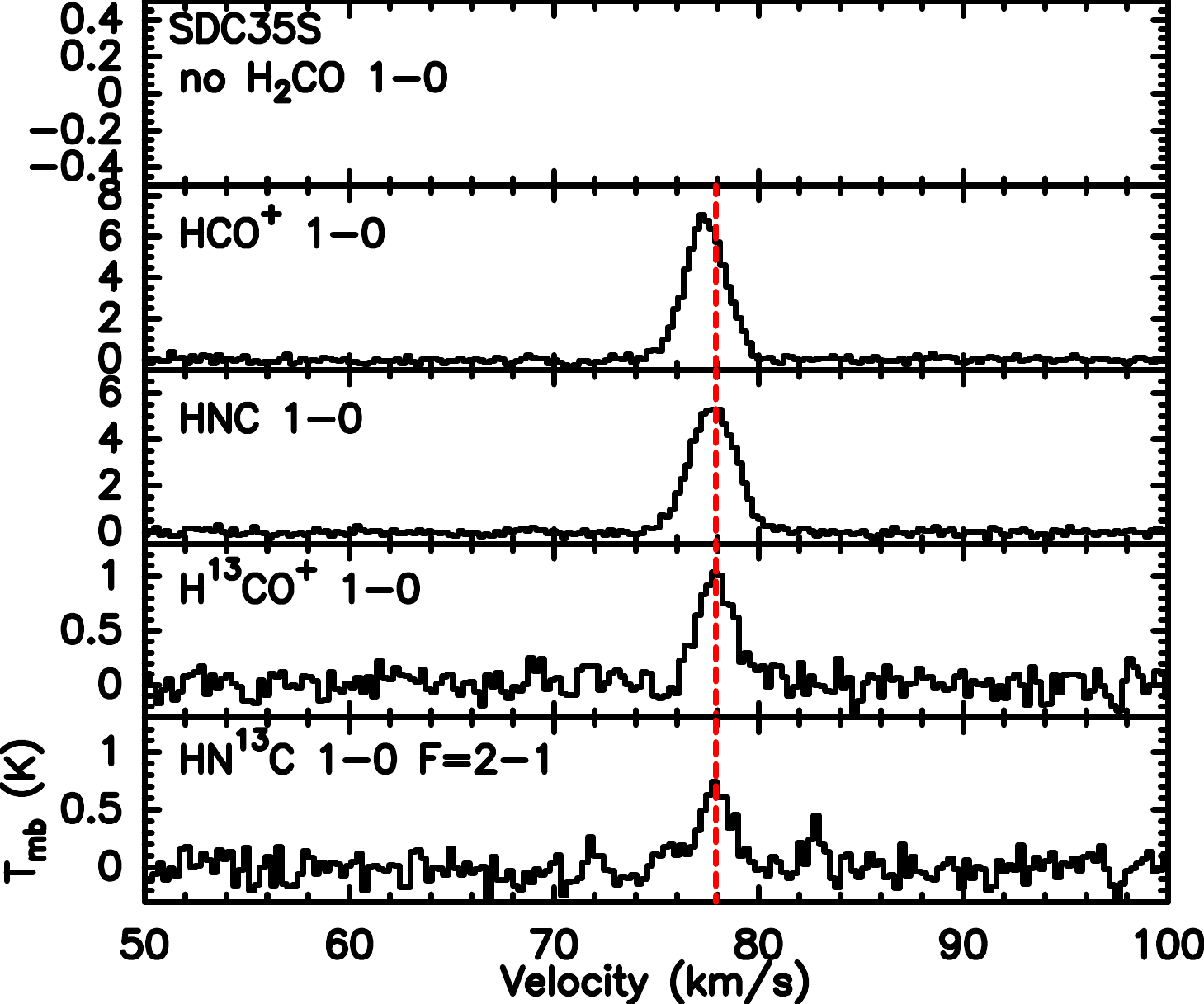}
\includegraphics[width=0.32\textwidth,height=0.39\textwidth,trim={0 0  0 0}, clip]{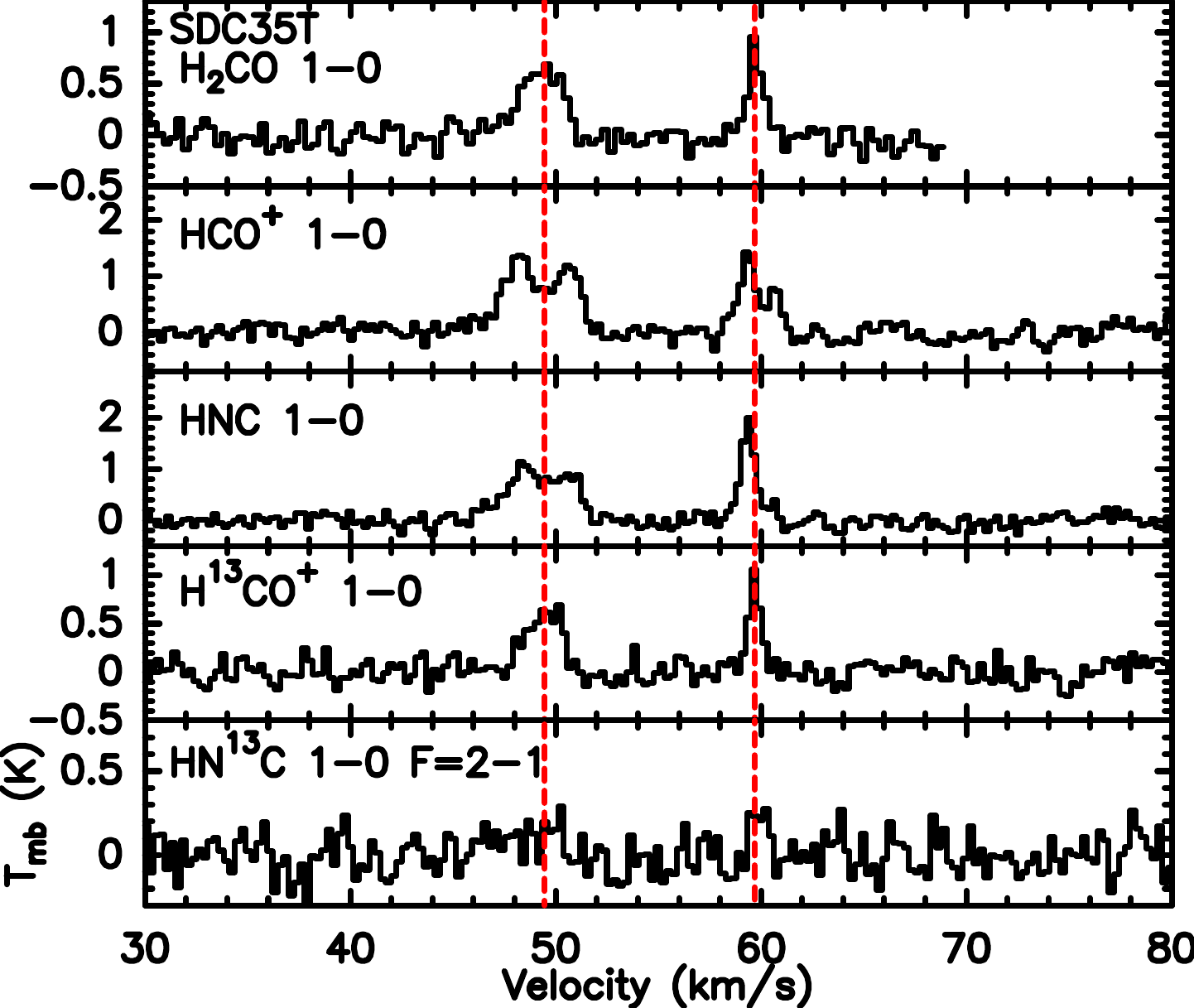}\\[1mm]

\includegraphics[width=0.32\textwidth,height=0.39\textwidth,trim={0 0  0 0}, clip]{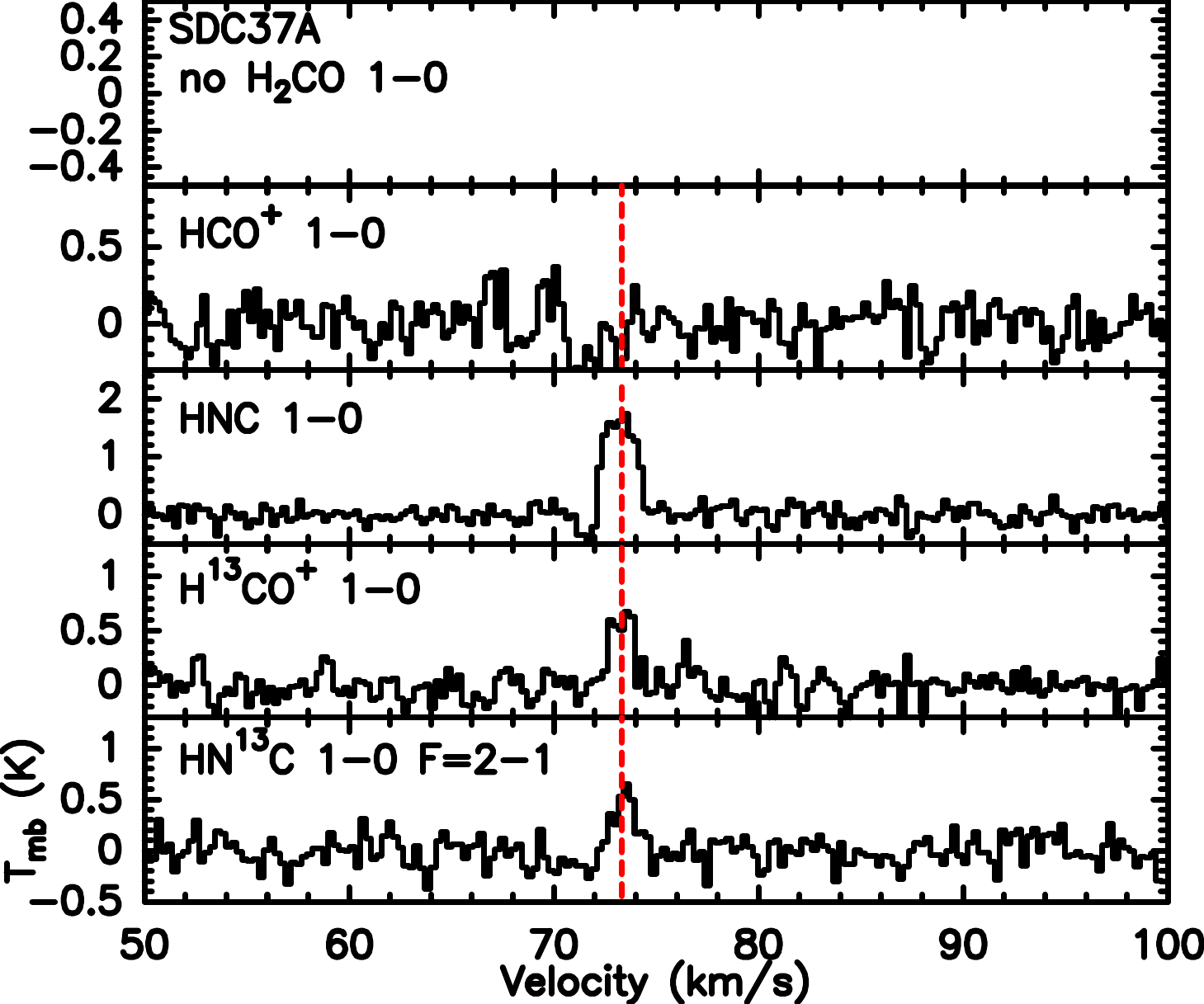}
\includegraphics[width=0.32\textwidth,height=0.39\textwidth,trim={0 0  0 0}, clip]{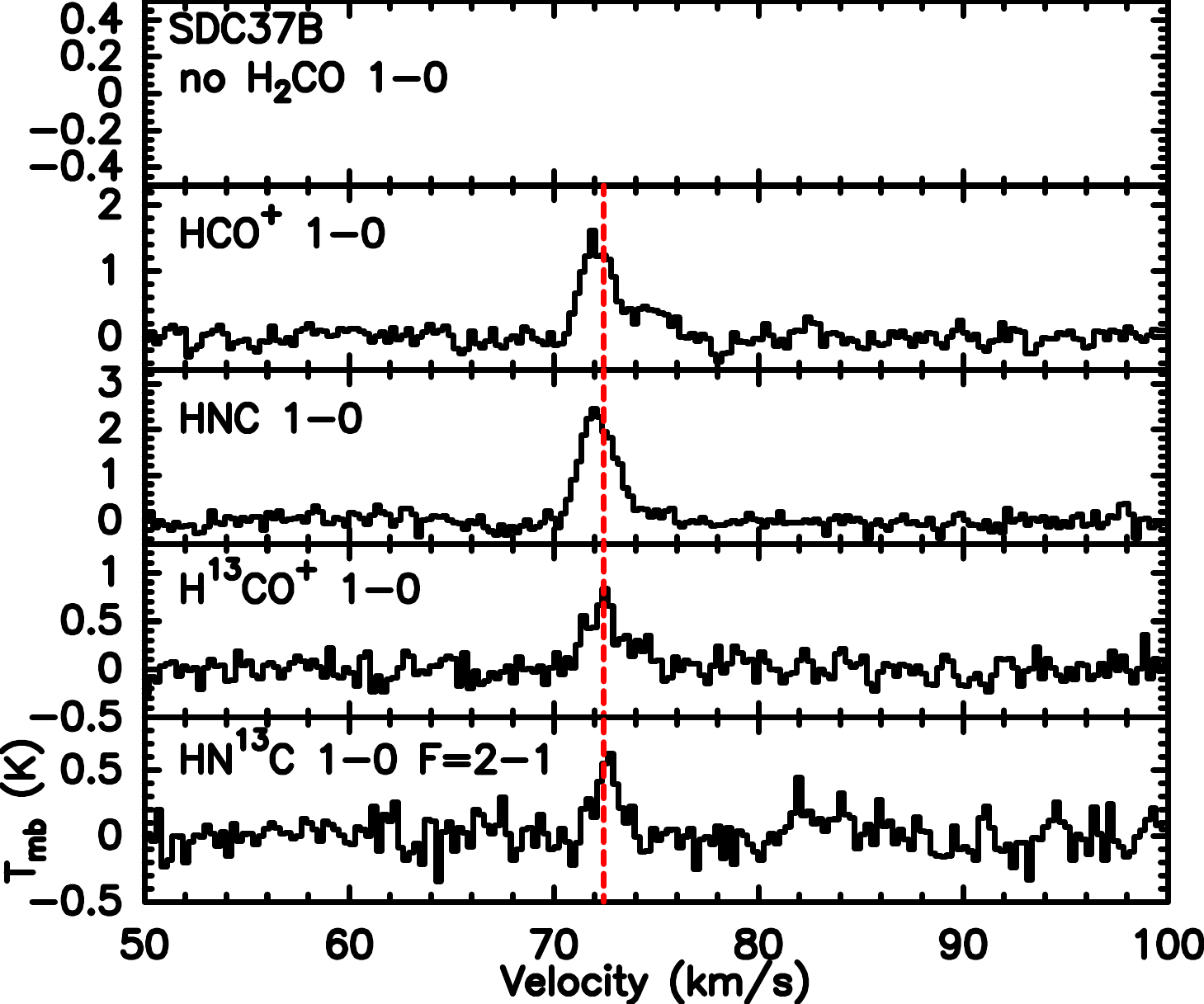}
\includegraphics[width=0.32\textwidth,height=0.39\textwidth,trim={0 0  0 0}, clip]{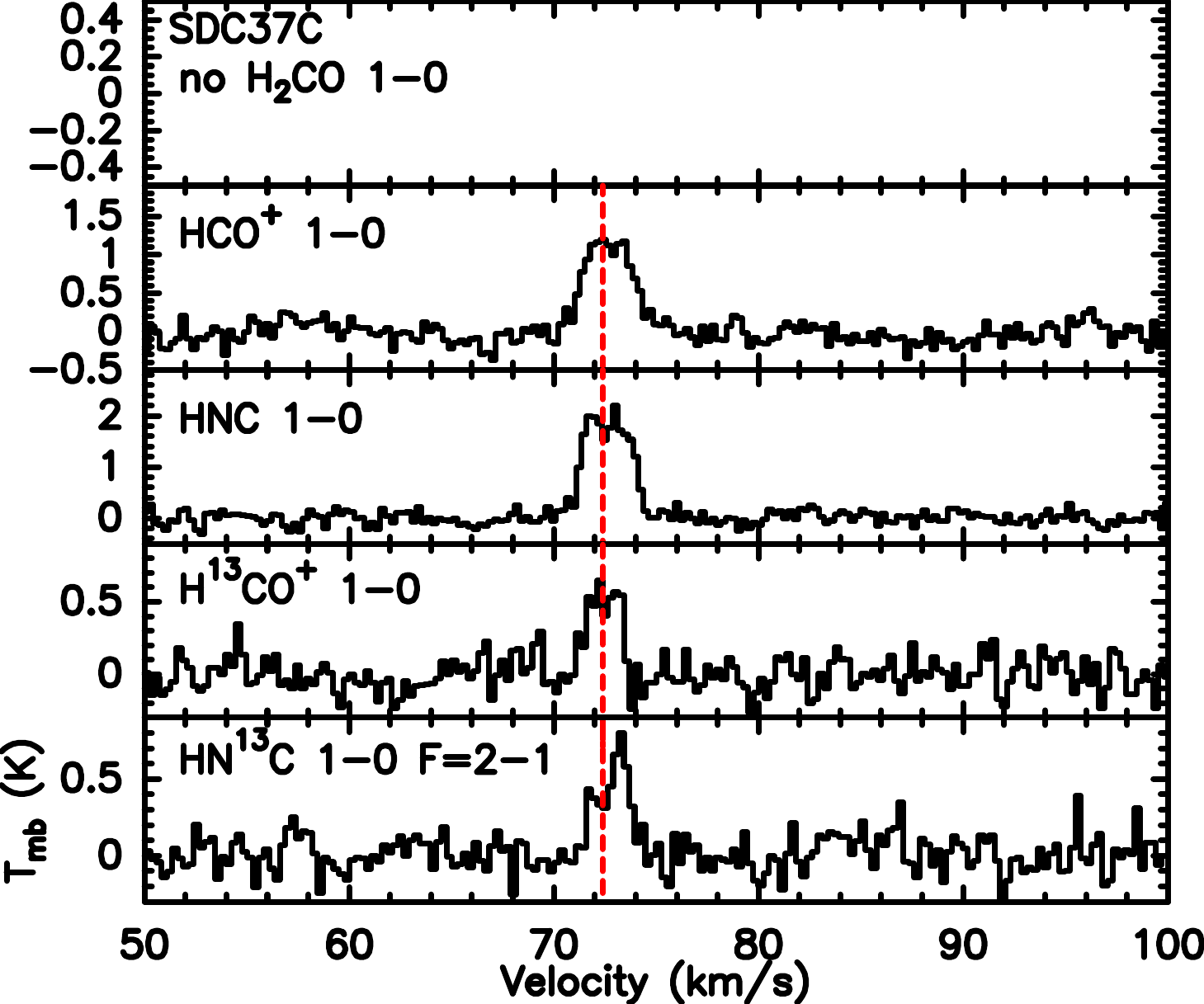}
\caption{Continued. Molecular-line spectra toward the ALOHA IRDCs clumps.
}
\label{fig:appendix_spectra_5}
\end{figure}

 \begin{figure}
  \centering
\includegraphics[width=0.32\textwidth,height=0.39\textwidth, trim={0 0  0 0}, clip]{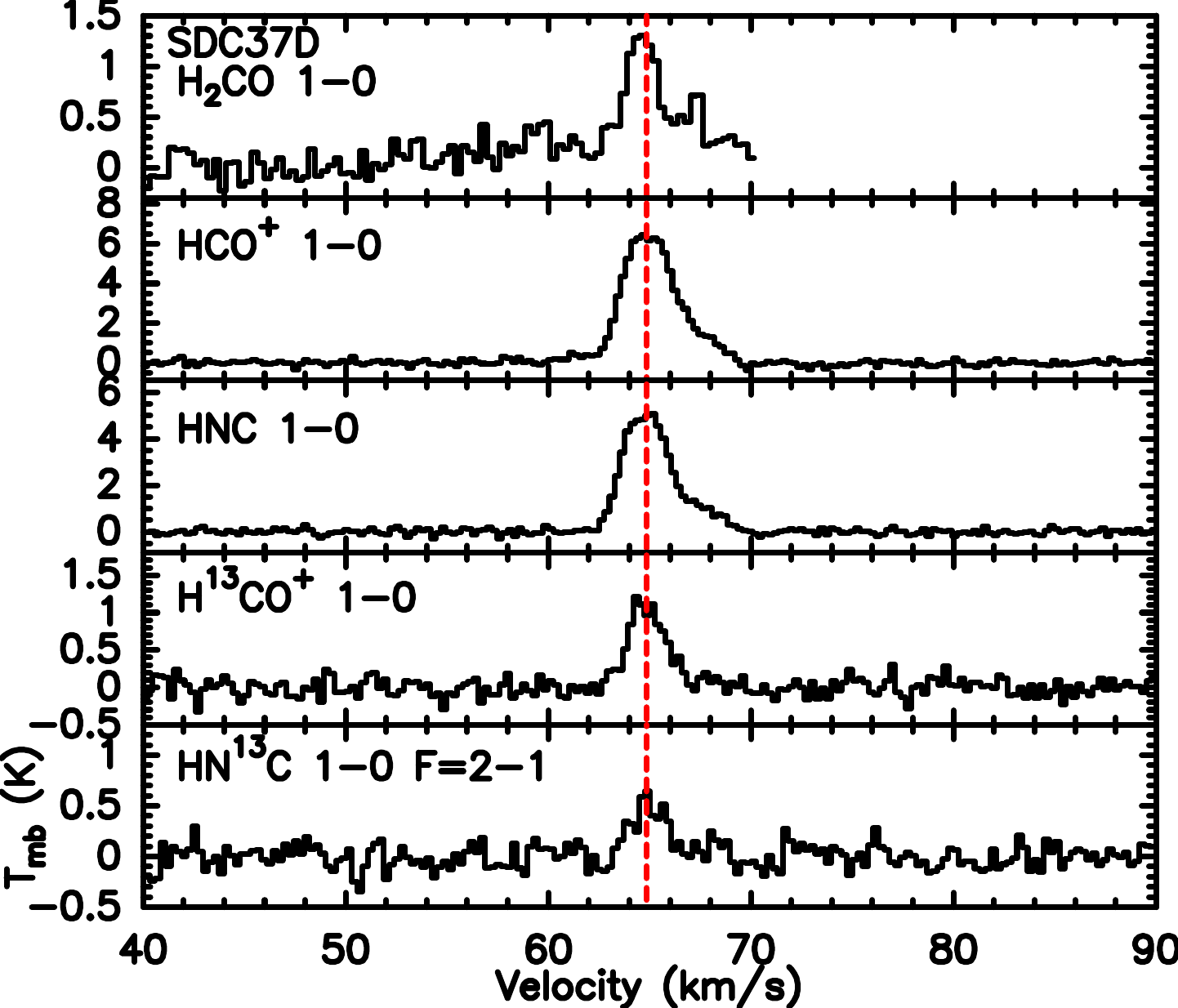}
\includegraphics[width=0.32\textwidth,height=0.39\textwidth,trim={0 0  0 0}, clip]{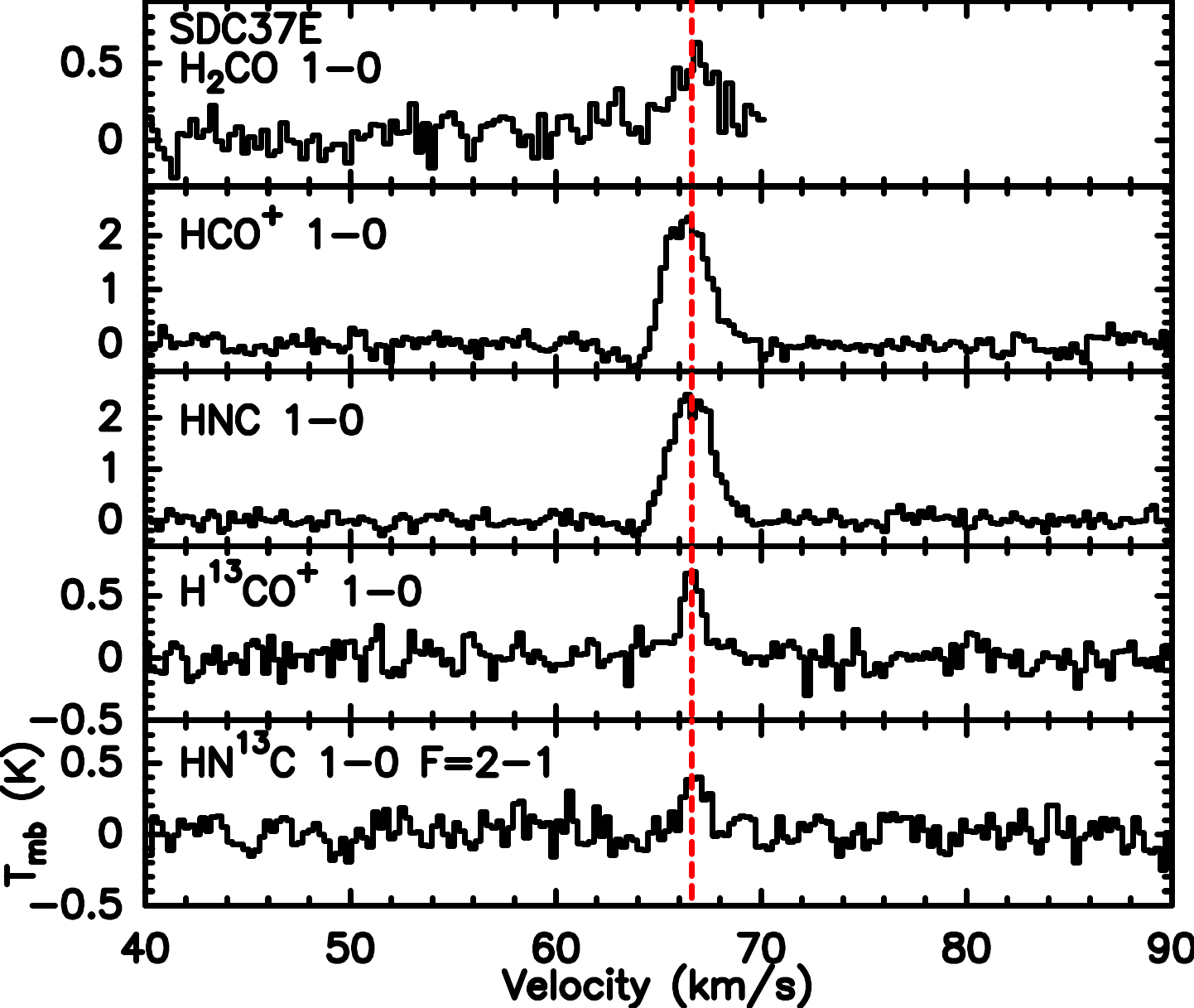}
\includegraphics[width=0.32\textwidth,height=0.39\textwidth,trim={0 0  0 0}, clip]{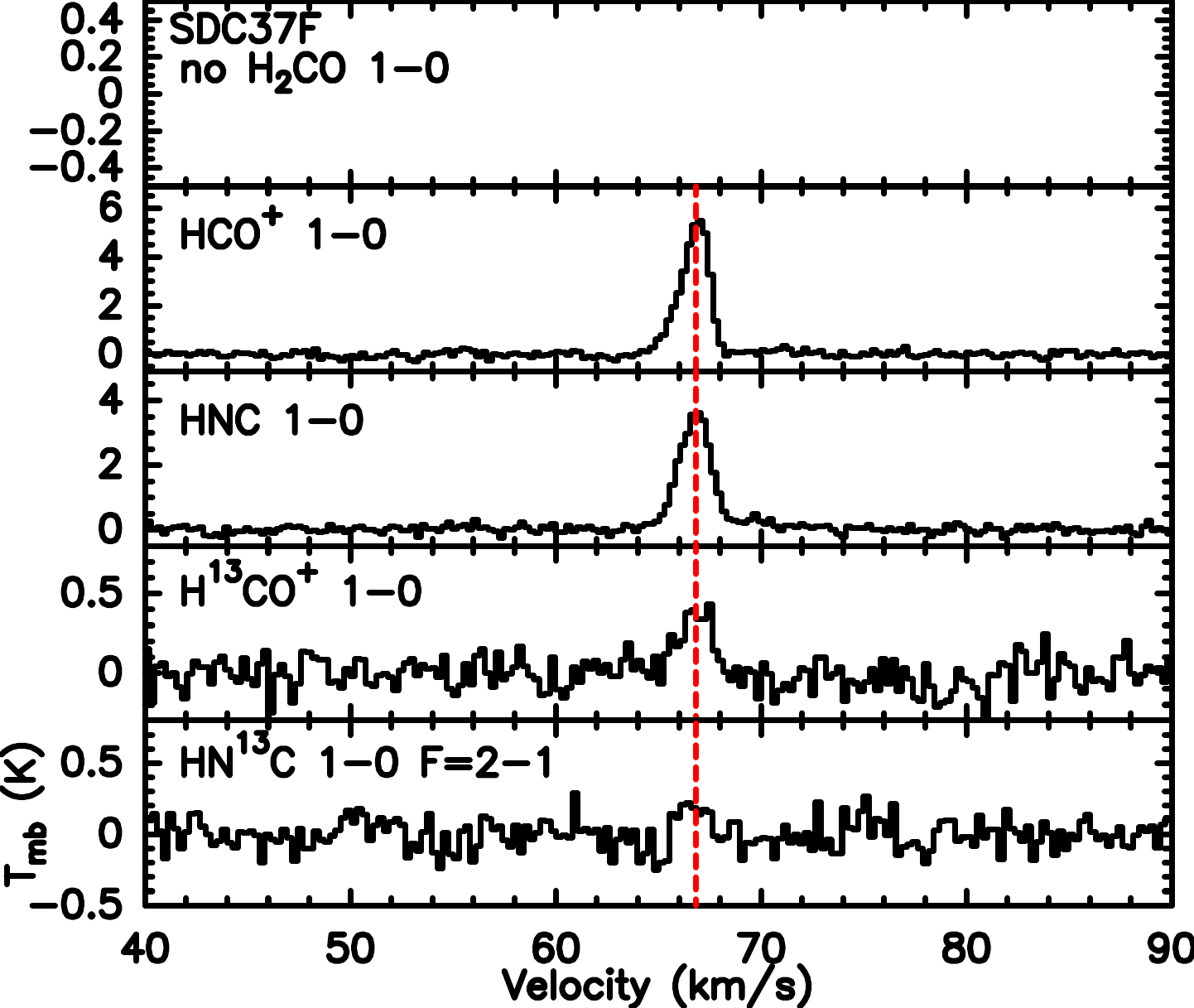}\\[1mm]

\includegraphics[width=0.32\textwidth,height=0.39\textwidth,trim={0 0  0 0}, clip]{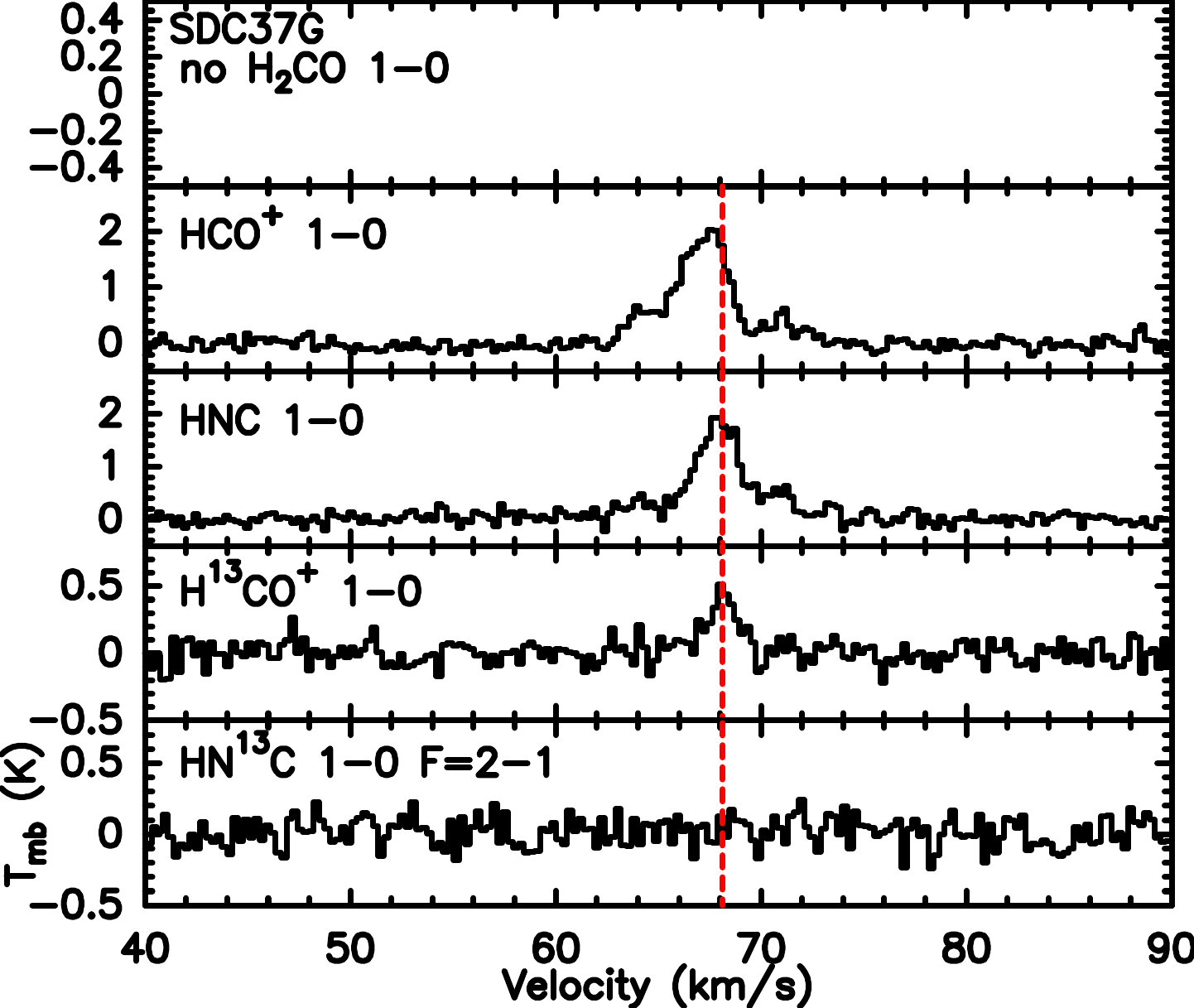}
\includegraphics[width=0.32\textwidth,height=0.39\textwidth,trim={0 0  0 0}, clip]{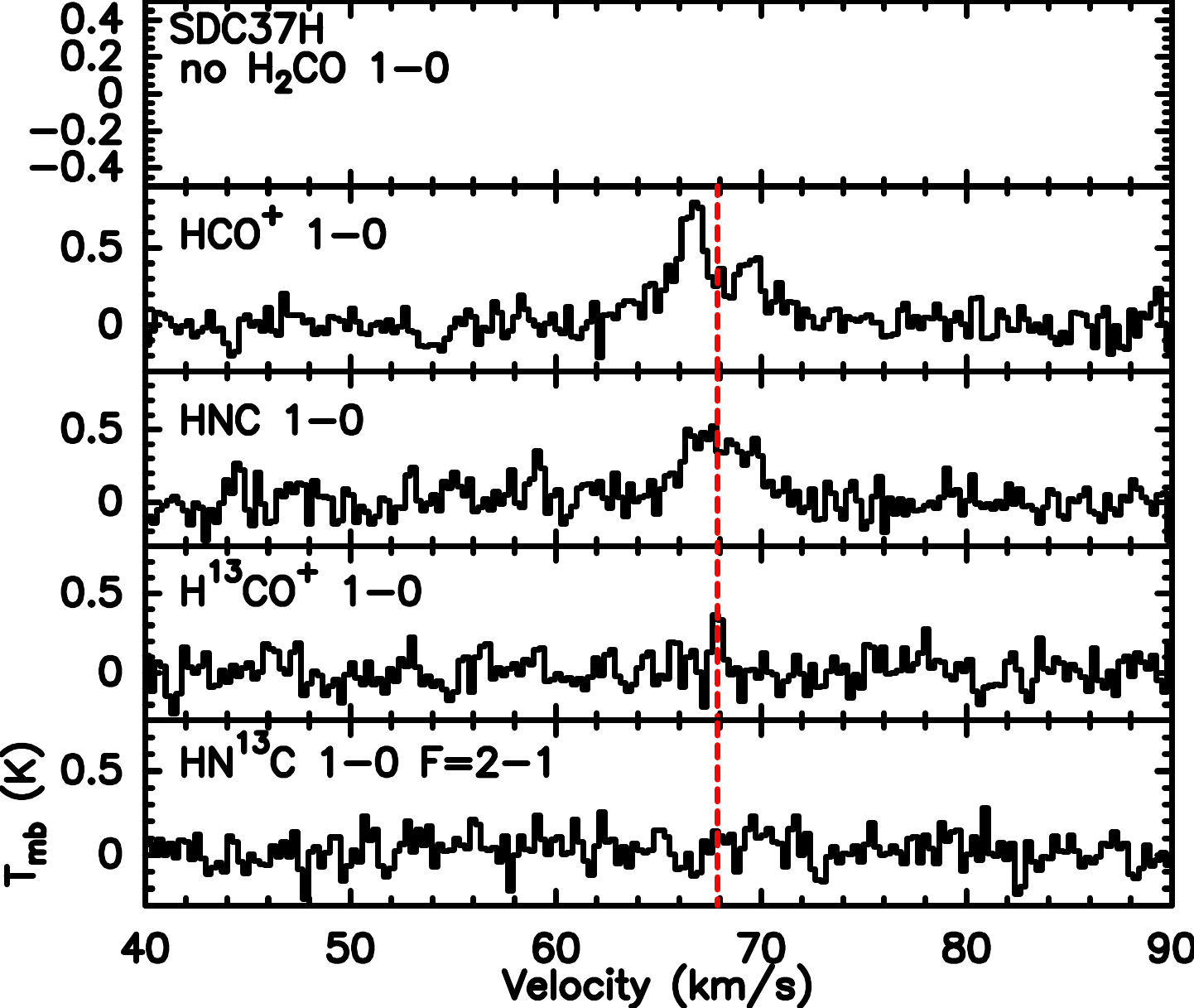}
\includegraphics[width=0.32\textwidth,height=0.39\textwidth,trim={0 0  0 0}, clip]{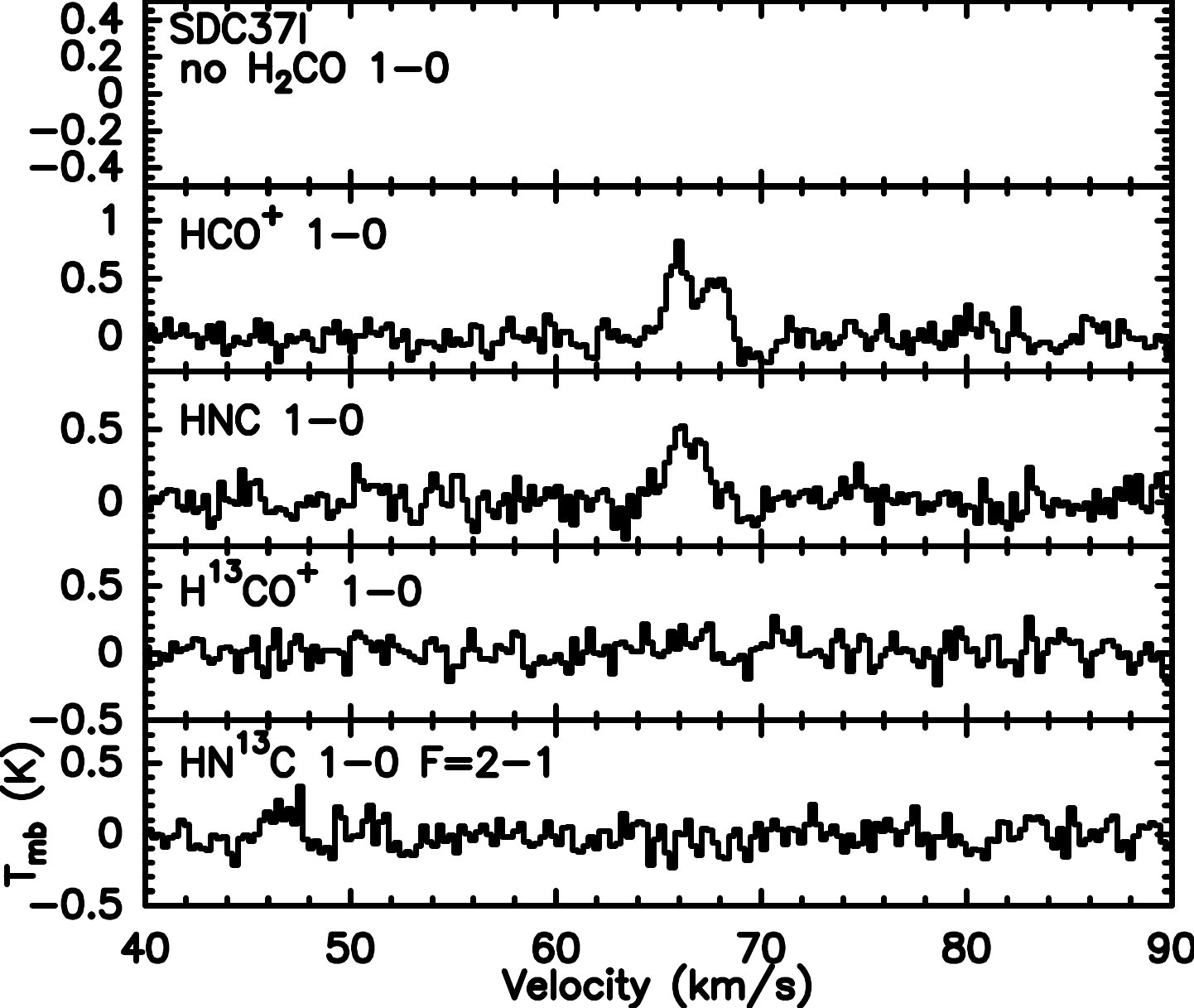}\\[1mm]

\includegraphics[width=0.40\textwidth,height=0.39\textwidth,trim={0 0  0 0}, clip]{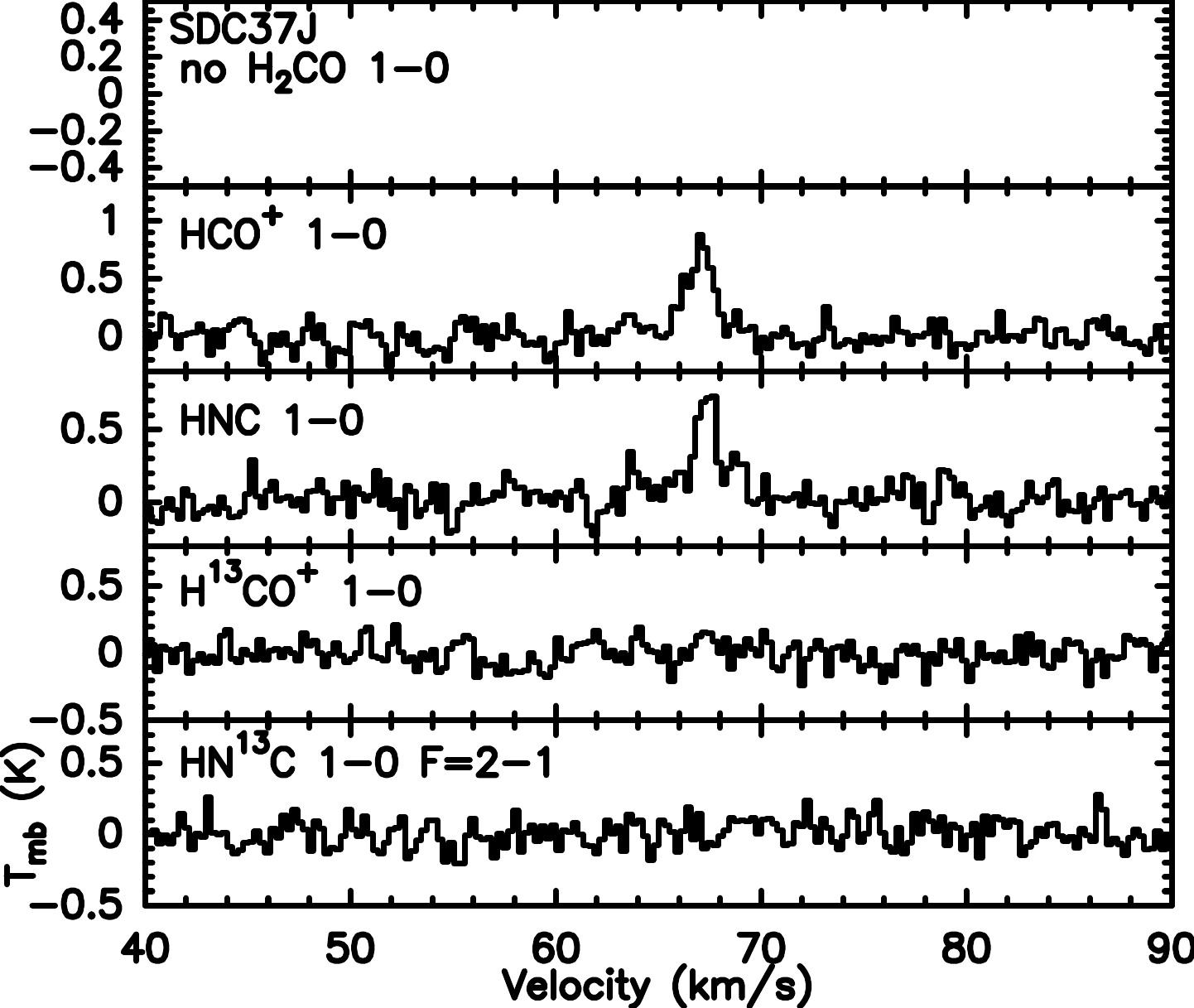}
\includegraphics[width=0.40\textwidth,height=0.39\textwidth,trim={0 0  0 0}, clip]{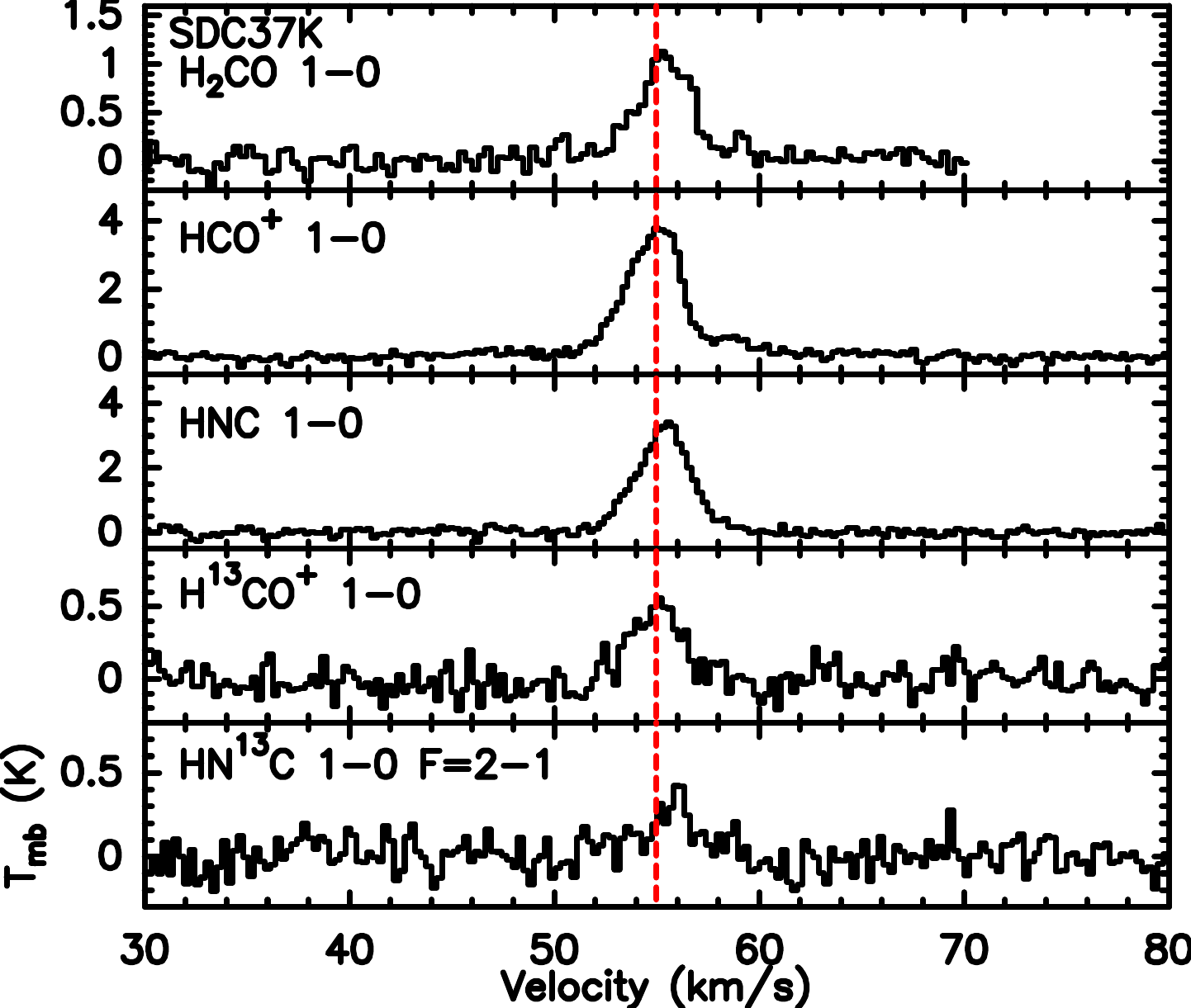}
\caption{Continued. Molecular-line spectra toward the ALOHA IRDCs clumps.
}
\label{fig:appendix_spectra_6}
\end{figure}

\clearpage
\section{Observed infall profile parameters from Fuller's measurements}
\label{app:profile_parameters}

The full line profile parameter table is provided in Table~\ref{tab:fuller_settled}. It lists $P_{\rm r}$, $v_{\rm sep}$, $w_{1/2}$, and $D_{\rm r}$ for H$_2$CO, HCO$^+$, and HNC. \begingroup
\footnotesize
\setlength{\tabcolsep}{2.2pt}
\renewcommand{\arraystretch}{1.0}
\setlength{\LTleft}{\fill}
\setlength{\LTright}{\fill}

\begin{longtable}{@{}l*{12}{c}@{}}
\caption{Observed infall profile parameters from Fuller's measurements.}
\label{tab:fuller_settled}\\
\hline
Source
& \multicolumn{4}{c}{H$_2$CO}
& \multicolumn{4}{c}{HCO$^+$}
& \multicolumn{4}{c}{HNC}\\
& $P_{\rm r}$ & $v_{\rm sep}$ & $w_{1/2}$ & $D_{\rm r}$
& $P_{\rm r}$ & $v_{\rm sep}$ & $w_{1/2}$ & $D_{\rm r}$
& $P_{\rm r}$ & $v_{\rm sep}$ & $w_{1/2}$ & $D_{\rm r}$\\
\hline
SDC19A & 1.76(14) & 2.40(16) & 5.08(16) & 0.49(5) & -- & -- & -- & -- & -- & -- & -- & --\\
SDC19C & 1.58(22) & 1.40(31) & 3.44(27) & 0.58(8) & -- & -- & -- & -- & -- & -- & -- & --\\
SDC19E & 1.21(8) & 2.00(47) & 7.55(19) & 0.78(6) & 1.50(12) & 5.10(13) & 16.05(24) & 0.50(4) & 1.85(8) & 5.30(38) & 10.55(22) & 0.46(3)\\
SDC19H & -- & -- & -- & -- & 2.60(22) & 3.80(26) & 3.97(12) & 0.27(5) & -- & -- & -- & --\\
SDC33A & -- & -- & -- & -- & -- & -- & -- & -- & 1.83(36) & 2.40(38) & 2.43(28) & 0.27(9)\\
SDC33B & -- & -- & -- & -- & -- & -- & -- & -- & 1.20(11) & 1.20(13) & 2.99(14) & 0.75(8)\\
SDC33C & -- & -- & -- & -- & 1.56(24) & 1.50(26) & 3.86(47) & 0.33(13) & -- & -- & -- & --\\
SDC33E & -- & -- & -- & -- & -- & -- & -- & -- & 1.07(9) & 0.70(25) & 3.10(12) & 0.87(7)\\
SDC33G[v2] & -- & -- & -- & -- & 1.71(21) & 3.00(26) & 5.57(43) & 0.33(8) & 1.75(20) & 2.40(13) & 3.57(21) & 0.21(9)\\
SDC33H[v1] & -- & -- & -- & -- & -- & -- & -- & -- & 1.44(10) & 2.60(25) & 4.28(9) & 0.54(5)\\
SDC33H[v2] & -- & -- & -- & -- & 2.78(74) & 3.59(26) & 4.07(20) & 0.42(8) & 2.28(55) & 1.77(25) & 4.45(23) & 0.50(10)\\
SDC33J & -- & -- & -- & -- & 1.59(25) & 2.80(51) & 4.37(52) & 0.42(15) & 1.50(32) & 2.40(50) & 3.27(30) & 0.50(13)\\
SDC33K & -- & -- & -- & -- & 2.56(200) & 1.54(13) & 2.59(31) & 0.81(28) & -- & -- & -- & --\\
SDC33O & -- & -- & -- & -- & 1.19(5) & 1.30(26) & 2.94(6) & 0.81(4) & 1.38(9) & 1.20(25) & 2.49(10) & 0.64(5)\\
SDC33R & -- & -- & -- & -- & 1.08(9) & 1.60(13) & 3.21(8) & 0.67(7) & 1.20(11) & 1.60(25) & 3.09(9) & 0.67(8)\\
SDC35E & -- & -- & -- & -- & 2.14(8) & 2.50(26) & 3.86(6) & 0.44(2) & 3.14(26) & 2.52(20) & 2.60(105) & 0.34(3)\\
SDC35G[v1] & -- & -- & -- & -- & 1.40(21) & 1.90(26) & 4.69(61) & 0.57(11) & 1.25(25) & 1.60(13) & 3.15(32) & 0.60(13)\\
SDC35G[v2] & -- & -- & -- & -- & 2.00(19) & 3.90(26) & 3.09(29) & 0.19(5) & 2.56(24) & 2.50(38) & 5.12(25) & 0.35(4)\\
SDC35I & -- & -- & -- & -- & 1.67(8) & 3.40(26) & 5.68(15) & 0.07(4) & 1.44(9) & 2.40(38) & 4.56(13) & 0.65(5)\\
SDC35K & 2.33(48) & 4.00(79) & 4.38(33) & 0.29(11) & -- & -- & -- & -- & -- & -- & -- & --\\
SDC35M & 1.29(18) & 4.10(94) & 7.03(32) & 0.56(7) & -- & -- & -- & -- & 6.05(151) & 6.81(38) & 4.71(9) & 0.23(4)\\
SDC35N & 1.15(10) & 1.70(31) & 5.43(16) & 0.80(6) & -- & -- & -- & -- & -- & -- & -- & --\\
SDC35P & -- & -- & -- & -- & 1.25(9) & 2.05(2) & 3.69(10) & 0.83(6) & 1.76(21) & 2.27(25) & 4.13(15) & 0.82(7)\\
SDC35Q & -- & -- & -- & -- & 1.15(10) & 2.80(38) & 4.51(14) & 0.33(6) & 1.57(19) & 2.80(25) & 3.45(17) & 0.45(7)\\
SDC35T[v1] & -- & -- & -- & -- & 1.08(11) & 2.60(26) & 4.02(16) & 0.57(6) & 1.33(13) & 2.30(76) & 3.78(17) & 0.58(8)\\
SDC35T[v2] & -- & -- & -- & -- & 1.88(25) & 1.20(13) & 1.82(13) & 0.33(7) & 5.00(114) & 1.30(0) & 0.86(6) & 0.15(5)\\
SDC37B & -- & -- & -- & -- & 3.40(47) & 3.10(38) & 1.97(13) & 0.18(6) & -- & -- & -- & --\\
SDC37G & -- & -- & -- & -- & 4.00(45) & 3.90(13) & 3.63(13) & 0.15(3) & 3.80(38) & 2.90(38) & 2.89(13) & 0.16(5)\\
SDC37H & -- & -- & -- & -- & 2.00(29) & 3.10(38) & 5.05(25) & 0.25(9) & -- & -- & -- & --\\
SDC37I & -- & -- & -- & -- & 1.60(24) & 2.00(26) & 2.71(17) & 0.37(10) & 1.25(31) & 1.00(25) & 2.01(22) & 0.60(17)\\
SDC37K & -- & -- & -- & -- & 6.00(57) & 3.80(38) & 2.70(4) & 0.14(2) & -- & -- & -- & --\\
\hline
\end{longtable}
\endgroup
\noindent Notes. $P_{\rm r}=T_{\rm blue}/T_{\rm red}$, $v_{\rm sep}=V_{\rm red}-V_{\rm blue}$, and $D_{\rm r}=T_{\rm dip}/T_{\rm blue}$. The width $w_{1/2}=V_{1/2,{\rm r}}-V_{1/2,{\rm b}}$ is the separation between the outer blue- and red-side crossings at $T_{\rm red}/2$. Values in parentheses give the $1\sigma$ uncertainty in the last quoted digit(s). The parameters $P_{\rm r}$ and $D_{\rm r}$ are dimensionless, whereas $v_{\rm sep}$ and $w_{1/2}$ are in km\,s$^{-1}$. A parenthesised zero indicates that the 16th and 84th percentiles take the same discrete value at the processed channel sampling.
\clearpage

\section{Gaussian fitting results for H$^{13}$CO$^+$ and HN$^{13}$C}
\label{app:gaussian}
The full Gaussian-fitting results for the optically thinner tracers are provided in Table \ref{tab:h13cop_hn13c_gaussian_parameters}. These include the integrated intensity, $V_{\rm lsr}$, $\Delta V$, and $T_{\rm peak}$ for the fitted velocity components.

\begingroup \scriptsize \setlength{\tabcolsep}{1.6pt} \renewcommand{\arraystretch}{0.88} \begin{longtable}{@{}>{\raggedright\arraybackslash}p{0.075\textwidth} *{8}{>{\centering\arraybackslash}p{0.096\textwidth}}@{}} \caption{Gaussian-fit parameters for H$^{13}$CO$^+$ and HN$^{13}$C.} \label{tab:h13cop_hn13c_gaussian_parameters}\\ \hline\hline Source & \multicolumn{4}{c}{H$^{13}$CO$^+$} & \multicolumn{4}{c}{HN$^{13}$C} \\ \cline{2-5}\cline{6-9} &
$W$ &
$V_{\rm LSR}$ &
$\Delta V$ &
$T_{\rm peak}$ &
$W$ &
$V_{\rm LSR}$ &
$\Delta V$ &
$T_{\rm peak}$ \\
&
(K km s$^{-1}$) &
(km s$^{-1}$) &
(km s$^{-1}$) &
(K) &
(K km s$^{-1}$) &
(km s$^{-1}$) &
(km s$^{-1}$) &
(K) \\
\hline
\endfirsthead

\caption[]{Gaussian-fit parameters for H$^{13}$CO$^+$ and HN$^{13}$C -- continued.}\\
\hline\hline
Source &
\multicolumn{4}{c}{H$^{13}$CO$^+$} &
\multicolumn{4}{c}{HN$^{13}$C} \\
\cline{2-5}\cline{6-9}
&
$W$ &
$V_{\rm LSR}$ &
$\Delta V$ &
$T_{\rm peak}$ &
$W$ &
$V_{\rm LSR}$ &
$\Delta V$ &
$T_{\rm peak}$ \\
&
(K km s$^{-1}$) &
(km s$^{-1}$) &
(km s$^{-1}$) &
(K) &
(K km s$^{-1}$) &
(km s$^{-1}$) &
(km s$^{-1}$) &
(K) \\
\hline
\endhead

SDC19A & 6.82(14) & 26.54(3) & 2.90(7) & 2.21 & 4.48(12) & 26.72(4) & 2.93(10) & 1.44 \\
SDC19B & 2.68(11) & 26.22(4) & 2.15(10) & 1.17 & 2.39(10) & 26.45(5) & 2.04(10) & 1.10 \\
SDC19C & 2.59(12) & 27.05(5) & 2.47(11) & 0.99 & 2.29(12) & 27.12(5) & 2.02(13) & 1.06 \\
SDC19D & 0.40(12) & 122.20(26) & 1.62(46) & 0.23 & -- & -- & -- & -- \\
SDC19E & 11.52(27) & 19.51(7) & 6.28(16) & 1.72 & 4.55(75) & 18.56(35) & 4.60(71) & 0.93 \\
SDC19F & 1.31(13) & 26.62(6) & 1.32(15) & 0.94 & -- & -- & -- & -- \\
SDC19G & 3.34(15) & 113.06(6) & 2.84(15) & 1.11 & 1.25(13) & 113.45(10) & 1.86(21) & 0.63 \\
SDC19H & 1.49(18) & 23.06(16) & 2.68(37) & 0.52 & 1.21(16) & 23.47(23) & 3.22(39) & 0.35 \\
SDC33A & 0.85(14) & 103.34(12) & 1.69(38) & 0.47 & 0.76(10) & 103.56(7) & 1.09(16) & 0.65 \\
SDC33B & 0.49(11) & 84.43(36) & 3.14(76) & 0.14 & -- & -- & -- & -- \\
SDC33C & 0.30(11) & 86.32(62) & 3.06(90) & 0.09 & -- & -- & -- & -- \\
SDC33E & 0.55(11) & 73.90(29) & 2.76(62) & 0.19 & -- & -- & -- & -- \\
SDC33G & 0.68(17) & 74.35(43) & 2.98(77) & 0.21 & -- & -- & -- & -- \\
SDC33H & 1.00(20) & 74.40(24) & 3.07(90) & 0.31 & 0.67(17) & 74.27(39) & 2.23(87) & 0.29 \\
SDC33M & 1.85(10) & 98.75(5) & 1.95(12) & 0.89 & 1.36(10) & 98.96(7) & 1.90(14) & 0.67 \\
SDC33N & 0.79(10) & 98.42(10) & 1.51(22) & 0.49 & -- & -- & -- & -- \\
SDC33O & 1.11(11) & 84.86(9) & 1.95(23) & 0.53 & -- & -- & -- & -- \\
SDC33P & 2.02(10) & 85.27(3) & 1.52(9) & 1.25 & 0.65(7) & 85.33(5) & 0.98(12) & 0.63 \\
SDC33Q & 2.99(8) & 9.31(1) & 0.98(3) & 2.86 & 0.65(8) & 9.42(8) & 1.20(19) & 0.51 \\
SDC33R & 1.66(13) & 83.24(8) & 2.07(18) & 0.75 & 1.25(16) & 83.29(12) & 2.15(42) & 0.55 \\
SDC35A & 1.90(18) & 28.02(6) & 1.39(17) & 1.28 & 1.15(18) & 28.31(11) & 1.42(25) & 0.76 \\
SDC35B & 0.53(13) & 45.19(11) & 0.92(29) & 0.53 & -- & -- & -- & -- \\
SDC35C[v1] & 0.94(15) & 50.13(14) & 1.78(33) & 0.50 & -- & -- & -- & -- \\
SDC35C[v2] & 0.74(12) & 53.64(10) & 1.18(23) & 0.59 & -- & -- & -- & -- \\
SDC35E & 2.91(10) & 45.83(4) & 2.10(9) & 1.30 & 2.09(10) & 46.05(6) & 2.26(12) & 0.87 \\
SDC35F[v1] & 1.30(11) & 42.83(6) & 1.42(15) & 0.85 & 0.75(9) & 42.83(7) & 1.19(15) & 0.59 \\
SDC35F[v2] & 2.00(11) & 45.37(5) & 1.64(10) & 1.14 & 1.45(10) & 45.69(5) & 1.64(12) & 0.87 \\
SDC35G & 1.91(11) & 55.51(7) & 2.56(17) & 0.70 & 0.61(9) & 55.43(11) & 1.58(24) & 0.36 \\
SDC35H & 2.03(13) & 53.61(7) & 2.33(17) & 0.82 & 0.98(14) & 53.68(20) & 2.79(39) & 0.33 \\
SDC35I & 2.27(13) & 64.41(9) & 3.19(21) & 0.67 & -- & -- & -- & -- \\
SDC35J & 0.96(12) & 54.04(13) & 2.14(29) & 0.42 & -- & -- & -- & -- \\
SDC35L & 4.86(19) & 52.54(10) & 5.09(22) & 0.90 & 1.49(16) & 52.62(22) & 4.02(49) & 0.35 \\
SDC35M[v1] & 2.76(21) & 53.31(13) & 3.53(31) & 0.73 & 1.39(13) & 53.43(17) & 3.34(42) & 0.39 \\
SDC35M[v2] & 1.31(20) & 57.56(21) & 2.88(47) & 0.43 & 0.26(7) & 57.86(9) & 0.68(24) & 0.35 \\
SDC35N & 3.45(16) & 50.05(10) & 4.45(25) & 0.73 & 2.01(14) & 50.08(14) & 3.91(33) & 0.49 \\
SDC35O & 1.48(10) & 95.14(7) & 2.03(15) & 0.68 & 0.99(10) & 95.31(11) & 2.05(21) & 0.46 \\
SDC35P & 1.45(11) & 92.56(8) & 2.20(21) & 0.62 & 0.51(9) & 92.98(8) & 1.10(26) & 0.43 \\
SDC35Q & 0.80(11) & 94.40(13) & 1.67(33) & 0.45 & -- & -- & -- & -- \\
SDC35R & 0.99(10) & 76.12(13) & 2.47(28) & 0.38 & -- & -- & -- & -- \\
SDC35S & 2.20(13) & 77.92(6) & 2.07(14) & 1.00 & 1.32(13) & 77.91(8) & 1.86(22) & 0.67 \\
SDC35T[v1] & 1.30(11) & 49.45(9) & 1.92(18) & 0.64 & -- & -- & -- & -- \\
SDC35T[v2] & 0.71(8) & 59.68(3) & 0.67(8) & 1.00 & -- & -- & -- & -- \\
SDC37A & 0.95(11) & 73.32(7) & 1.31(16) & 0.68 & 0.75(11) & 73.47(9) & 1.17(19) & 0.60 \\
SDC37B & 1.55(13) & 72.47(10) & 2.41(27) & 0.60 & 0.75(11) & 72.62(9) & 1.29(24) & 0.55 \\
SDC37C & 1.13(10) & 72.41(8) & 1.71(15) & 0.62 & 1.15(11) & 73.01(9) & 1.81(19) & 0.60 \\
SDC37D & 2.59(12) & 64.85(5) & 2.21(12) & 1.10 & 1.19(12) & 64.96(11) & 2.06(24) & 0.55 \\
SDC37E & 0.89(10) & 66.64(6) & 1.22(19) & 0.69 & 0.55(9) & 66.78(9) & 1.24(21) & 0.41 \\
SDC37F & 0.85(11) & 66.81(12) & 1.91(28) & 0.41 & -- & -- & -- & -- \\
SDC37G & 0.78(9) & 68.13(9) & 1.67(23) & 0.44 & -- & -- & -- & -- \\
SDC37H & 0.22(5) & 67.88(6) & 0.50(13) & 0.42 & -- & -- & -- & -- \\
SDC37K & 1.59(12) & 54.98(10) & 2.91(25) & 0.51 & 1.02(15) & 55.77(22) & 2.47(77) & 0.28 \\
\hline
\multicolumn{9}{@{}p{0.9\textwidth}@{}}{\footnotesize Notes. $W$ is the integrated intensity. Values in parentheses indicate the uncertainty in the last quoted digit(s). A dash indicates that no reliable Gaussian fit is available for the corresponding molecular tracer.}\\
\end{longtable}
\endgroup

\clearpage

\section{Source-by-source $\delta V$ classifications}
\label{app:deltaV}
The source-by-source $\delta V$ classifications are listed in Table~\ref{tab:deltaV_by_source_revised}. Sources without a velocity-matched optically thinner reference component are not included in this table.

\begingroup
\scriptsize
\setlength{\tabcolsep}{2pt}
\renewcommand{\arraystretch}{0.88}
\setlength{\LTleft}{0.18\textwidth}
\setlength{\LTright}{\fill}
\begin{longtable}{@{} >{\raggedright\arraybackslash}p{0.18\textwidth} >{\centering\arraybackslash}p{0.16\textwidth} >{\centering\arraybackslash}p{0.16\textwidth} >{\centering\arraybackslash}p{0.16\textwidth} @{}}
\caption{Normalised velocity differences used in the blue-excess census.}
\label{tab:deltaV_by_source_revised}\\
\hline\hline
Source & H$_2$CO & HCO$^+$ & HNC \\ \hline
SDC19A & $-0.32(6)$ (B) & -- & -- \\
SDC19B & $0.26(8)$ (R) & -- & -- \\
SDC19C & $-0.14(7)$ (N) & -- & -- \\
SDC19E & $-0.27(3)$ (B) & $0.51(3)$ (R) & $-0.17(9)$ (N) \\
SDC19G & -- & $-0.09(5)$ (N) & $-0.22(9)$ (N) \\
SDC19H & $0.05(8)$ (N) & $-0.32(9)$ (B) & $-0.07(8)$ (N) \\
SDC33A & -- & -- & $-0.79(18)$ (B) \\
SDC33B & -- & $0.08(12)$ (N) & -- \\
SDC33C & -- & $-0.10(21)$ (N) & -- \\
SDC33E & -- & $0.17(12)$ (N) & -- \\
SDC33G[v1] & -- & $-0.13(15)$ (N) & -- \\
SDC33H[v1] & -- & $-0.07(9)$ (N) & $-0.30(22)$ (B) \\
SDC33M & -- & $0.94(9)$ (R) & $0.01(8)$ (N) \\
SDC33N & -- & $0.47(13)$ (R) & -- \\
SDC33O & -- & $-0.29(9)$ (B) & -- \\
SDC33P & -- & $0.12(9)$ (N) & $0.06(14)$ (N) \\
SDC33Q & $0.02(16)$ (N) & $0.12(13)$ (N) & $-0.09(13)$ (N) \\
SDC33R & -- & $-0.31(8)$ (B) & $-0.37(11)$ (B) \\
SDC35A & $0.13(12)$ (N) & $0.49(12)$ (R) & $-0.06(12)$ (N) \\
SDC35B & $-0.32(23)$ (B) & $-0.40(22)$ (B) & -- \\
SDC35C[v1] & -- & $-0.20(11)$ (N) & -- \\
SDC35C[v2] & -- & $-0.21(14)$ (N) & -- \\
SDC35E & $0.00(8)$ (N) & $-0.44(7)$ (B) & $-0.38(7)$ (B) \\
SDC35F[v1] & -- & $0.10(10)$ (N) & $-0.19(12)$ (N) \\
SDC35F[v2] & $-0.71(11)$ (B) & $-0.45(9)$ (B) & $-0.53(9)$ (B) \\
SDC35G[v2] & $-0.13(7)$ (N) & $-0.55(7)$ (B) & $-0.46(13)$ (B) \\
SDC35H & $-0.30(8)$ (B) & $-0.42(7)$ (B) & $-0.07(9)$ (N) \\
SDC35I & $-0.08(6)$ (N) & $-0.54(6)$ (B) & -- \\
SDC35J & $0.30(10)$ (R) & $0.02(9)$ (N) & -- \\
SDC35L & $0.11(4)$ (N) & -- & $0.20(7)$ (N) \\
SDC35M[v1] & $-0.06(6)$ (N) & $-0.04(5)$ (N) & $-0.04(6)$ (N) \\
SDC35M[v2] & -- & $1.36(24)$ (R) & -- \\
SDC35N & $-0.06(4)$ (N) & $-0.43(4)$ (B) & $-0.21(5)$ (N) \\
SDC35O & -- & $0.21(7)$ (N) & $0.01(8)$ (N) \\
SDC35P & -- & $-0.44(8)$ (B) & $-1.07(29)$ (B) \\
SDC35Q & -- & $-0.36(13)$ (B) & -- \\
SDC35R & -- & $-0.07(7)$ (N) & -- \\
SDC35S & -- & $-0.25(7)$ (N) & $-0.07(8)$ (N) \\
SDC35T & $-0.08(9)$ (N) & -- & -- \\
SDC35T[v1] & -- & $-0.81(11)$ (B) & -- \\
SDC35T[v2] & -- & $-0.27(20)$ (B) & -- \\
SDC37A & -- & -- & $-0.17(14)$ (N) \\
SDC37B & -- & $-0.24(7)$ (N) & $-0.36(14)$ (B) \\
SDC37C & -- & $0.15(9)$ (N) & $-0.23(9)$ (N) \\
SDC37D & $-0.10(7)$ (N) & $0.07(6)$ (N) & $-0.04(8)$ (N) \\
SDC37E & $0.11(14)$ (N) & $-0.20(12)$ (N) & $-0.13(13)$ (N) \\
SDC37F & -- & $0.01(9)$ (N) & -- \\
SDC37G & -- & $-0.50(12)$ (B) & -- \\
SDC37H & -- & $-2.56(72)$ (B) & -- \\
SDC37K & $0.15(7)$ (N) & $0.08(6)$ (N) & $-0.16(11)$ (N) \\
\hline
\end{longtable}
\vspace{-1.5ex}
\noindent\hspace*{0.17\textwidth}%
\begin{minipage}{0.70\textwidth}
\footnotesize Notes. Entries give $\delta V=(V_{\rm thick}-V_{\rm thin})/\Delta V_{\rm thin}$. Values in parentheses give the propagated $1\sigma$ uncertainty in the last quoted digit(s), including the Gaussian-fit uncertainties in $V_{\rm thin}$ and $\Delta V_{\rm thin}$ and an uncertainty of half a smoothed spectral channel in $V_{\rm thick}$. B, R, and N denote blue, red, and neutral profiles, respectively. Classifications are based on the central values. The optically thinner reference-line velocities and linewidths are listed in Table~\ref{tab:h13cop_hn13c_gaussian_parameters}. For H$_2$CO, H$^{13}$CO$^+$ is used as a proxy optically thinner reference tracer. A dash indicates that the corresponding tracer is not included in the census.
\end{minipage}
\endgroup

\clearpage


\bsp	
\label{lastpage}
\end{document}